\documentclass[11pt]{amsart}
 
\usepackage{subcaption}
\usepackage[export]{adjustbox}
\usepackage{setspace,kantlipsum}
\DeclareMathOperator{\diag}{diag}
\usepackage{bm}
\usepackage{tikz}
\usepackage{multirow, makecell}
\usepackage{subcaption}
\usepackage{booktabs}
\usepackage[flushleft]{threeparttable}
\usepackage[foot]{amsaddr}
\usepackage{mathptmx}
\usepackage{setspace}
\usepackage{color,colortbl}
\usepackage{lineno}
\theoremstyle{definition}

\theoremstyle{remark}

\numberwithin{equation}{section}

\begin{document}

\title{Direct and indirect transactions and requirements}
%\headers{}{}

%    Remove any unused author tags.

%    author one information
\author{Husna Betul Coskun}
\address{Department of Economics, University of Georgia, Athens, GA 30602}
%\address{Drexel University, LeBow College of Business, School of Economics, 3141 Chestnut Street, Philadelphia, PA 19104}
\curraddr{}
\email{BetulCoskun@uga.edu}
%\email{betul.coskun@drexel.edu}
\thanks{}

%    author two information
\author{Huseyin Coskun}
%\address{}
%\curraddr{}
\email{hsyncskndr@gmail.com}
\thanks{}

%\subjclass[2010]{Primary }

\keywords{
system decomposition theory; system partitioning; \texttt{diact} flows and storages; direct, indirect, and total transactions; direct, indirect, and total requirements coefficients; input-output economics; general equilibrium theory; mathematical economics; econometrics; multiplier theory}
%\keywords{
%system decomposition theory, system partitioning, complex systems theory, compartmental systems, \texttt{diact} flows and storages, direct, indirect, and total transactions, direct, indirect, and total requirements coefficients, input-output analysis, input-output economics, general equilibrium theory, theoretical economics, mathematical economics, econometrics, macroeconomics, ecological economics, socioeconomic systems, ecosystem ecology, quantitative finance, financial systems, ecological networks, control theory, multiplier theory, graph theory, information theory}
% JEL: C02, C67, C68, D57, D85, Q57, R15
% 91B74; 91B76; 37N40; 93C15; 92B20; 92D30; 92D40; 92C42; 94A15
%The proposed methodology solves an age-old open problem in economic system analysis. It defines the indirect transactions between any two sectors of the system. This novel concept is completely different than the current indirect effect formulations.

\date{}

\dedicatory{}

\begin{abstract}
%\onehalfspacing
%\doublespacing 
The indirect transactions between sectors of an economic system has been a long-standing open problem. There have been numerous attempts to conceptually define and mathematically formulate this notion in various other scientific fields in literature as well. The existing \textit{direct} and \textit{indirect effects} formulations, however, can neither determine the \textit{direct} and \textit{indirect transactions} separately nor quantify these transactions between two individual sectors of interest in a multisectoral economic system. The novel concepts of the \textit{direct, indirect} and \textit{transfer (total) transactions} between any two sectors are introduced, and the corresponding \textit{requirements matrices} and \textit{coefficients} are systematically formulated relative to both final demands and gross outputs based on the system decomposition theory in the present manuscript. It is demonstrated theoretically and through illustrative examples that the proposed requirements matrices accurately define and correctly quantify the corresponding direct, indirect, and total interactions and relationships. The proposed requirements matrices for the US economy using aggregated input-output tables for multiple years are then presented and briefly analyzed.
\end{abstract}

\maketitle

%\onehalfspacing
%\doublespacing

\section{Introduction}
\label{sec:intro}

The analysis of observable direct transactions is relatively straightforward even in complex economic systems. The indirect transactions, however, is a complicated concept that has derived attention in many scientific fields, such as economics, ecology, graph theory, and network theory, for the last several decades. We define \textit{direct relationships} as pairwise immediate interactions between two sectors in an economic system, and \textit{indirect relationships} as pairwise interactions between two sectors through other sectors. With the directness and indirectness throughout the manuscript, we will constantly be referring to these fundamental definitions. The main focus of this manuscript is to introduce the novel \textit{direct, indirect}, and \textit{transfer (total) transactions} concepts defined pairwise between any two sectors of a multisectoral economic system and explicitly formulate the corresponding \textit{requirements matrices} and \textit{coefficients} relative to both final demands and gross outputs.

Modeling interactions among industries and their interconnectedness goes back to the concept of the ``circular flow'' in an economy~\cite{Murphy1993}. This idea is related to Petty's concept of the interdependence of industries. Fran\c{c}ois Quesnay created the Tableau \'Economique (economic table) in which he depicted the idea of the economy as a circular flow of income and output among economic sectors. The table is known for its diagrammatic representation of how transactions can systematically be traced through an economic system~\cite{Meek1962,Stokes1992}. Achille-Nicholas Isnard is known to be the first person to represent the circular flow of income and expenditure as an algebraic system of equations~\cite{Kurz2000,Miller2009}. 

The Tableau \'Economique is considered the first method for the explicit conceptualization of the nature of economic equilibrium. It is also hailed as a forerunner of general equilibrium theory pioneered by L\'eon Walras~\cite{Walras1874,Meek1962,Stokes1992}. Walras used production coefficients that compared the required resources for a product and its total production~\cite{Miller2009}. Leontief's empirical economic studies were based on Quesnay's table and Walras's formulations of general equilibrium, although his conclusion was that an economy is never in equilibrium. He made the circular flow transactions into a table which then led to the founding of the analytical tool called the input-output model~\cite{Leontief1936}. The input-output analysis as we know today with contributions of many other economists analyzes intersectoral interactions in economic systems.

It is generally accepted that the input-output economics derive its significance largely from the fact that the total requirements coefficients measuring the combined effects of the direct and indirect repercussions of a change in final demands can easily be determined~\cite{Steenge1990}. There have been numerous definitions and corresponding mathematical formulations of the \textit{direct} and \textit{indirect effects} concepts separately in the literature for about a century since the development of the input-output model~\cite{Fath1999,Miller2009,ece1972,Parikh1975,Carter1970,Lancaster1968,Leontief1966,Alterman1965,BEA1963,Leontief1951,Jeong1984,Gim1998,Horowitz2009,Sancho2012}. Each of these attempts, however meaningful, does not seem to be accurately describing and correctly quantifying the indirect transactions. The existing approaches use the total effects formulation to define the direct and indirect effects. The total effects, however, can neither determine the \textit{direct} and \textit{indirect transactions} separately nor quantify these transactions pairwise between any two sectors individually.

A mathematical theory, known as the \textit{system decomposition theory}, and associated methodologies for the analysis of dynamic nonlinear compartmental systems was recently introduced by~\cite{Coskun2017DCSAM,Coskun2017NDP,Coskun2017DESM}. The static version of this theory has also been developed recently~\cite{Coskun2017SCSA,Coskun2017SESM}. The system decomposition partitions the system into subsystems, each of which separately represents all economic activities induced by an individual sector within the system.

The system decomposition theory improves and refines the current static, linear, sectoral level compartmental system analysis to the dynamic, nonlinear, subsectoral level. The system decomposition enables tracking the evolution of initial stocks, external inputs, and arbitrary intercompartmental flows of currency, goods, and services, as well as the associated storages derived from these stocks, inputs, and flows individually and separately within the system. The \textit{transient} and the \textit{direct, indirect, cycling, acyclic}, and \textit{transfer (total)} flows and associated storages along any given flow path or from one sector to another\textemdash along all paths\textemdash are also systematically formulated. The novel production (residence) time concept also incorporates the real time dimension into the analysis~\cite{Coskun2017SCSA,Coskun2017SESM}. In the present manuscript, we use these \textit{direct} and \textit{indirect flows} and \textit{distributions} notions to conceptually redefine and mathematically formulate the \textit{direct, indirect} and \textit{transfer (total) transactions} and \textit{requirements} in the context of multisectoral economic systems.

The conceptualization and redefinition of directness and indirectness are inseparable. We first define the \textit{direct transaction} between any two sectors as the total immediate pairwise flows between these sectors in an economic system. The total pairwise flows from one sector indirectly through other sectors to another will then be defined as the \textit{indirect transaction} between these two sectors. Therefore, unlike the direct and indirect effects notions, the direct and indirect transactions concepts introduce a direction from the source to the destination.
%~\cite{Coskun2021ESA}. 
For example, since steel is used in car production, each car purchase \textit{directly} from the automotive sector includes an \textit{indirect purchase} from the steel sector. The proposed methodology can separately quantify this indirect purchase specifically from the steel sector and, as a matter of fact, from any other individual sector of interest. The existing formulations in the literature, however, cannot determine and quantify such indirect purchases and transactions.
 
An immediate application of the proposed direct and indirect transactions and requirements concepts is the impact analysis. The direct and indirect requirements  matrices can respectively be used to separately determine how a disaggregated segment of final demand for the output of one sector directly and indirectly affects every other individual sector in the system. Considering a hypothetical economic system, if the final demand for the output of the automotive industry is cut in half, the direct and indirect implications of half the demand can separately be determined for the output of any individual industry in the car production chain through the proposed methodology. The existing formulations, however, cannot quantify how half the demand for the cars affects the direct and indirect purchases from the steel or any other individual sector of interest.

Following on the same hypothetical model, the corresponding amount of steel to make the cars in the first step, coal to produce the steel in the second step, energy to extract the coal in the third step, and so on, can be calculated, based on the given arbitrary segment of the final demand. In the existing formulations, the transaction between the steel and automotive industries in the first step is considered as the \textit{direct effect} of the change in the final demand. On the other hand, when this production chain cycles back to steel at any later production step again, the current methodologies consider the value of steel used to make cars at that step as an \textit{indirect effect} of the change (see Fig.~\ref{fig:hypothetical_the}). That is, the transactions between the same two sectors, steel and automotive industries, are inconsistently classified as the direct and indirect effects, solely based on the step number or order of propagation of the final demand within the system. 

From a different perspective, the proposed definitions of the direct and indirect transactions are formulated based on the sectoral interactions and the existing definitions of the direct and indirect effects are in reference to final demands. It can be seen that the existing indirect effects are formulated without actually defining the indirect transactions between any two sectors in an economic system. The indirect effects are considered to be the total transaction carried by all subsequent steps after the first entrance of goods and services into each sector in the existing formulations. Therefore, even the immediate transactions between two sectors of interest after the first step in their interactions are considered as indirect in these approaches. The indirect effects are, therefore, implicitly defined microscopic quantities with limited practical use and cannot quantify indirect sectoral interactions. In contrast, fundamentally different from this classification, the system decomposition theory defines and explicitly formulates the direct and indirect transactions as separate measurable physical quantities based on the nature of sectoral relationships. The direct and indirect transactions disregard the order of propagation in intersectoral interactions in potentially circular production chains.

The critical theoretical differences in the proposed and existing formulations lie in the conceptualization of the notion of propagation and sectoral interactions. The existing formulations consider the propagation steps as simultaneous and synchronous rounds of production, as exemplified earlier. Since the total effects are formulated using geometric series with infinite terms, and the production times of sectors are different from each other, this prevalent interpretation is inaccurate and unrealistic. Needless to say, the idea of the infinite rounds of production in a base year and the assumption of the same production time for the service and automotive or agriculture sector are not plausible. The system decomposition theory, however, incorporates the real time dimension into the analysis through the production time concept and assumes different production time for each sector. Moreover, unlike the input-output economics, the system decomposition theory interprets the steps of propagation as steps in ``computational time'', rather than in real time.

The system decomposition theory defines the direct and indirect flows as complementary flows. The transfer (total) flows are accordingly defined as the sum of the direct and indirect flows. The corresponding direct, indirect, and transfer (total) flow distributions are defined relative to gross outputs in the context of the system decomposition theory~\cite{Coskun2017SCSA}. In economic system analysis, however, the requirements matrices relative to final demands are more desirable for many cases. The \textit{direct, indirect}, and \textit{total requirements matrices}, as well as the corresponding \textit{transactions matrices} at both sectoral and subsectoral levels are systematically defined and explicitly formulated relative to both gross outputs and final demands in the present manuscript for the first time in literature. These requirements matrices are expressed in terms of the make-use framework as well. 

The requirements matrices are defined as the scaled versions of the transactions matrices. The \textit{direct, indirect}, and \textit{total transactions matrices} represent the corresponding demand distributions induced by gross outputs and final demands. The requirements matrices relative to gross outputs are called \textit{composite requirements matrices}, and the ones relative to final demands are called the \textit{simple requirements matrices}. The elements of these requirements matrices will be called the \textit{direct, indirect}, and \textit{transfer (total) requirements coefficients}. The simple requirements matrices can be considered as the measures for the exogenous impacts on the sectors, while their composite counterparts as those for the intersectoral dynamics. 

There are two requirements matrices explicitly formulated and widely used in economic system analysis: the direct and total requirements matrices. The simple transfer (total) requirements matrix is different from the existing total requirements matrix in a way that it covers the internal workings of the system by considering only the producing sectors and excluding final demands from the coefficients. Similar suggestions in regard to such exclusion are also made in the literature~\cite{Mesnard2002b,Dietzenbacher2005b}. Both the simple transfer (total) and existing total requirements matrices are formulated relative to final demands. The composite transfer (total) flow concept is equivalent to the total flow definition introduced by~\cite{Szyrmer1987}. In the context of the total flows, the economic implications of the total requirements matrix relative to gross outputs instead of final demands is also discussed in the literature~\cite{Jeong1984,Milana1985,Szyrmer1992}.

The composite direct requirements matrix and the existing direct requirements or coefficient matrix are also the same. The difference between the simple direct requirements matrix and the coefficient matrix in terms of propagation is that the former yields the total direct transaction between any two sectors of the system regardless of the order of propagation, while the latter yields one step propagation within the system relative to final demands. The simple direct requirements matrix is defined relative to final demands, and the existing direct requirements matrix is defined relative to gross outputs.

The requirements tables are mainly used for the \textit{impact} and \textit{policy analyses} as exemplified earlier. The economic repercussions of a given portion of final demands within the system is critical information for policy and business planning. The existing formulations are designed for only the gross, lump-sum impacts of the given segments on the system. The proposed simple and composite direct and indirect requirements matrices, however, separately determine the total direct and indirect responses by an individual sector of interest to a disaggregated segment of a specific sector's final demand or gross output within an economy.

It is worth to emphasize that the system decomposition theory constructs the impact analysis as the repercussions of disaggregated segments of final demands and gross outputs on individual sectors rather than those of changes in these demands and outputs. Since assuming the constancy of coefficients while changing final demands is contradictory, the prevalent interpretation of the impact analysis in the context of the system response to exogenous changes is unrealistic for static systems. The system decomposition theory introduces the impact analysis in the context of the influence of exogenous changes on the system in the dynamic setting.

The simple direct and indirect coefficients can, for example, be used in emergency planning, such as separately estimating the direct and indirect effects that a portion of the petroleum demand would have on the production in each sector individually. A standard policy analysis problem is to investigate the implications of a new governmental policy change that impacts final demands for an economy in terms of interindustry production generated in response to the change. The simple direct and indirect requirements coefficients enable the analysis of the separate direct and indirect impacts of such programs targeting a specific sector on any other individual sector of interest. For example, the direct and indirect impacts of the government tax policy aimed at smaller consumer demand for a particular product individually on any other product within the system can separately be determined using the simple direct and indirect coefficients. Such thorough and in depth analyses are not possible through the state-of-the-art techniques.

The United Nations and most of the governments of the industrialized countries including the United States are currently using the input-output data to measure and analyze their national economic systems. The case studies at the end of the manuscript demonstrate that the proposed direct, indirect, and total transactions and requirements concepts accurately capture the corresponding interactions and relationships between sectors of economic systems and provide additional critical statistics that is not available through the existing formulations. The proposed concepts are applied to the aggregated US input-output data for multiple years to demonstrate their practicality and efficiency. The numerical results and their graphical representations for the simple direct, indirect, and total requirements matrices using these real data sets are also presented and briefly analyzed in the case studies.

\section{Methods}
\label{sec:methods}

In this section, the fundamental relationships of input-output economics are summarized. The conceptual developments and methodological advancements brought by the system decomposition theory are also briefly discussed. The novel \textit{simple} and \textit{composite direct, indirect}, and \textit{transfer (total) transactions} and \textit{requirements matrices} and \textit{coefficients} are then systematically introduced based on this theory. The differences between the proposed and existing concepts and formulations are theoretically explained and methodically demonstrated. The potential applications of the proposed methodology are outlined at the end of the section.
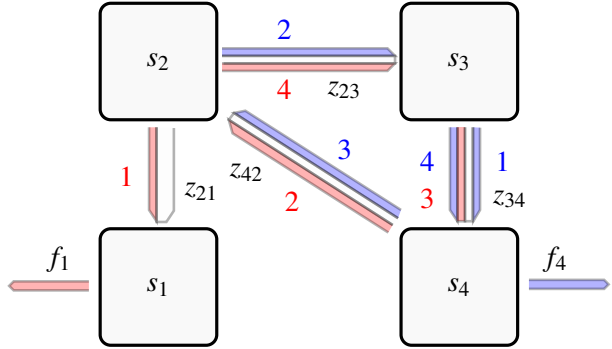
\begin{figure}[t]
\centering
\begin{tikzpicture}
\draw[very thick,  fill=gray!5, draw=black, rounded corners] (-.05,-.05) rectangle
node(R1) {$s_1$} (1.5,1.5) ;
\draw[very thick,  fill=gray!5, draw=black, rounded corners] (3.95,-.05) rectangle 	node(R2) {$s_4$} (5.5,1.5) ;     
\draw[very thick,  fill=gray!5, draw=black, rounded corners] (-.05,2.95) rectangle 	node(R3) {$s_2$} (1.5,4.5) ;  
\draw[very thick,  fill=gray!5, draw=black, rounded corners] (3.95,2.95) rectangle node(R3) {$s_3$} (5.5,4.5) ;    
\draw[thick, fill=red, line width=1pt, draw=black, opacity=.3]  (-.2,.8) -- ++ (-1,0) -- ++ (-.06,-.06) -- ++ (.06,-.06) -- ++ (1,0);
 \node (p) [text=black, opacity=1] at (-.6,1.1) {$f_1$};
\draw[thick, fill=blue, line width=1pt, draw=black, opacity=.3]  (5.65,.7) -- ++ (1,0) -- ++ (.06,.06) -- ++ (-.06,.06) -- ++ (-1,0);	 
\node (p) [text=black, opacity=1] at (6,1.1) {$f_4$};
\draw[thick, fill=blue, line width=1pt, draw=black, opacity=.3]  (4.6,2.85) -- ++ (0,-1.1) -- ++ (.1,-.15)  -- ++ (0,1.25) ; %z43 far left
\node (p) [text=blue, opacity=1] at (4.3,2.4) {$4$};
\draw[thick, fill=red, line width=1pt, draw=black, opacity=.3] (.6,2.85) -- ++(0,-1.1) -- ++ (.1,-.15) -- ++ (0,1.25) ;%z21 left
\draw[thick, fill=gray!5, line width=1pt, draw=black, opacity=.3] (.7,2.85) -- ++(0,-1.26) -- ++ (.13,0) -- ++ (.1,.15)  -- ++ (0,1.1) ;%z21 right
\node (p) [text=red, opacity=1] at (0.3,2.2) {$1$};	
\draw[thick, fill=blue, line width=1pt, draw=black, opacity=.3]  (1.57,3.9) -- ++ 		(2.2,0) -- ++ (.1,-.1) -- ++ (-2.3,0) ; %z32 top
\node (p) [text=blue, opacity=1] at (2.4,4.15) {$2$};
\draw[thick, fill=red, line width=1pt, draw=black, opacity=.3] (3.84, 1.55) -- ++ (-2.18, 1.4)  --++ (0.05,-.18) -- ++ (2.07,-1.32) ; %z42 bottom
\node (p) [text=red] at (2.5,1.87) {$2$};  
\draw[thick, fill=blue, line width=1pt, draw=black, opacity=.3]  (3.95,1.75) -- ++ 		(-2.07,1.32) -- ++ (-0.15,-.03) -- ++ (2.18,-1.4) ; %z42 top
\node (p) [text=blue] at (3.2,2.55) {$3$};               
\draw[thick, fill=red, line width=1pt, draw=black, opacity=.3]  (4.7,2.85) -- ++(0,-1.25) -- ++ (.1,0) -- ++ (0,1.25)  ;  %z43 middle left
\node (p) [text=red, opacity=1] at (4.3,1.9) {$3$};    
\draw[thick, fill=red, line width=1pt, draw=black, opacity=.3]  (1.57,3.7) -- ++ 		(2.3,0) -- ++ (-.1,-.1) -- ++ (-2.2,0);	%z32 bottom 
\node (p) [text=red, opacity=1] at (2.4,3.35) {$4$};	 
\draw[thick, fill=blue, line width=1pt, draw=black, opacity=.3]  (4.91,2.85) -- ++ 	(0,-1.25) -- ++ (.09,.15)-- ++ (0,1.095)  ;  %z43 far right
\node (p) [text=blue, opacity=1] at (5.3,2.4) {$1$};              
\draw[thick, fill=gray!5, line width=1pt, draw=black, opacity=.3]  (4.8,2.85) -- ++ 	(0,-1.25) -- ++(.1,0) -- ++ (0,1.25)   ;  %z43 middle right
\node (p) [text=black, opacity=1] at (5.4,1.9) {$z_{34}$}; 
\node (p) [text=black] at (1.3,2) {$z_{21}$};
\draw[thick, fill=gray!5, line width=1pt, draw=black, opacity=.3]  (1.57,3.8) -- ++ (2.3,0) -- ++  (0,-.1) -- ++ (-2.28,0);	%z32 middle 
\node (p) [text=black] at (3.2,3.35) {$z_{23}$};
\draw[thick, fill=gray!5, line width=1pt, draw=black, opacity=.3]  (3.91,1.64) -- ++ (-2.18, 1.4) -- ++ (-.07,-.09) -- ++ (2.18,-1.4) ; %z24 middle
\node (p) [text=black] at (1.9,2.25) {$z_{42}$};  
\end{tikzpicture}
\caption{Schematic representation of the indirect transactions and indirect effects. The sector $i$ is denoted by $s_i$ in the figure. The numbers next to arrows represent the step numbers or order, $n$, in the geometric series expansion of the total requirements matrix, $L$, given in Eq.~\ref{eq:fd_rounds}. In the existing formulations, the flow segments labeled with the power of the first order term, $A {f}$, $n=1$, in both colors are generally considered as the direct effects, and with the powers of all higher order terms, $A^n {f}$, $n>1$, as the indirect effects. The unshaded flow segments represent the sum of the flow segments generated by all the remaining higher order terms of propagation, $n>4$. In the context of the system decomposition theory, however, $z_{ik}$ represents the composite direct transactions from sector $i$ to $k$. Both the blue-shaded flow segments labeled with $1$ and $4$ (cycling flow at $s_3$) within $z_{34}$ represent the portions of the simple direct transaction from $s_3$ to $s_4$. A flow segment initiated at a sector and transmitted through other sectors to another is then defined as the indirect transaction. Only the initial flow segments at each step are depicted. The red-shaded flow segment labeled with 4, for example, reaches sector 1 to exit as a portion of $f_1$ in 4 steps: $s_2 \rightarrow s_3 \rightarrow s_4 \rightarrow s_2 \rightarrow s_1$. The red-shaded flow segments labeled with $2$, $3$, and $4$ within $z_{42}$, $z_{34}$, and $z_{23}$ are the portions of the indirect transactions from $s_4$ to $s_1$, $s_3$ to $s_1$, and $s_2$ to $s_1$, respectively.}
\label{fig:hypothetical_the}
\end{figure} 

The standard mathematical representation of the flow regime of a multi-sectoral economic system can be expressed as follows:
\begin{equation}
\label{eq:model}
x = Z \, \pmb{1} + f
\end{equation}
where $x = [x_1,\ldots,x_n]'$ is the vector of the \textit{gross outputs}, $f = [f_1,\ldots,f_n]'$ is the vector of the \textit{final demands}, $Z = (z_{ik})$ is the \textit{transactions matrix} representing the intermediate flows of products between the sectors, $\pmb{1}$ is the vector whose entries are all one, and the prime symbol represents the matrix transpose~\cite{Coskun2017SCSA,Miller2009}. More specifically, the $(i,k)-$element of the transactions matrix, $z_{ik}$, represents the direct total input from sector $i$ to $k$; the $i$th element of the final demands vector, $f_i$, quantifies the total final demand from sector $i$; and the $i$th element of the gross outputs vector, $x_i$, provides the total input to or output from sector $i$.

Let $\hat{x} = \diag(x)$ be the diagonal matrix whose diagonal elements are the corresponding elements of vector $x$, and $\hat{L}$ be the diagonal matrix whose diagonal elements are the diagonal elements of matrix $L$. The \textit{direct requirements} or \textit{(technical) coefficients matrix} is then defined as
\begin{equation}
\label{eq:dr}
A = Z \, \hat{x}^{-1} .
\end{equation}
The direct requirements matrix shows the amount of direct inputs from industries in each row an industry in a column needs in order to produce one dollar of its output. This matrix is called the \textit{composite direct distribution matrix} in the context of the system decomposition theory~\cite{Miller2009,Coskun2017SCSA}.

The \textit{total requirements matrix} can be formulated based on the direct requirements matrix as a geometric series:
\begin{equation}
\label{eq:tr}
L = (1-A)^{-1} = I + A + A^2 + A^3 + \cdots + A^n + \cdots 
\end{equation}
where $I$ is the identity matrix. The derivation of $L$, sometimes called the Leontief's inverse, can be found in~\cite{Miller2009}. The total requirements matrix is also formulated with a different rationale and called the \textit{cumulative flow distribution matrix} in the context of the system decomposition theory~\cite{Coskun2017SCSA}. The total requirements matrix represents the total inputs from industries in each row an industry in a column needs to satisfy one dollar of final demand for its products. The relationship between the gross outputs and final demands vectors can be expressed as 
\begin{equation}
\label{eq:tgo} 
x = L \, f .
\end{equation}

The gross outputs vector can be expressed using the geometric series expansion of $L$ as
\begin{equation}
\label{eq:fd_rounds}
x = L \, {f} =  I {f} + A {f} + A^2  {f} + \cdots + A^n {f} + \cdots
\end{equation}
The terms in this geometric series are generally interpreted in terms of propagations of final demands throughout the system in the existing methodologies as follows. The final demands, ${f} = I {f}$, generate a need for inputs from the productive sectors. These inputs are satisfied by the outputs of the first step that is represented by the direct requirements matrix $A$, $A f$. These outputs themselves, however, generate a need for additional inputs for the functioning of the economic system. The additional inputs are satisfied by the outputs of the second step that is represented by the second order term of $L \, f$, $A^2 f$, and so on. The steps are ordered by the power of the direct requirements matrix, $n$. In general, the first step in the geometric series, $A f$, is considered as the \textit{direct effects}, and the subsequent steps, $A^n f$, $n>1$, as the \textit{indirect effects} in the input-output economics literature. Consequently, the existing methodologies formulate the direct and indirect effects in reference to final demands.

There are several other indirect effects formulations proposed in the literature with slight differences but still in line with the notions of directness and indirectness defined above~\cite{Parikh1975}. Some of these indirect requirements matrices are listed below: 
\begin{equation}
\label{eq:id}
\begin{aligned}
E_1 & = L-I  = A + A^2 + A^3 + \cdots + A^n + \cdots \\
E_2 & = L-A  = I + A^2 + A^3 + \cdots + A^n + \cdots \\
E_3 & = L-I-A = A^2 + A^3 + \cdots + A^n + \cdots \\
E_4 &= L-\hat{L} = A - \diag(A) + A^2  - \diag(A^2) + \cdots 
+ A^n  - \diag(A^n) + \cdots 
\end{aligned}
\end{equation}
The left-multiplication of these indirect requirements matrices by $f$ yields the corresponding indirect effects vectors, $E_i f$, $i=1,\ldots,4$. They represent the way in which final demands are transmitted as gross outputs through the productive sectors of an economic system, generally after the first transactions. 

The first indirect effects formulation, $E_1 f$, represents the impact of the total effects, direct and indirect, less that of final demands, $f$~\cite{Parikh1975,Carter1970}. This formulation excludes only the impacts of final demands, and all intersectoral intermediate transactions are counted as indirect. It cannot distinguish the direct and indirect transactions and, consequently, cannot quantify the indirect interactions.

The second indirect effects formulation, $E_2 f$, removes only the direct effects, $A f$, from the total effects~\cite{Miller2009,Parikh1975,ece1972,Lancaster1968,BEA1963}. It was used by the US Department of Commerce and a variation of it by UK input-output analysts~\cite{Parikh1975}. This formulation cannot quantify the indirect transactions either, as it includes direct transactions generated by the higher order terms of propagation due to cycling (see Fig.~\ref{fig:hypothetical_the}).

Yet, the third formulation, $E_3$, removes the impacts of both final demands and direct effects simultaneously~\cite{Parikh1975,Alterman1965}. This indirect effects formulation is used by the Bureau of Economic Analysis (BEA) of the US and also widely used in ecological network analyses~\cite{Horowitz2009,Fath1999}. The shortcomings of the first two formulations persist in this formulations as well. 

The last indirect effects formulation, $E_4 f$, is reported by~\cite{Parikh1975} referring to~\cite{Leontief1951}. This formulation seems to be an attempt at removing final demands and cycling effects from the total effects (see Fig.~\ref{fig:hypothetical_the}). The cycling effects, however, cannot be determined by only the diagonal entries of the total requirements matrix, $\hat{L}$, as the off-diagonal entries of $L$ also account for cycling flows. A detailed derivation of the cycling flows through the system decomposition theory is introduced recently by~\cite{Coskun2017SCSA}.

There are various other disadvantages and shortcomings of the existing effects formulations. First of all, the input-output analysis is essentially at the sectoral level based on vector equations, such as Eq.~\ref{eq:model} and \ref{eq:fd_rounds}. These equations provide only the lump-sum total, direct and indirect, repercussions of final demands within the system. They can not quantify the direct and indirect transactions separately, let alone the direct and indirect transactions between any two compartments of the system.

The static systems are independent of time, by definition. Unlike the prevalent interpretations in input-output economics, each step of the propagation\textemdash which is represented by a term in the geometric series expansion of Eq.~\ref{eq:fd_rounds} and \ref{eq:id}\textemdash can accordingly be considered as a discrete ``computational time'' step, rather than the real time step. Moreover, the infinitely many steps in the series expansion are taking place simultaneously for static systems. On the other hand, the rounds of production in an economy are carried out in real time and each sector has a different production time. The production time of the service sector, for example, is different from that of the agriculture or automotive sector. Therefore, the economic activities of sectors cannot be simultaneous. Needless to say, there can not be infinitely many rounds of production in a single base year either. Therefore, the existing direct and indirect effects notions are essentially computational concepts and microscopic quantities rather than physical measures in real time.

The system decomposition theory addresses all these incompetencies and disadvantages in the existing methodologies. The theory improves and refines the existing static, linear, sectoral level analysis into dynamic, nonlinear, subsectoral level analysis based on matrix equations, such as Eqs.~\ref{eq:supp_dist}, \ref{eq:fd}, \ref{eq:indirect2}, and \ref{eq:supp_simp}. In this context, the directness and indirectness notions are conceptualized based on the sectoral interactions and relationships, rather than the order of propagation or, in other words, in reference to final demands. Unlike the existing approaches, the proposed subsectoral level analysis enables the explicit and separate formulations of the direct and indirect transactions pairwise between any two individual sectors of interest within both dynamic and static systems. The residence time concept to quantify the production times for each sector is also incorporated into the analysis through this theory~\cite{Coskun2017SCSA,Coskun2017SESM}.

The system decomposition theory defines the \textit{composite direct transaction} from sector $i$ to $k$, ${\tau}_{ik}^\texttt{d}$, as the total pairwise  immediate intersectoral flow from sector $i$ to $k$, regardless of the order of propagation of products in their potentially circular interactions within the system\textemdash that is, whether the products are cycling at sector $i$ and reentering sector $k$ multiple times through $i$ (see Fig.~\ref{fig:hypothetical_the}). The \textit{composite indirect transaction} from sector $i$ to $k$, ${\tau}_{ik}^\texttt{i}$, is then defined as the total pairwise intersectoral flow of products from sector $i$ indirectly through other sectors to $k$. The simple counterparts of the composite transactions can be described similarly, except that these transactions are induced by the individual final demands. More specifically, the \textit{simple direct transaction} from sector $i$ to $k$, ${\tau}_{i_k}^\texttt{d}$, is defined as the total pairwise  immediate intersectoral flow from sector $i$ to $k$, to satisfy the final demand from sector $k$. The \textit{simple indirect transaction} from sector $i$ to $k$, ${\tau}_{i_k}^\texttt{i}$, is then defined as the total pairwise intersectoral flow of products from sector $i$ indirectly through other sectors to $k$, to satisfy the final demand from sector $k$.

The system decomposition theory formulates the \textit{simple} and \textit{composite direct, indirect}, and \textit{transfer (total) flows} and \textit{storages} relative to gross outputs. The \textit{composite indirect distribution} and \textit{flow matrices} relative to gross outputs, $\pmb{N}^\texttt{i} = ({n}_{ik}^\texttt{i})$ and $\pmb{T}^{\texttt{i}} = ({\tau}_{ik}^\texttt{i}) $, are formulated as follows:
\begin{equation}
\label{eq:indirect}
\pmb{N}^\texttt{i} = (L-I) \,  \hat{L}^{-1} - A
\quad \mbox{and} \quad
\pmb{T}^{\texttt{i}} = \pmb{N}^\texttt{i} \, \hat{x} .
\end{equation}
by~\cite{Coskun2017SCSA}. These composite indirect distribution and flow matrices will respectively be called the \textit{composite indirect requirements} and \textit{transactions matrices} relative to gross outputs in the present study. The $(i,k)-$elements of $\pmb{N}^\texttt{i}$ and $\pmb{T}^\texttt{i}$ represent the total purchases from sector $i$ indirectly by $k$ to produce a dollar's worth of its output and its gross output, respectively, to satisfy both the intermediate and final demands.

The composite direct and indirect flows are defined as complementary flows:
\begin{equation}
\label{eq:suppl}
\pmb{T}^\texttt{t} = \pmb{T}^\texttt{d} + \pmb{T}^\texttt{i}
\quad \Rightarrow \quad
\pmb{T}^\texttt{i} = \pmb{T}^\texttt{t} - \pmb{T}^\texttt{d}
\quad \mbox{and} \quad 
\pmb{N}^\texttt{i} = \pmb{N}^\texttt{t} - \pmb{N}^\texttt{d}
\end{equation}
where the \textit{composite transfer (total)} and \textit{direct distribution matrices} relative to gross outputs, $\pmb{N}^\texttt{t} = ({n}_{ik}^\texttt{t})$ and $\pmb{N}^\texttt{d} = ({n}_{ik}^\texttt{d})$, and the corresponding \textit{flow matrices}, $\pmb{T}^\texttt{t} = ({\tau}_{ik}^\texttt{t})$ and $\pmb{T}^\texttt{d} = ({\tau}_{ik}^\texttt{d})$, are formulated as
\begin{equation}
\label{eq:supp_dist}
\pmb{N}^\texttt{t} = (L-I) \,  \hat{L}^{-1} 
\, \, \, \, \Rightarrow \, \, \, \,
\pmb{T}^\texttt{t} = \pmb{N}^\texttt{t}  \, \hat{x}
\quad \mbox{and} \quad 
\pmb{N}^\texttt{d} = A
\, \, \, \, \Rightarrow \, \, \, \, 
Z = \pmb{T}^\texttt{d}  = \pmb{N}^\texttt{d} \, \hat{x} 
\end{equation}
through the system decomposition theory~\cite{Coskun2017SCSA}. These composite direct and total distribution and flow matrices will respectively be called the \textit{composite direct} and \textit{total requirements} and \textit{transactions matrices} relative to gross outputs in the present manuscript.

The composite direct requirements matrix is the same as the existing direct requirements or technical coefficients matrix, $\pmb{N}^\texttt{d} = A$. The $(i,k)-$elements of $\pmb{N}^\texttt{d}$ and $\pmb{T}^\texttt{d} = Z$ represent the total purchases from sector $i$ directly by $k$ to produce a dollar's worth of its output and its gross output, respectively, to satisfy both the intermediate and final demands. The $(i,k)-$elements of $\pmb{N}^\texttt{t}$ and $\pmb{T}^\texttt{t}$ represent the total purchases from sector $i$ by $k$ to produce a dollar's worth of its output and its gross output, respectively, to satisfy both the intermediate and final demands. 

Although derived with different rationale, the composite transfer (total) flow matrix of~\cite{Coskun2017SCSA} is equivalent to the \textit{total flow matrix} of~\cite{Szyrmer1987} and the \textit{total output-to-output multiplier} of~\cite{Miller2009} after slight modifications. The total or net multipliers has been a topic of scholarly conversations for the last three decades~\cite{Dietzenbacher2005b,Oosterhaven2007,Mesnard2007}. In the context of the total flows, the economic implications of the total requirements matrix relative to gross outputs instead of final demands is also detailed in the literature~\cite{Jeong1984,Milana1985,Szyrmer1992}.

The system decomposition theory represents the cumulative demand distribution within the system through the \textit{subthroughflow matrix}, $T = ({\tau}_{i_k})$. It is defined as follows:
\begin{equation}
\label{eq:fd}
T = L \, \hat{f} \quad \mbox{with} \quad x = T \, \mathbf{1} = L \, {f} 
\end{equation}
based on physical conservation principles~\cite{Coskun2017SCSA}. The subthroughflow matrix represents the total inputs from industries in each row, an industry in a column needs to satisfy the final demand for its output. It is important to note that this subsectoral level system partitioning is not possible through the existing methodologies. The second, vector equation in Eq.~\ref{eq:fd} integrates the system partitioning methodology and input-output economics by combining two different formulations of the gross output, $x$.

The \textit{simple indirect flow matrix} is formulated relative to subthroughflows as $T^\texttt{i} = \pmb{N}^\texttt{i} \, \mathsf{T}$ where the diagonal matrix $\mathsf{T}$ is defined to be $\mathsf{T} = \diag{(T)}$~\cite{Coskun2017SCSA}. The analysis of economic systems relative to final demands, however, is more desirable for many cases. We, therefore, formulate the \textit{simple indirect transactions}, $T^\texttt{i} = ({\tau}_{i_k}^\texttt{i})$, and introduce the corresponding \textit{simple indirect requirements matrix}, $N^{\texttt{i}} = ({n}_{i_k}^\texttt{i})$, relative to final demands as follows:
\begin{equation}
\label{eq:indirect2}
N^\texttt{i} = \pmb{N}^\texttt{i} \, \hat{L} =  L-I - A \,  \hat{L}
\quad \mbox{and} \quad
T^{\texttt{i}} = N^\texttt{i} \, \hat{f} 
\end{equation}
using Eqs.~\ref{eq:fd} and~\ref{eq:indirect}, as well as the relationship $\mathsf{T} = \hat{L} \, \hat{f}$. The $(i,k)-$elements of ${N}^\texttt{i}$ and ${T}^\texttt{i}$ represent the total purchases from sector $i$ indirectly by $k$ to satisfy a dollar's worth of final demand and final demand for its products, respectively.

The simple direct and indirect transactions are also complementary. Using Eq.~\ref{eq:indirect2}, this relationship can be expressed as follows: 
\begin{equation}
\label{eq:suppl2}
{T}^\texttt{t} = {T}^\texttt{d} + {T}^\texttt{i}
\, \, \, \Rightarrow \, \, \,
{T}^\texttt{i} = {T}^\texttt{t} - {T}^\texttt{d}
\quad \mbox{and} \quad 
{N}^\texttt{i} = {N}^\texttt{t} - {N}^\texttt{d}
\end{equation}
where the \textit{simple transfer (total)} and \textit{direct requirements matrices} relative to final demands, ${N}^\texttt{t} = ({n}_{i_k}^\texttt{t})$ and ${N}^\texttt{d} = ({n}_{i_k}^\texttt{d})$, and the corresponding \textit{simple transactions matrices}, ${T}^\texttt{t} = ({\tau}_{i_k}^\texttt{t})$ and ${T}^\texttt{d} = ({\tau}_{i_k}^\texttt{d})$, can be formulated as
\begin{equation}
\label{eq:supp_simp}
{N}^\texttt{t} = L-I
\, \, \, \Rightarrow \, \, \,
{T}^\texttt{t} = {N}^\texttt{t}  \, \hat{f}
\quad \mbox{and} \quad 
{N}^\texttt{d} = A \, \hat{L}
\, \, \, \Rightarrow \, \, \,
{T}^\texttt{d}  = {N}^\texttt{d} \, \hat{f} .
\end{equation}
The simple direct and total requirement matrices, ${N}^\texttt{d}$ and ${N}^\texttt{t}$, are different from the existing direct and total requirements matrices, $A$ and $L$, as detailed below.

The $(i,k)-$elements of ${N}^\texttt{t}$ and ${T}^\texttt{t}$ represent the total purchases from sector $i$ by $k$ to satisfy a dollar's worth of final demand and final demand for its products, respectively. Therefore, the simple total requirements matrix represents internal workings of the system by excluding the final demands, while the existing formulation, $L$, represents all transactions including sales to final demands. The simple transfer (total) requirements and the existing total requirements matrices are both defined relative to final demands. Similar suggestions in regard to excluding final demands from the total requirements coefficients are also proposed in the literature~\cite{Mesnard2002b,Dietzenbacher2005b}.

The $(i,k)-$elements of ${N}^\texttt{d}$ and ${T}^\texttt{d}$ represent the total purchases from sector $i$ directly by $k$ to satisfy a dollar's worth of final demand and final demand for its products, respectively. On the other hand, the $(i,k)-$elements of $\pmb{N}^\texttt{d} = A$ and $\pmb{T}^\texttt{d}$ represent the total purchases from sector $i$ directly by $k$ for a dollar's worth of its gross output and output, respectively. In terms of propagations, while the proposed simple direct requirements matrix yields the total immediate pairwise direct transactions regardless of the order of propagation, the existing direct requirements provides one step lump-sum repercussions of final demands within the system, $A \, f$, as given in Eq.~\ref{eq:fd_rounds}.

The simple requirements matrices, $N^\texttt{*}$, are defined relative to final demands, $\hat{f}$, and their composite counterparts, $\pmb{N}^\texttt{*}$, relative to gross outputs, $\hat{x}$, as detailed above, where the superscript $(^\texttt{*})$ represents any of the \texttt{d}, \texttt{i}, and \texttt{t} symbols. From a different perspective, the difference between the simple and composite direct, indirect, and transfer (total) transactions from sector $i$ to $k$, $\tau_{i_k}^\texttt{*}$ and ${\tau}_{ik}^\texttt{*}$, is that the simple transactions quantify the corresponding flows from sector $i$ to $k$ as their final destination, and the composite transactions quantify the corresponding flows from $i$ to $k$ regardless of their external destinations.

The simple and composite direct, indirect, and transfer (total) transactions matrices can be interpreted as the simple and composite direct, indirect, and transfer (total) demand distributions throughout the system. The simple direct, indirect, and transfer (total) requirements matrices represent the corresponding demand distributions induced by unit final demands. Their composite counterparts can then be interpreted as the corresponding demand distributions induced by unit gross outputs. The \textit{simple direct, indirect}, and \textit{transfer (total) gross outputs vectors}, $x^{\texttt{d}}$, $x^{\texttt{i}}$, and $x^{\texttt{t}}$ can also be compactly expressed as follows:
\begin{equation}
\label{eq:igo}
x^{\texttt{*}} = T^{\texttt{*}} \, \pmb{1} = N^{\texttt{*}} \, {f} 
\quad \mbox{and} \quad 
\pmb{x}^{\texttt{*}} = \pmb{T}^{\texttt{*}} \, \pmb{1} = \pmb{N}^{\texttt{*}} \, {x} 
\end{equation}
similar to the linear relationship between $x$, $T$, and $f$ in Eq.~\ref{eq:fd}.

Graph theoretically, the sign of a simple indirect transaction or requirement coefficient shows the existence of an indirect path between the corresponding two sectors in the system. That is, if the $(i,k)-$element of $T^{\texttt{i}}$ is positive, $\tau^{\texttt{i}}_{i_k} >0$, then there is a demand chain from sector $i$ indirectly to $k$. It is also worth emphasizing that the diagonal elements of the simple indirect transactions matrix, ${T}^\texttt{i}$, represent the \textit{simple cycling transactions} from the corresponding subsectors reflexively back into themselves indirectly through other sectors. The composite indirect transactions matrix has the same properties as well~\cite{Coskun2017SCSA}. The cycling dynamics in the system can be used as an economic indicator for the efficiency and a measure for the intensity of the circular economy.
%~\cite{Coskun2021ESA}.

The \textit{impact analysis} in the context of the input-output economics can be described as the determination of the lump-sum impact of exogenous changes, that is the alterations in final demands, on sectors using the total requirements matrix. On the other hand, assuming constancy of requirements coefficients while allowing final demands to change is contradictory. The system decomposition theory extends and reinterprets the impact analysis using the proposed requirements matrices in terms of system response to disaggregated segments of both final demands and gross outputs.

The impact of a disaggregated segment or ``change'' of final demands, $\Delta f$, on producing sectors, $\Delta T^{\texttt{*}}$, can be determined as follows:
\begin{equation}
\label{eq:deltax}
\begin{aligned}
\Delta T = L \, \Delta \hat{f} \, \, \, \Rightarrow \, \, \, \Delta x = L \, \Delta f 
\quad \mbox{and} \quad 
\Delta T^{\texttt{*}} = N^{\texttt{*}} \, \Delta \hat{f} \, \, \, \Rightarrow \, \, \, \Delta x^{\texttt{*}} = N^{\texttt{*}} \, \Delta f 
\end{aligned}
\end{equation}
due to the linearity of the relationships given in Eq.~\ref{eq:fd} and~\ref{eq:igo}. Here, $\Delta \hat{f}$ is the diagonal matrix whose diagonal elements are the corresponding elements of $\Delta {f}$. In other words, a disaggregated segment of final demands, $\Delta f$, corresponds to the simple direct, indirect, transfer (total), and cumulative demand distributions, $\Delta T^{\texttt{d}}$, $\Delta T^{\texttt{i}}$, $\Delta T^{\texttt{t}}$, and $\Delta T$, as well as the associated gross outputs, $\Delta x^{\texttt{d}}$, $\Delta x^{\texttt{i}}$, $\Delta x^{\texttt{t}}$, and $\Delta x$. The composite counterparts of these relationships relative to gross outputs, $\hat{x}$, can similarly be formulated:
\begin{equation}
\label{eq:deltaxC}
\begin{aligned}
\Delta \pmb{T}^{\texttt{*}} = \pmb{N}^{\texttt{*}} \, \Delta \hat{x} 
\, \, \, \Rightarrow \, \, \, \Delta \pmb{x}^{\texttt{*}} = \pmb{N}^{\texttt{*}} \, \Delta x 
\end{aligned}
\end{equation} 
where $\Delta \hat{x}$ is the diagonal matrix whose diagonal elements are the corresponding elements of $\Delta {x}$. That is, a disaggregated segment of gross outputs, $\Delta x$, corresponds to the composite direct, indirect, and transfer (total) demand distributions, $\Delta \pmb{T}^{\texttt{d}}$, $\Delta \pmb{T}^{\texttt{i}}$, and $\Delta \pmb{T}^{\texttt{t}}$, as well as the associated gross outputs, $\Delta \pmb{x}^{\texttt{d}}$, $\Delta \pmb{x}^{\texttt{i}}$, and $\Delta \pmb{x}^{\texttt{t}}$.
\begin{table}[t!] 
     \centering
     \caption{The simple and composite direct, indirect, and transfer (total) requirements matrices in terms of the make-use framework}
     \label{tab:reqs}
\resizebox{\linewidth}{!}{%      
     \begin{tabular}{ l l l }
     \toprule
\multicolumn{1}{c|} {Type} & \multicolumn{1}{c|} {Simple} & \multicolumn{1}{c} {Composite} \\ 
\midrule
     \noalign{\vskip 2pt} 
Direct
&
$
\begin{array}{c}
N^\texttt{d} = D \, B \, \diag{((I - DB)^{-1})}^{-1}
\end{array}
$
& 
$
\begin{array}{c}
\pmb{N}^\texttt{d} = D \, B %= A
\end{array}
$
\\
\midrule
Indirect
&
$
\begin{array}{c}
N^\texttt{i} = {N}^{\texttt{t}} - {N}^{\texttt{d}}
\end{array}
$
& 
$
\begin{array}{c}
\pmb{N}^{\texttt{i}} = \pmb{N}^{\texttt{t}} - \pmb{N}^{\texttt{d}}
\end{array}
$
\\
\midrule
Transfer
&
$
\begin{array}{c}
N^{\texttt{t}} = (I - DB)^{-1} - I  %= L - I
\end{array}
$
& 
$
\begin{array}{c}
\pmb{N}^{\texttt{t}} = \left( (I - D \, B)^{-1} - I \right) \, \diag{((I - D \, B)^{-1})}^{-1}
\end{array}
$
\\
\noalign{\vskip 1pt} 
\bottomrule
     \end{tabular}
     } 
\end{table}

Lastly, we provide the simple and composite direct, indirect, and transfer (total) requirements matrices in terms of the make-use framework in Table~\ref{tab:reqs}. The $D$ and $B$ matrices used in the table are defined as follows:
\begin{equation}
\label{eq:DB}
D = V \, \diag({V' \, \pmb{1}})^{-1}
\quad \mbox{and} \quad 
B = U \, \diag{(V \, \pmb{1})}^{-1}
\quad \Rightarrow \quad 
A = D \, B 
\end{equation}
where $U$ and $V$ are the use and make matrices, respectively~\cite{Miller2009,Horowitz2009,Ritz1980}. The aggregated real US input-output data for multiple years are briefly analyzed based on these formulations in Case study~\ref{sec:real}. The requirements matrices given in the table are in the industry-by-industry terms under the industry-based technology assumption. They can similarly be formulated in the commodity-by-commodity, industry-by-commodity, and commodity-by-industry terms as well.

\section{Results}
\label{sec:results}

A hypothetical economic system is analyzed for illustrative purposes in the first case study of this section to elucidate the uses and meanings of the proposed direct, indirect, and transfer (total) transactions and coefficients, as well as to communicate the differences between these coefficients and their existing counterparts.

In the second case study, the real US input-output data aggregated to seven sectors are analyzed for 15 years. Since the main focus of the present manuscript is to introduce the proposed novel concepts and formulations, the numerical results and their graphical representations are briefly analyzed in this section essentially to demonstrate their effectiveness and wide applicability. Once the proposed methodology is accessible to the community, it is expected to be used for a comprehensive analysis of specific economic systems.

\subsection{Case study} 
\label{sec:hypo} 

In this case study, a simple hypothetical model is analyzed as an application of the proposed methodology to demonstrate the accuracy of the novel concepts and formulations introduced in the present manuscript and outline the incompetencies in their existing counterparts.
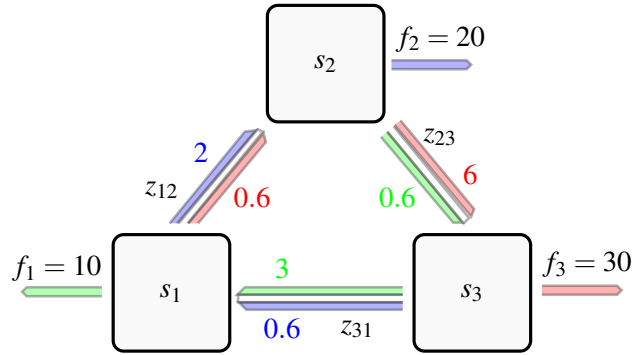
\begin{figure}[t]
\begin{center}
\begin{tikzpicture}
\centering
   \draw[very thick,  fill=gray!5, draw=black, rounded corners] (-.05,-.05) rectangle node(R1) {$s_1$} (1.5,1.5) ;
   \draw[very thick,  fill=gray!5, draw=black, rounded corners] (3.95,-.05) rectangle node(R3) {$s_3$} (5.5,1.5) ;   
   \draw[very thick,  fill=gray!5, draw=black, rounded corners] (2,3) rectangle node(R2) {$s_2$} (3.55,4.55) ;      
       \node (p) [text=black, opacity=1] at (3.2,0.25) {$z_{31}$};	    
       \node (p) [text=black, opacity=1] at (.6,2.1) {$z_{12}$};	    
       \node (p) [text=black, opacity=1] at (4.3,2.8) {$z_{23}$};	                  
       \node (p) [text=black, opacity=1] at (6.25,1.1) {$f_{3} = 30$};	    
       \node (p) [text=black, opacity=1] at (-.8,1.05) {$f_{1} = 10$};	           
	 \node (p) [text=black, opacity=1] at (4.3,4.1) {$f_{2} = 20$};	                      
        %z23
	 \draw[thick, fill=green, line width=1pt, draw=black, opacity=.3]  (3.51,2.8) -- ++ (.96,-1.14) -- ++ (.12,0) -- ++ (-.99,1.21);
	 \draw[thick, fill=gray!5, line width=1pt, draw=black, opacity=.3]  (3.59,2.87) -- ++ (1.02,-1.22) -- ++ (.07,.07) -- ++ (-1,1.22);
	 \draw[thick, fill=red, line width=1pt, draw=black, opacity=.3]  (3.69,2.96) -- ++ (1.02,-1.22) -- ++ (.04,.1) -- ++ (-.98,1.18);
       \node (p) [text=red, opacity=1] at (4.7,2.3) {$6$};	 	 
       \node (p) [text=green!100, opacity=1] at (3.73,2) {$0.6$};	 	        
         %z12
	 \draw[thick, fill=blue, line width=1pt, draw=black, opacity=.3]  (.7,1.65) -- ++ (1.0,1.22) -- ++ (.15,.02) -- ++ (-1.05,-1.24);
       \node (p) [text=blue, opacity=1] at (1.12,2.6) {$2$};
       \node (p) [text=red, opacity=1] at (1.8,2) {$0.6$};   	            
	 \draw[thick, fill=gray!5, line width=1pt, draw=black, opacity=.3]  (.82,1.65) -- ++ (1.05,1.25) -- ++ (0.07,-0.03) -- ++ (-1,-1.22);	 
	 \draw[thick, fill=red, line width=1pt, draw=black, opacity=.3]  (.95,1.65) -- ++ (1,1.2) -- ++ (0.03,-0.12) -- ++ (-.9,-1.08);	 
       % z31
	 \draw[thick, fill=green, line width=1pt, draw=black, opacity=.3]  (3.8,0.8) -- ++ (-2.1,0) -- ++ (-.1,-.1) -- ++ (2.2,0);
       \node (p) [text=green!100, opacity=1] at (2.2,1.05) {$3$};   	               	 
	 \draw[thick, fill=gray!5, line width=1pt, draw=black, opacity=.3]  (3.8,0.7) -- ++ (-2.19,0) -- ++ (0,-.1) -- ++ (2.19,0);	 
	 \draw[thick, fill=blue, line width=1pt, draw=black, opacity=.3]  (3.8,0.6) -- ++ (-2.18,0) -- ++ (.1,-.1) -- ++ (2.08,0);       
       \node (p) [text=blue, opacity=1] at (2.2,.25) {$0.6$};   	               	 	 
	 %f3 and f2
	 \draw[thick, fill=red, line width=1pt, draw=black, opacity=.3]  (5.65,.7) -- ++ (1,0) -- ++ (.06,.06) -- ++ (-.06,.06) -- ++ (-1,0);	 
	 \draw[thick, fill=blue, line width=1pt, draw=black, opacity=.3]  (3.65,3.7) -- ++ (1,0) -- ++ (.06,.06) -- ++ (-.06,.06) -- ++ (-1,0);		 
	 \draw[thick, fill=green, line width=1pt, draw=black, opacity=.3]  (-.2,.8) -- ++ (-1,0) -- ++ (-.06,-.06) -- ++ (.06,-.06) -- ++ (1,0);
\end{tikzpicture}
\end{center}
\caption{Schematic representation of the hypothetical economic model. The gross outputs at the first two steps are $Af = [2, 6, 3]'$ and $A^2f = [0.6,0.6, 0.6]'$. Only the initial flow segments at each step are depicted. (Case Study~\ref{sec:hypo}).}
\label{fig:hypothetical_ex}
\end{figure}

Let the sectors of a three-sector economic system model be agriculture (sector 1), manufacturing (sector 2), and services (sector 3). Let also the technical coefficients matrix be given as
\begin{equation}
\label{eq:Acoeff} 
\small
\begin{aligned}
A = \left[
\begin{array}{ccc} 
0 & 0.1 & 0\\ 0 & 0 & 0.2\\ 0.3 & 0 & 0
\end{array}
\right] .
\end{aligned} 
\end{equation}
Using ``row-to-column'' convention, this indicates that the production of a dollar's worth of products in the agriculture sector requires a direct input of \$0.30 from the service sector. Similarly, the production of a dollar's worth of manufacturing requires a direct input of \$0.10 from the agriculture sector, and that of the service sector requires \$0.20 direct input from the manufacturing sector.

For the  final demands $f = [10, 20, 30]'$, in million dollars, the gross outputs from each sector required to satisfy this demand becomes $x = L \, f$. That is,
\begin{equation}
\small
\begin{aligned}
f = \left[
\begin{array}{c} 
10 \\ 20 \\ 30
\end{array}
\right] 
\quad \mbox{then} \quad
x = L \, f  = \left[
\begin{array}{c} 
   12.6761 \\
   26.7606 \\
   33.8028
\end{array}
\right] .
\end{aligned}
\nonumber
\end{equation}
The composite direct transactions between the sectors of the system can then be expressed as follows:
\begin{equation}
\label{eq:Z}
\small
\begin{aligned}
Z = A \, \hat{x} = \left[
\begin{array}{ccc} 
         0 & 2.6761    &   0 \\
         0    &    0 & 6.7606 \\
    3.8028    &    0    &    0 
\end{array}
\right] 
\end{aligned}
\end{equation}
as formulated in Eq.~\ref{eq:dr}. That is, $z_{12} = \$2.6761$ million worth of products are transferred from the agriculture to the manufacturing sector, $z_{23} = \$6.7606$ million worth of products are transferred from the manufacturing to the service sector, and $z_{31} = \$3.8028$ million worth of service are transferred from the service to the agriculture sector (see Fig.~\ref{fig:hypothetical_ex}).

The total requirements matrix becomes
\begin{equation}
\label{eq:L}
\small
\begin{aligned}
L = (1-A)^{-1} = I + A + \cdots + A^n + \cdots = \left[
\begin{array}{ccc} 
    1.0060 & 0.1006 & 0.0201 \\
    0.0604 & 1.0060 & 0.2012 \\
    0.3018 & 0.0302 & 1.0060
\end{array}
\right] 
\end{aligned}
\end{equation}
based on the formulation given in Eq.~\ref{eq:tr}. A step-by-step computation shows that the terms in the series expansion of $L$ diminish quickly: 
\begin{equation}
\small
\begin{aligned}
A^2 & = \left[
\begin{array}{ccc} 
0 & 0 & 0.02\\ 0.06 & 0 & 0\\ 0 & 0.03 & 0
\end{array}
\right], \quad \quad \quad  \, \, 
A^3 = \left[
\begin{array}{ccc} 
0.006 & 0 & 0\\ 0 & 0.006 & 0\\ 0 & 0 & 0.006
\end{array}
\right] , \\
A^4 &= \left[
\begin{array}{ccc} 
0 & 0.0006 & 0\\ 0 & 0 & 0.0012\\ 0.0018 & 0 & 0
\end{array}
\right] , \, \, \, 
A^5 = \left[
\begin{array}{ccc} 
0 & 0 & 1.2\,{10}^{-4}\\ 3.6\,{10}^{-4} & 0 & 0\\ 0 & 1.8\,{10}^{-4} & 0 
\end{array}
\right] .
\end{aligned}
\nonumber
\end{equation}
These matrices can be interpreted as described in the text. As an example, the $(2,3)-$entry of $L$, $\ell_{23} = 0.2012$, indicates that, for a dollar's worth of service to meet its final demand, the service sector requires a total\textemdash direct and indirect\textemdash input of \$0.2012 from the manufacturing sector. Similarly, the $(2,3)-$entries of $A^n$ give the value of products that, for a dollar's worth of service, the service sector requires directly from the manufacturing sector at the $n$th round of production. All the other entries of these matrices can be interpreted similarly.

The gross outputs for the first step then become:
\begin{equation}
\small
\begin{aligned}
A \hat{f} = \left[
\begin{array}{ccc} 
     0 & 2 & 0 \\
     0 & 0 & 6 \\
     3 & 0 & 0
\end{array}
\right] 
\quad \Rightarrow \quad 
A f = \left[
\begin{array}{c} 
     2 \\
     6 \\
     3
\end{array}
\right] .
\end{aligned}
\nonumber 
\end{equation}
This computation indicates that to satisfy the final demands of \$10 million worth of products from the agriculture sector, \$20 million worth of products from the manufacturing sector, and \$30 million worth of products from the service sector, the manufacturing sector needs \$2 million worth of products from the agriculture sector, the service sector needs \$6 million worth of products from the manufacturing sector, and the agriculture sector needs \$3 million worth of products from the service sector. In order to satisfy these intermediate demands from each sector in the first step, the gross outputs in the second step should be:
\begin{equation}
\small
\begin{aligned}
A^2 \hat{f} = \left[
\begin{array}{ccc} 
         0   &   0  &  0.6 \\
      0.6   &   0  &  0 \\
         0   &  0.6  &  0
\end{array}
\right] 
\quad \Rightarrow \quad 
A^2 f = \left[
\begin{array}{c} 
    0.6 \\
    0.6 \\
    0.6
\end{array}
\right] .
\end{aligned}
\nonumber
\end{equation}
All the subsequent steps can be calculated and interpreted similarly (see Fig.~\ref{fig:hypothetical_ex}).

The subthroughflow matrix that represents the cumulative demand distribution throughout the system to satisfy the final demands can then be calculated as follows:
\begin{equation}
\label{eq:T}
\small
\begin{aligned}
T = L \, \hat{f} = \left[
\begin{array}{ccc} 
   10.0604 & 2.0121 & 0.6036 \\
    0.6036 & 20.1207 & 6.0362 \\
    3.0181 & 0.6036 &  30.1811
\end{array}
\right] 
\quad \mbox{with} \quad
x = T \, \pmb{1} = \left[
\begin{array}{c} 
   12.6761 \\
   26.7606 \\
   33.8028
\end{array}
\right] 
\end{aligned}
%\nonumber
\end{equation}
based on the formulation given in Eq.~\ref{eq:fd}. The $(2,3)-$entry of $T$, $\tau_{2_3} = 6.0362$, indicates that \$6.0362 million worth of products in the manufacturing sector is consumed by the service sector to satisfy the final demand for its products. All the other entries of $T$ can be interpreted similarly.

There are several indirect requirements matrices proposed in the literature as partially listed in Eq.~\ref{eq:id}. For this hypothetical economic system, they become
\begin{equation}
\small
\begin{aligned}
E_1 &= \left[
\begin{array}{ccc} 
    0.0060 & 0.1006 & 0.0201 \\
    0.0604 & 0.0060 & 0.2012 \\
    0.3018 & 0.0302 & 0.0060
\end{array}
\right],
\quad 
E_2 = \left[
\begin{array}{ccc}
    1.0060 & 0.0006 & 0.0201 \\
    0.0604 & 1.0060 & 0.0012 \\
    0.0018 & 0.0302 & 1.0060
\end{array}
\right] \\
E_3 & = \left[
\begin{array}{ccc}
    0.0060 & 0.0006 & 0.0201 \\
    0.0604 & 0.0060 & 0.0012 \\
    0.0018 & 0.0302 & 0.0060
\end{array}
\right] ,
\quad
E_4 = \left[
\begin{array}{ccc}
         0 & 0.1006 & 0.0201 \\
    0.0604 &      0 & 0.2012 \\
    0.3018 & 0.0302 &      0
\end{array}
\right] 
\end{aligned}
\nonumber
\end{equation}
The proposed simple indirect requirements matrix for this hypothetical system, however, is
\begin{equation}
\label{eq:howcome}
\small
\begin{aligned}
N^\texttt{i} = \left[
\begin{array}{ccc}
    0.0060 & 0 & 0.0201 \\
    0.0604 & 0.0060 &      0 \\
    0 & 0.0302 & 0.0060
\end{array}
\right]
\end{aligned}
%\nonumber
\end{equation}
as given in Eq.~\ref{eq:indirect2}. The $(1,3)-$entry of $N^\texttt{i}$, $n^\texttt{i}_{1_3} = 0.0201$, indicates that, for a dollar's worth of service to meet its final demand, the service sector requires an indirect input of \$0.0201 from the agriculture sector through other sectors. All the other entries can be interpreted similarly.

The incompetency of the existing indirect effects formulations is theoretically demonstrated above, in the discussion following Eq.~\ref{eq:id}. The inaccuracy of these formulations can numerically be shown for this hypothetical economic system as well. As seen from Fig.~\ref{fig:hypothetical_ex}, there are no indirect transactions from $s_1$ to $s_2$, from $s_2$ to $s_3$, and from $s_3$ to $s_1$. Therefore, the corresponding simple indirect coefficients in the proposed indirect requirements matrix are zero: $n^\texttt{i}_{1_2} = n^\texttt{i}_{2_3} = n^\texttt{i}_{3_1}  = 0$. More specifically, the relationship $n^\texttt{i}_{1_2} = 0$, for example, indicates that there are no products in the agriculture sector allocated for the consumption of the manufacturing sector indirectly through the service sector to satisfy one unit of final demand for its products. Graph theoretically, this implies that there is no indirect path from sector 1 to 2, which can easily be verified from Fig~\ref{fig:hypothetical_ex}. Since the existing indirect effects formulations, $E_1$ to $E_4$, cannot exclude the cycling effects at the sectors along the path, they all have nonzero values in these entries (see Figs.~\ref{fig:hypothetical_the} and~\ref{fig:hypothetical_ex}). This indicates that the existing formulations cannot correctly quantify the indirect transactions within the system.

The indirect transactions and requirements matrices can also be used as a measure for the intensity of a circular economy as outlined in the text. The diagonal entries of $N^\texttt{i}$ represent reflexive indirect, that is cycling, transactions. Since there is one closed path in this system, $s_1 \rightarrow s_2 \rightarrow s_3 \rightarrow s_1$, the cycling flow is the same at each sector along the path (see Fig.~\ref{fig:hypothetical_ex}). Consequently, the diagonal simple indirect coefficients are all equal to each other: $n^\texttt{i}_{1_1} = n^\texttt{i}_{2_2} = n^\texttt{i}_{3_3} = 0.0060$. These coefficients indicate that for a dollar worth of their final demands, the sectors self-demand \$0.006 for their products. The small values of these coefficients may be interpreted as indications for a less efficient and weaker circular economy.

The simple indirect transactions matrix, which represents the indirect demand distribution throughout the system, and the indirect gross outputs vector can then be expressed as follows: 
\begin{equation}
\label{eq:howcome2}
\small
\begin{aligned}
T^{\texttt{i}} = N^\texttt{i} \, \hat{f} = \left[
\begin{array}{ccc} 
    0.0604 & 0 & 0.6036 \\
    0.6036 & 0.1207  &   0 \\
         0 & 0.6036 & 0.1811
\end{array}
\right] ,
\quad
x^{\texttt{i}} = T^{\texttt{i}} \, \pmb{1} =  \left[
\begin{array}{c} 
    0.6640 \\
    0.7243 \\
    0.7847
\end{array}
\right] 
\end{aligned}
%\nonumber
\end{equation}
based on the formulations in Eq.~\ref{eq:indirect2}. The nonzero entries of $T^{\texttt{i}}$ represent the indirect transactions between sectors that satisfy the corresponding final demands. For example, $\tau^{\texttt{i}}_{3_2} = 0.6036$ indicates that \$0.6036 million worth of products from the service sector are purchased by the manufacturing sector indirectly through the agriculture sector to satisfy the final demand for its products. Three indirect transactions, however, are zero: $\tau^{\texttt{i}}_{1_2} = \tau^{\texttt{i}}_{2_3} = \tau^{\texttt{i}}_{3_1} = 0$. More specifically, the relationship $\tau^{\texttt{i}}_{1_2} = 0$ indicates, for example, that there are no products in the agriculture sector allocated for the consumption of the manufacturing sector indirectly through the service sector to satisfy the final demand for its products. The absence of indirect transactions in the given directions in $T^{\texttt{i}}$ is consistent with the values of the corresponding indirect coefficients in $N^{\texttt{i}}$, as discussed above. The other simple indirect transactions can be interpreted similarly.

The existing direct and total requirements matrices are given in Eqs.~\ref{eq:Acoeff} and~\ref{eq:L}. The proposed simple direct and transfer (total) requirements matrices are also presented below for comparison:
\begin{equation} 
\label{eq:NdNt}
\small
\begin{aligned}
N^\texttt{d} & = \left[
\begin{array}{ccc}
         0 & 0.1006 &      0 \\
         0 &      0 & 0.2012 \\
    0.3018 &      0 &      0
\end{array}
\right] ,
\quad
N^\texttt{t} = \left[
\begin{array}{ccc}
    0.0060 & 0.1006 & 0.0201 \\
    0.0604 & 0.0060 & 0.2012 \\
    0.3018 & 0.0302 & 0.0060
\end{array}
\right] 
\end{aligned}
%\nonumber
\end{equation}
based on the formulations in Eq.~\ref{eq:supp_simp}. As seen from these results, the nonzero coefficients of $N^\texttt{d}$ and $\pmb{N}^\texttt{d}=A$ are different. For example, the $(1,2)-$element of $N^\texttt{d}$ and $\pmb{N}^\texttt{d}$, $n_{1_2}^\texttt{d}$ and ${n}_{12}^\texttt{d}$, respectively represent the total purchases from sector $1$ directly by $2$ to satisfy a dollar's worth of final demand for its products and a dollar's worth of its gross output. Any other coefficients of the direct requirements matrices can be interpreted similarly. The only difference between the proposed simple transfer (total) and existing total requirements matrices, $N^\texttt{t} = L-I$ and $L$, is that the diagonal transfer coefficients are one less than the corresponding total coefficients. Since the simple transfer (total) requirements matrix covers only the internal workings of the producing sectors, it excludes the final demands from the coefficients on the main diagonal.

The proposed composite direct requirements matrix is equivalent to the existing direct requirements matrix, that is $\pmb{N}^\texttt{d} = A$, as formulated in Eq.~\ref{eq:supp_dist}. The composite indirect and total requirements matrices can be computed as follows: 
\begin{equation}
\label{eq:NiNt}
\small
\begin{aligned}
\pmb{N}^\texttt{i} & = \left[
\begin{array}{ccc}
    0.0060 & 0 & 0.0200 \\
    0.0600 & 0.0060 &      0 \\
         0 & 0.0300 & 0.0060 \\
\end{array}
\right] ,
\quad
\pmb{N}^\texttt{t} = \left[
\begin{array}{ccc}
    0.0060 & 0.1000 & 0.0200 \\
    0.0600 & 0.0060 & 0.2000 \\
    0.3000 & 0.0300 & 0.0060 \\
\end{array}
\right] 
\end{aligned}
%\nonumber
\end{equation}
using Eqs.~\ref{eq:indirect} and~\ref{eq:supp_dist}. The $(3,2)-$element of $\pmb{N}^\texttt{i}$, that is the composite indirect coefficient ${n}_{32}^\texttt{i} = 0.03$, quantifies the total indirect purchases at the amount of \$0.03 from the service sector indirectly through the agriculture sector by the manufacturing sector to produce a dollar's worth of its gross output. However, the corresponding simple and composite direct coefficients are zero, due to the absence of the direct demand chains or transactions in this direction, ${n}_{3_2}^\texttt{d} = {n}_{32}^\texttt{d} = 0$. Similarly, the $(3,2)-$element of $\pmb{N}^\texttt{t}$, that is the transfer coefficient ${n}_{32}^\texttt{t} = 0.03$, measures the total purchases at the amount of \$0.03 from the service sector by the manufacturing sector to produce a dollar's worth of its gross output. Since the direct and indirect flows are complementary, as formulated in Eq.~\ref{eq:suppl}, the relationship ${n}_{32}^\texttt{i} = {n}_{32}^\texttt{t} = 0.03$ indicates that there is no direct composite flow in the given direction. This observation agrees with the corresponding composite direct coefficient or transaction, that is $z_{32} = a_{32} = {n}_{32}^\texttt{d} = 0$. All the other composite direct, indirect, and transfer (total) coefficients can be interpreted similarly.

As seen from these illustrative results, the proposed requirements matrices provide critical novel statistics to comprehensively analyze complex economic systems at both sectoral and subsectoral levels. The results demonstrate that the proposed directness and indirectness concepts accurately capture the direct, indirect, and transfer (total) relationships and interactions between sectors and the proposed matrix measures correctly quantify these interactions. Such accurate and detailed analyses are not possible through the existing methodologies.

\subsection{Case study} 
\label{sec:real} 

In this case study, the real US input-output data for 15 years are briefly analyzed using the aggregated use and make tables, $U$ and $V$, provided by~\cite{Miller2009}. These $U$ and $V$ matrices are expressed in millions of US current year dollars. The sectors in these aggregated data sets for the US economy are as follows: Agriculture (1), Mining (2), Construction (3), Manufacturing (4), Trade, Transport \& Utilities (5), Services (6), and Other (7).

The requirements matrices are computed using the formulations listed in Table~\ref{tab:reqs} for the make-use framework. The results are in the industry-by-industry terms under the industry technology assumption. The numerical results for the composite indirect and total requirements matrices, $\pmb{N}^\texttt{i}$ and $\pmb{N}^\texttt{t}$, have already been presented in the context of ecosystem analysis using real data by~\cite{Coskun2017SCSA}. Therefore, no numerical tables and graphical representations for these matrices are provided in the present manuscript.

The numerical results for the simple direct, indirect, and transfer (total) requirements matrices, $N^\texttt{d}$, $N^\texttt{i}$, and $N^\texttt{t}$, are presented in Tables~\ref{tab:2006}-\ref{tab:1919} and their graphical representations are presented in Figs.~\ref{fig:d_dist}-\ref{fig:t_dist} in the Appendix. Since the graphs of $N^\texttt{t} = L - I$ and $L$ differ only along their main diagonals, the graphs for $L$ can easily be predicted from those of $N^\texttt{t}$, and so, are also omitted. The graphical representations of the composite direct requirements matrices, $\pmb{N}^\texttt{d} = A$, are presented for a comparison in Fig.~\ref{fig:A_dist} for these 15 years as well.

The proposed requirements coefficients provide detailed novel statistics and enable a comprehensive analysis of the internal workings of economic systems. We use some coefficients from Table~\ref{tab:2006} for 2006 below to exemplify the interpretations of the numerical results provided in the tables. The simple indirect coefficient $n^{\texttt{i}}_{3_4}(2006) = 0.0051$ represents the total value of products purchased from sector $3$ at the amount of \$0.0051 indirectly through other sectors by $4$ in 2006 to satisfy a dollar's worth of final demand for its products. There is only one zero simple indirect coefficient in all simple indirect requirements matrices presented in Tables~\ref{tab:2006}-\ref{tab:1919}: $n^{\texttt{i}}_{3_7}(1919) = 0$. This indicates that there was no indirect transaction from the construction sector to the other sector (sector 7) in 1919. The composite indirect coefficient ${n}^{\texttt{i}}_{34}(2006) = 0.0032$ (not shown in the tables) quantifies the total value of products purchased from sector $3$ at the amount of \$0.0032 indirectly through other sectors by $4$ in 2006 to produce a dollar's worth of its products.

The simple direct coefficient $n^{\texttt{d}}_{3_4}(2006) = 0.0030$ gives the total value of products purchased from sector $3$ at the amount of \$0.0030 directly by $4$ in 2006 to satisfy a dollar's worth of final demand for its products. The composite direct coefficient ${n}^{\texttt{d}}_{34}(2006) = 0.0019$ (not shown in the tables) then measures the total value of products purchased from sector $3$ at the amount of \$0.0019 directly by $4$ in 2006 to produce a dollar's worth of its products. Similarly, the simple total coefficient $n^{\texttt{t}}_{3_4}(2006) = 0.0081$ determines the total value of products purchased from sector $3$ at the amount of \$0.0081 by $4$ in 2006 to satisfy a dollar's worth of final demand for its products. The composite total coefficient ${n}^{\texttt{t}}_{34}(2006) = 0.0051$ (not shown in the tables) ascertains the total value of products purchased from sector $3$ at the amount of \$0.0051 by $4$ in 2006 to produce a dollar's worth of its products.

Manufacturing seems to be the backbone of the US economy in terms of both direct and indirect transactions, as can be verified from both the numerical results in Tables~\ref{tab:2006}-\ref{tab:1919} and their graphical representations in Figs.~\ref{fig:A_dist}-\ref{fig:t_dist}. Interestingly, although the manufacturing sector directly contributes more to itself, $n_{4_4}^\texttt{d}$, it indirectly contributes more to the construction sector, $n_{4_3}^\texttt{i}$, in almost every year. The diagonal indirect coefficients quantify the cycling transactions\textemdash the reflexive circular flow from a sector back into itself after possibly being transmitted throughout the system\textemdash as outlined in the text. The diagonal simple indirect coefficients in Fig.~\ref{fig:i_dist} show that the cycling transactions are generally not significant for most years in the US economy, except for the manufacturing sector. This negligible cycling dynamics indicates the noncircular nature of the economy and may potentially imply a weaker economic efficiency.

The simple indirect coefficients indicate that the US industries indirectly rely increasingly more on the service sector (highlighted 6th rows in Fig.~\ref{fig:i_dist}). Intriguingly, the indirect contributions of the service sector to the US economy are even more than those of the manufacturing sector in recent years (4th rows in Fig.~\ref{fig:i_dist}). The 6th rows in Figure~\ref{fig:d_dist} indicate that the service sector also directly supplies all industries increasingly more but these increments are at a lower scale relative to the increments of other direct transactions. This increasing tendency in the US economy towards being service-oriented is discussed in some BEA reports and in the literature as well~\cite{Streitwieser2010}. On the other hand, the indirect contributions of the trade, transportation, and utilities sector (5th rows in Fig.~\ref{fig:i_dist}) to the other industries are gradually decreasing, following an increase from 1919 to 1939.

The proposed formulations enable determination of the impact of disaggregated segments of final demands and gross outputs on sectors, as formulated in Eq.~\ref{eq:deltax}. A portion of final demands of industries at the amount of $\Delta f = [1,0,0,0,0,2,0]'$, in million dollars, in 2006 corresponds to the following indirect demand distribution throughout the system:
\begin{equation}
\small
\label{eq:deltas}
\begin{aligned}
\Delta T^{\texttt{i}} = \left[
\begin{array}{ccccccc} 
    0.0154 & 0 & 0 & 0 & 0 & 0.0148 & 0 \\
    0.0444 & 0 & 0 & 0 & 0 & 0.0311 & 0 \\
    0.0045 & 0 & 0 & 0 & 0 & 0.0029 & 0 \\
    0.1792 & 0 & 0 & 0 & 0 & 0.1238 & 0 \\
    0.0693 & 0 & 0 & 0 & 0 & 0.0472 & 0 \\
    0.1871 & 0 & 0 & 0 & 0 & 0.0850 & 0 \\
    0.0210 & 0 & 0 & 0 & 0 & 0.0122 & 0 
\end{array}
\right] 
\quad
\mbox{and}
\quad
\Delta x^{\texttt{i}} =
\left[
\begin{array}{c} 
    0.0302 \\
    0.0755 \\
    0.0074 \\
    0.3030 \\
    0.1164 \\
    0.2721 \\
    0.0332
\end{array}
\right] 
\end{aligned}
\nonumber
\end{equation}
where the simple indirect transactions and gross outputs are formulated in Eqs.~\ref{eq:indirect2} and~\ref{eq:deltax}. These results indicate that a portion of final demands at the amount of \$1 million from the agriculture sector and \$2 million from the service sector yield the indirect demand distribution throughout the system as specified by $\Delta T^{\texttt{i}}$. For example, since $\Delta \tau^{\textit{i}}_{3_1} = 0.0045$, $\Delta \tau^{\textit{i}}_{3_6} = 0.0029$, and $\Delta x^{\textit{i}}_{3} = 0.0074$, the agriculture and service sectors end up buying \$4500 and \$2900 (\$7400 in total) worth of products, respectively, from the construction sector indirectly through the other sectors to meet the specified portion of final demands. The construction sector has the least level of indirect response to these portions of final demands. The sector that has the highest level of indirect response is the manufacturing sector with a correspondence in its gross output at the amount of \$303,000 ($\Delta x^{\textit{i}}_{4} = 0.3030$). 

The main focus of the present manuscript is to introduce the proposed methodology rather than the detailed analysis of a certain model in a given year. The analysis of all the other coefficients would further elucidate various other aspects of the US economy. The unique perspectives the proposed requirements coefficients bring to the understanding of the internal workings of economic systems, intersectoral interactions, and exogenous impacts on the system and individual sectors, are not available through the state-of-the-art techniques. The accuracy of the interpretations of economic activities increases with the disaggregation of the industries.

\section{Discussion}
\label{sec:discussion}

There have been numerous attempts in literature for about a century to define and formulate the indirect transactions between sectors of an economic system, species in an ecological system, or any two compartments in various other scientific fields. None of these formulations can accurately describe the indirect interactions and relationships and correctly quantify the indirect transactions.

The existing indirect effects are formulated by modifications of the total requirements matrix, $L$, at different levels. These approaches have several disadvantages and shortcomings. The idea has essentially been to remove some terms in the geometric series expansion of $L$ to distinguish the direct and indirect effects. The flow segments are classified as direct or indirect in reference to final demands, based on the order of propagation they satisfy these demands. In general, the flow segments satisfying final demands in the first step are considered as direct effects, and all the subsequent steps as indirect in these formulations.

Unrealistically, infinitely many steps of propagation, or rounds of production, is assumed in a single base year in these formulations. Moreover, all sectors are assumed to be acting synchronously and producing simultaneously within the same period of production time at each round. Consequently, the existing direct and indirect effects notions are mainly microscopic, computational concepts rather than physical measures. The existing coefficients, consequently, cannot even quantify the direct and indirect transactions separately, let alone the direct and indirect transactions between any two sectors of interest.

In the context of the system decomposition theory the directness and indirectness are determined based on the nature of interactions and relationships between sectors. The pairwise immediate transactions from one sector directly to another are called the direct transactions and pairwise transactions from one sector to another through other sectors are called the indirect transactions. The direct and indirect flows between any two compartments within a compartmental system have recently been conceptualized and mathematically formulated relative to gross outputs~\cite{Coskun2017SCSA}. Based on this theory, the composite and simple direct, indirect, and transfer (total) transactions between any two sectors, as well as the associated requirements coefficients for multisectoral economic systems are defined and explicitly formulated relative to both gross outputs and final demands in the present manuscript for the first time. 

The simple and composite direct, indirect, and transfer (total) requirements coefficients are defined as the scaled versions of the corresponding transactions. The requirements matrices are expressed in terms of the make-use framework in the text as well. The simple coefficients are the scaled versions of the corresponding transactions relative to final demands, while the composite coefficients are relative to gross outputs. Therefore, the simple and composite direct, indirect, and transfer (total) requirements matrices represent the direct, indirect, and transfer (total) intermediate demand distributions throughout the system per unit final demand and gross output, respectively. The direct, indirect, and transfer (total) transactions matrices then represent the corresponding intermediate demand distributions throughout the system induced by these outputs and demands.

There have been essentially two main statistics for economic system analysis in the input-output economics: the direct and total requirements matrices. These existing direct and total requirements matrices are defined relative to gross outputs and final demands, respectively. Therefore, the proposed composite direct requirements matrix is equivalent to the existing direct requirements or coefficient matrix. The difference between the simple direct requirements matrix and coefficient matrix in terms of propagations is that while the former provides total direct transactions pairwise between any two sectors regardless of the order of propagation in their potentially circular interactions, the latter provides only one step propagation relative to final demands. The difference between the proposed and existing total requirements matrices is that while the former provides coefficients for total transactions between only producing sectors by excluding final demands to quantify the internal workings of the system, the latter provides coefficients for total transactions including sales to final demands.

The simple requirements coefficients can also be considered as measures for exogenous impacts on each sector, while their composite counterparts for sectoral impacts on one another. Therefore, the proposed matrices provide multiple measures for more accurate, rigorous, and detailed analyses of economic systems to address their full complexity, exogenous influence, and intersectoral dynamics. 

The existing direct and total requirements matrices are mainly used for impact and policy analysis in national and regional economic systems. The BEA publishes these requirements tables together with the annual US input-output data. The proposed simple and composite direct, indirect, and transfer (total) requirements statistics provide different unique perspectives about the system response to arbitrary segments of final demands and gross outputs that are not available through the state-of-the-art methodologies. In the context of the impact analysis, while the proposed coefficients separately ascertain the direct, indirect, and transfer (total) repercussions of the disaggregated segments of these demands or outputs from a single sector on any other individual sector of interest, the existing formulations provide only the lump-sum effects of the segments.

The existing methodologies in the input-output economics prevalently interpret the impact analysis in the context of the influence of exogenous changes on the system. Such interpretations in static systems is inaccurate and misleading. This is because of the fact that allowing alterations in final demands or gross outputs contradicts the constancy of the requirements coefficients. Unlike the existing methodologies, the system decomposition theory rather considers the impact analysis as the quantification of how the system corresponds arbitrarily given segments of final demands or gross outputs.

The accuracy and efficiency of the proposed simple and composite direct, indirect, and transfer (total) transactions and requirements matrices in capturing the corresponding interactions and relationships between sectors of an economic system is demonstrated through a hypothetical model in the first case study. In the second case study, the transactions and requirements matrices for the US economy, using the aggregated input-output data for 15 years, are presented with brief discussion and some implications. The numerical results for these real data sets and their graphical representations are also presented in the appendix. 

\section*{Declarations}

\subsection*{Authors' contributions}

HBC and HC conceived the ideas and designed the methodology; HBC analyzed and interpreted the data; HBC and HC wrote the manuscript.

\subsection*{Funding}

The authors received no specific funding for this work.

\subsection*{Competing Interests}

The authors declare that they have no competing interests.

\subsection*{Availability of data and materials}

%The data is included in the manuscript.
All data generated or analyzed during this study are included in this published article.

%\subsection*{Acknowledgments}
%%%The author would like to thank Svetlana Pashchenko for useful discussions and helpful comments.
%The author would like to thank Huseyin Coskun for introducing the system decomposition theory to the author and for useful comments.

%\bibliographystyle{chicago}
\bibliographystyle{amsplain}
%\bibliography{economics}
\bibliography{ditar}

\newpage
\appendix

\section{The US Requirements Tables}
\label{asec:tables}

The aggregated input-output tables in terms of the make-use framework ($U$ and $V$), as well as the direct and total requirements matrices ($A$ and $L$ in industry-by-industry terms) for the US economy are provided by~\cite{Miller2009} for 15 years (1919, 1929, 1939, 1947, 1958, 1963, 1967, 1972, 1977, 1982, 1987, 1992, 1997, 2002, 2006). The sectors in these aggregated real data sets are: Agriculture (1), Mining (2), Construction (3), Manufacturing (4), Trade, Transport \& Utilities (5), Services (6), and Other (7).

\begin{figure}[h]
    \centering
    \subcaptionbox{2006}{\includegraphics[width=0.32\textwidth]{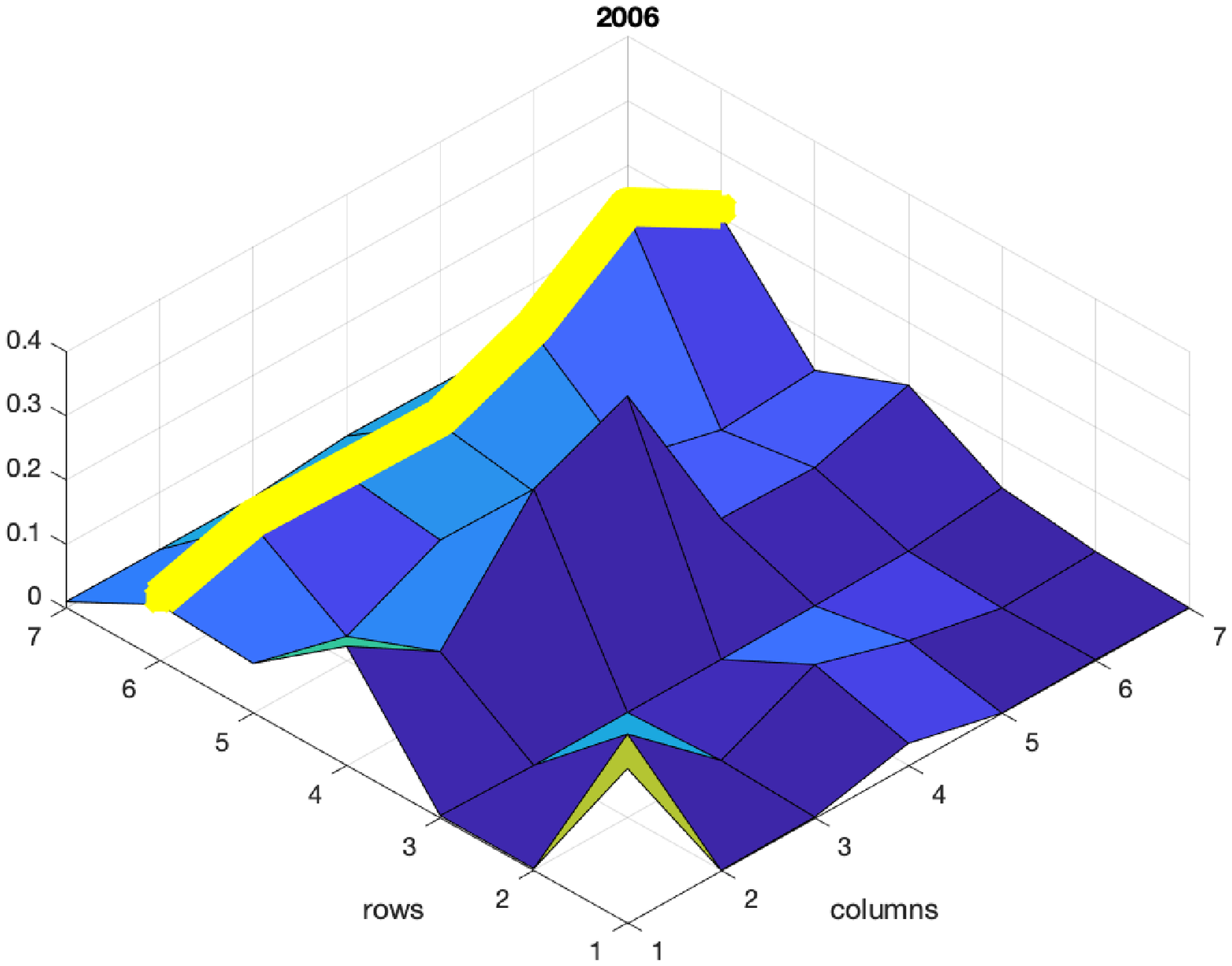} \label{fig:A_2006}}
    \subcaptionbox{2002}{\includegraphics[width=0.32\textwidth]{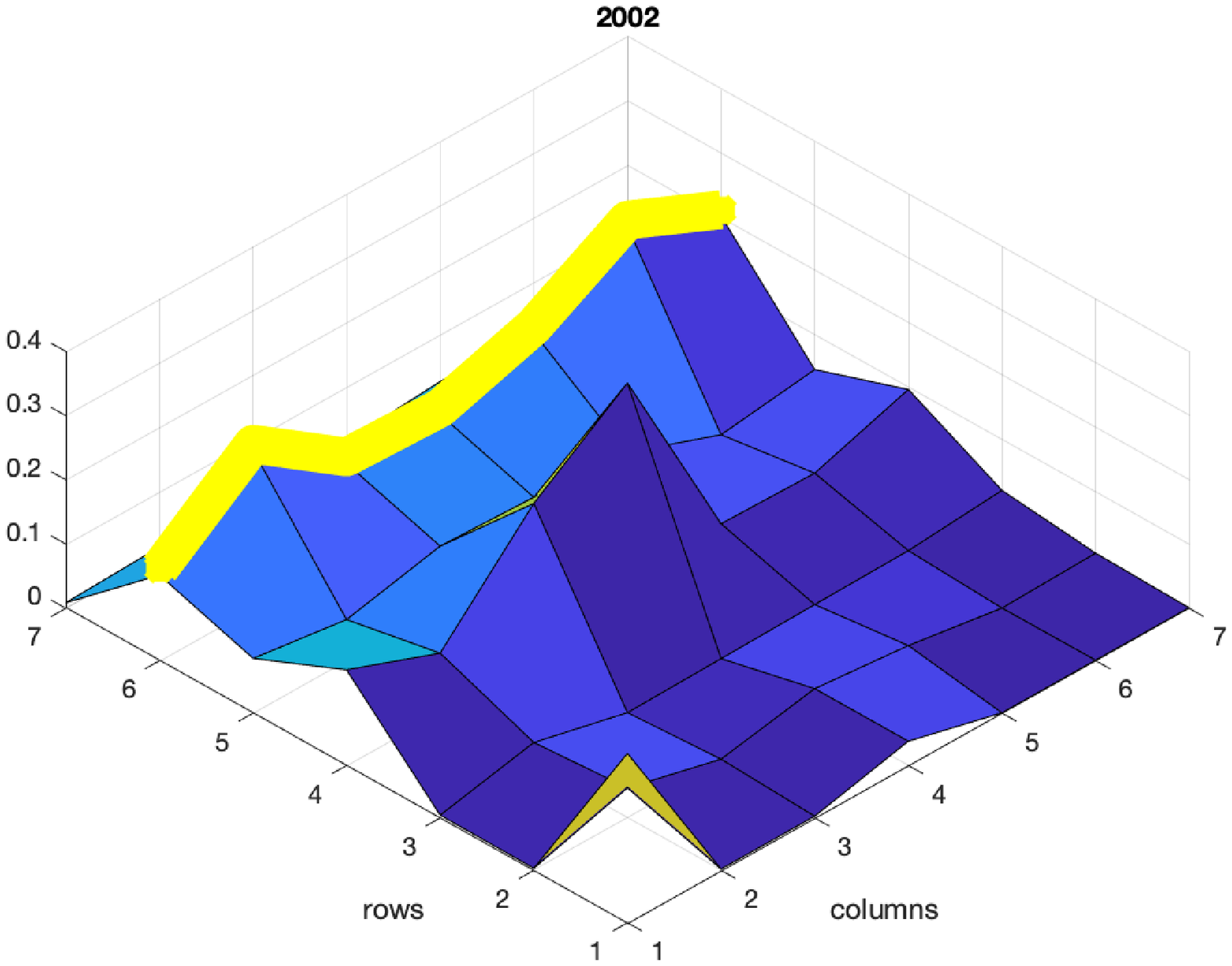} \label{fig:A_2002}}
    \subcaptionbox{1997}{\includegraphics[width=0.32\textwidth]{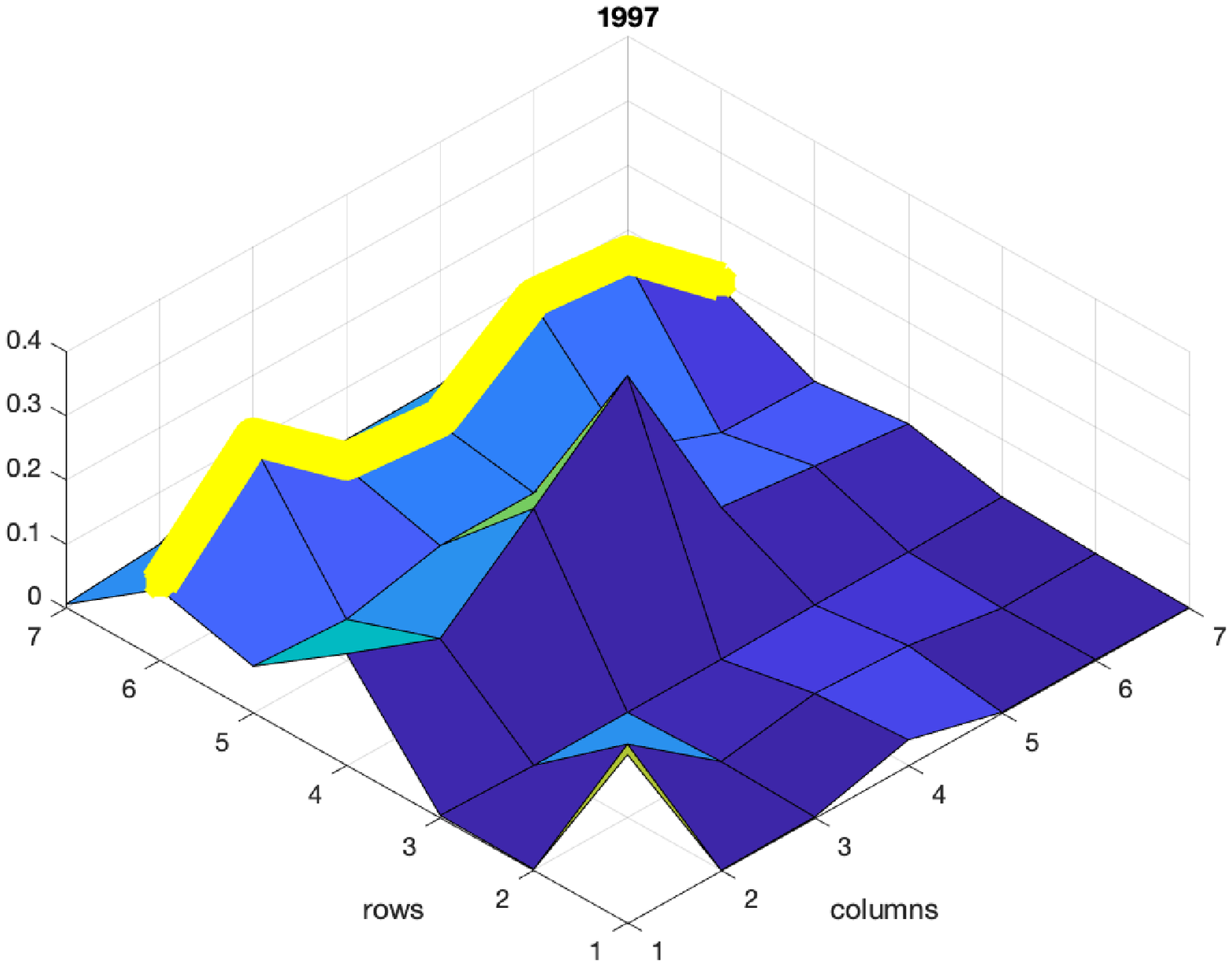} \label{fig:A_1997}}  \\
    \subcaptionbox{1992}{\includegraphics[width=0.32\textwidth]{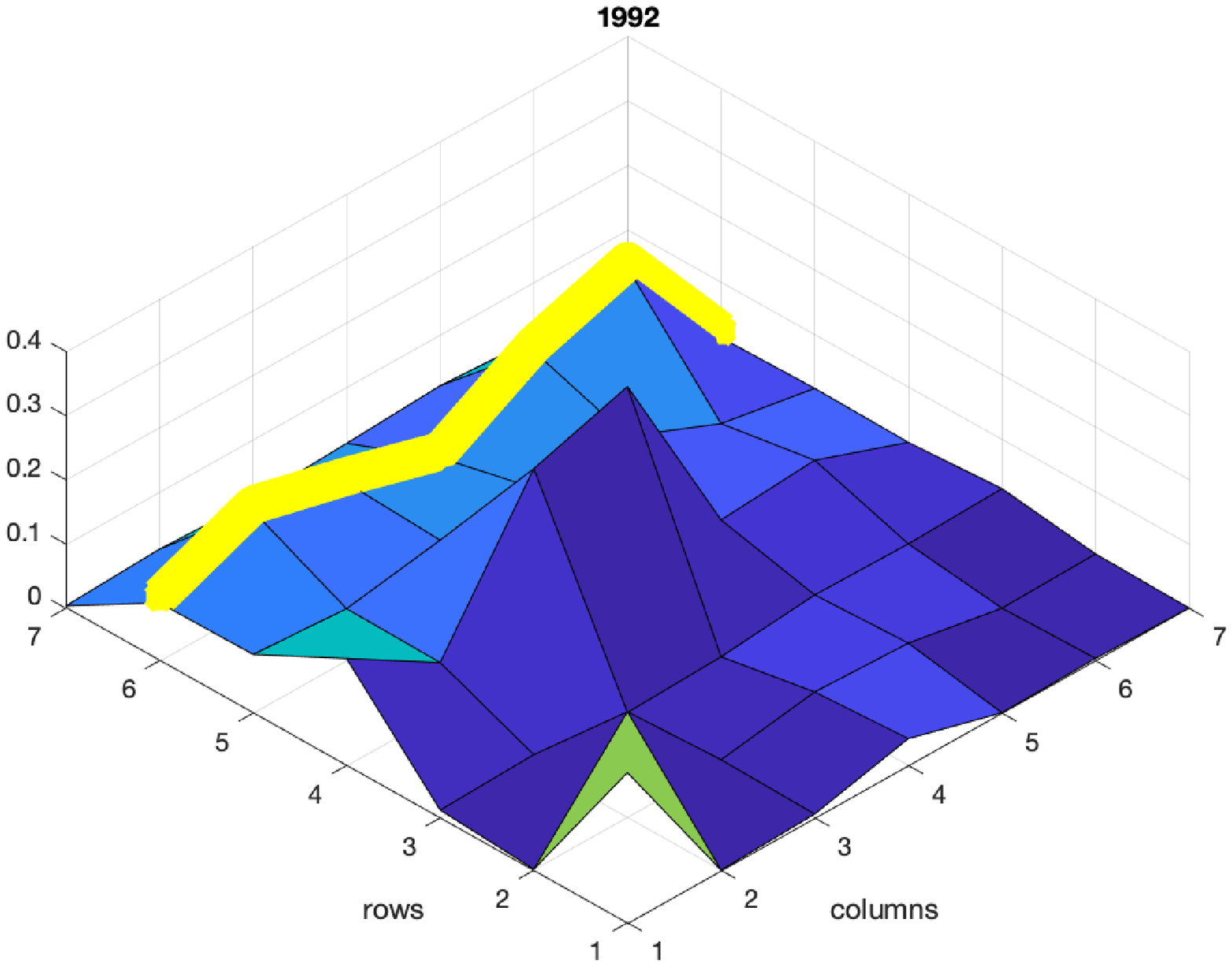} \label{fig:A_1992}}
    \subcaptionbox{1987}{\includegraphics[width=0.32\textwidth]{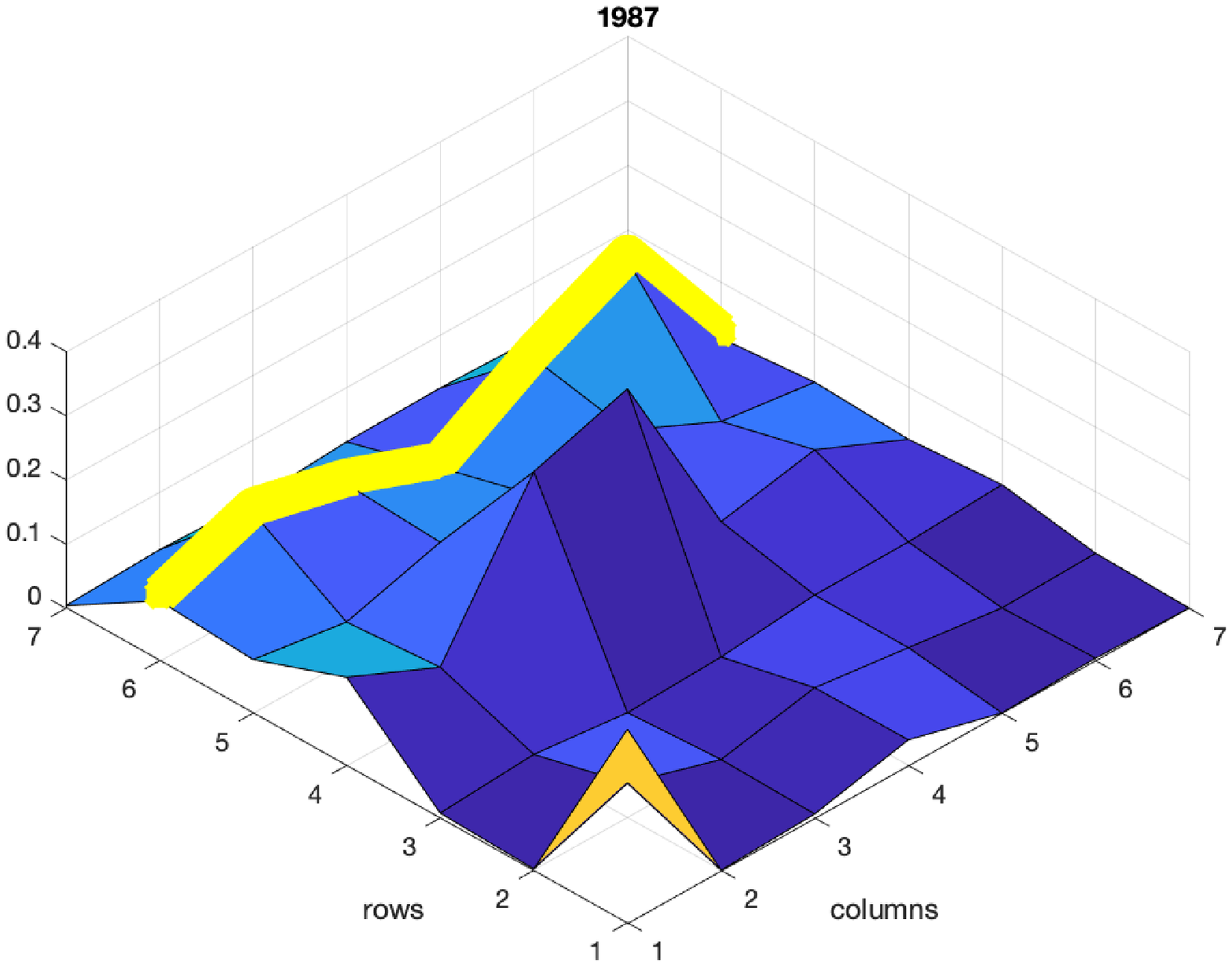} \label{fig:A_1987}}
    \subcaptionbox{1982}{\includegraphics[width=0.32\textwidth]{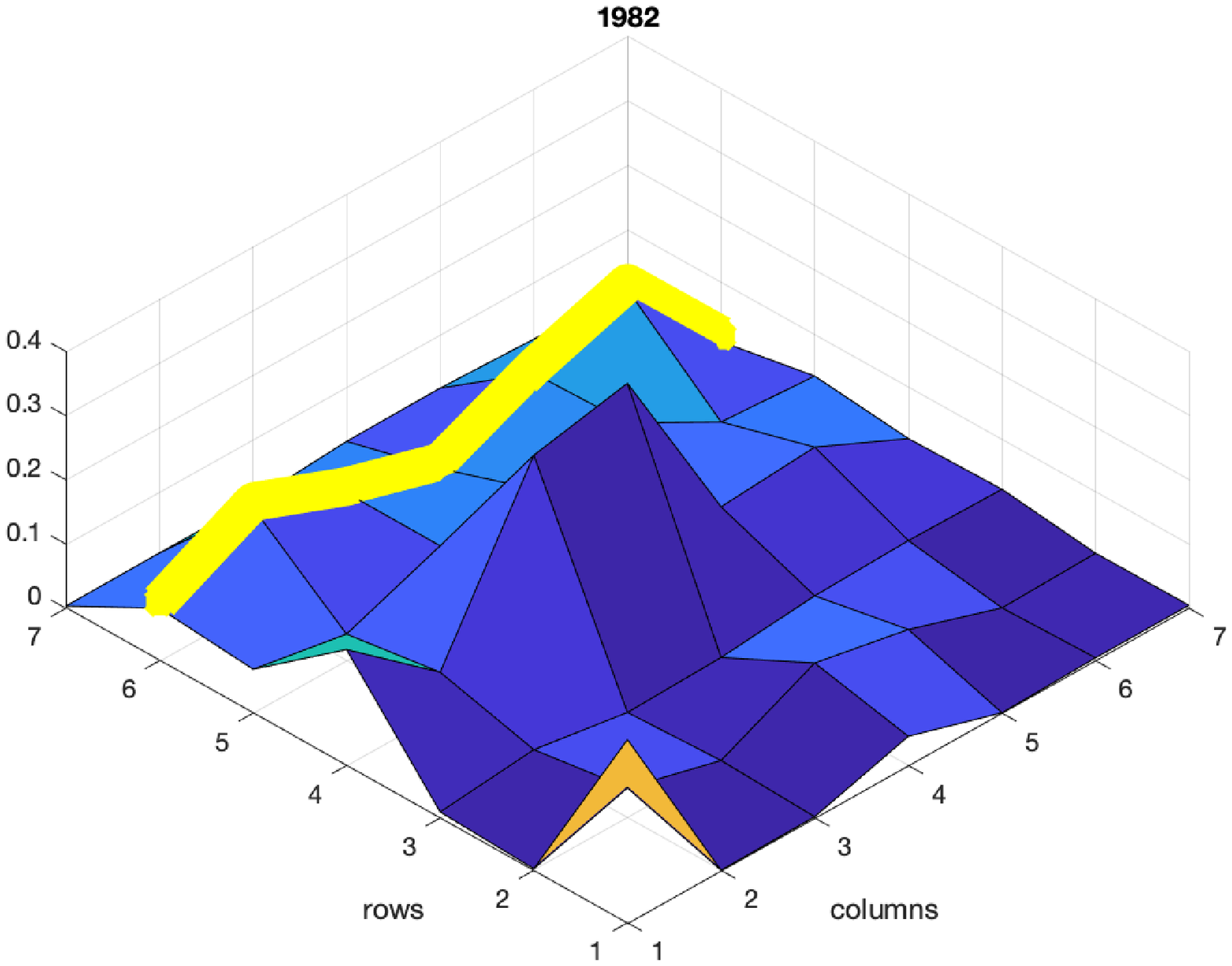} \label{fig:A_1982}}  \\
    \subcaptionbox{1977}{\includegraphics[width=0.32\textwidth]{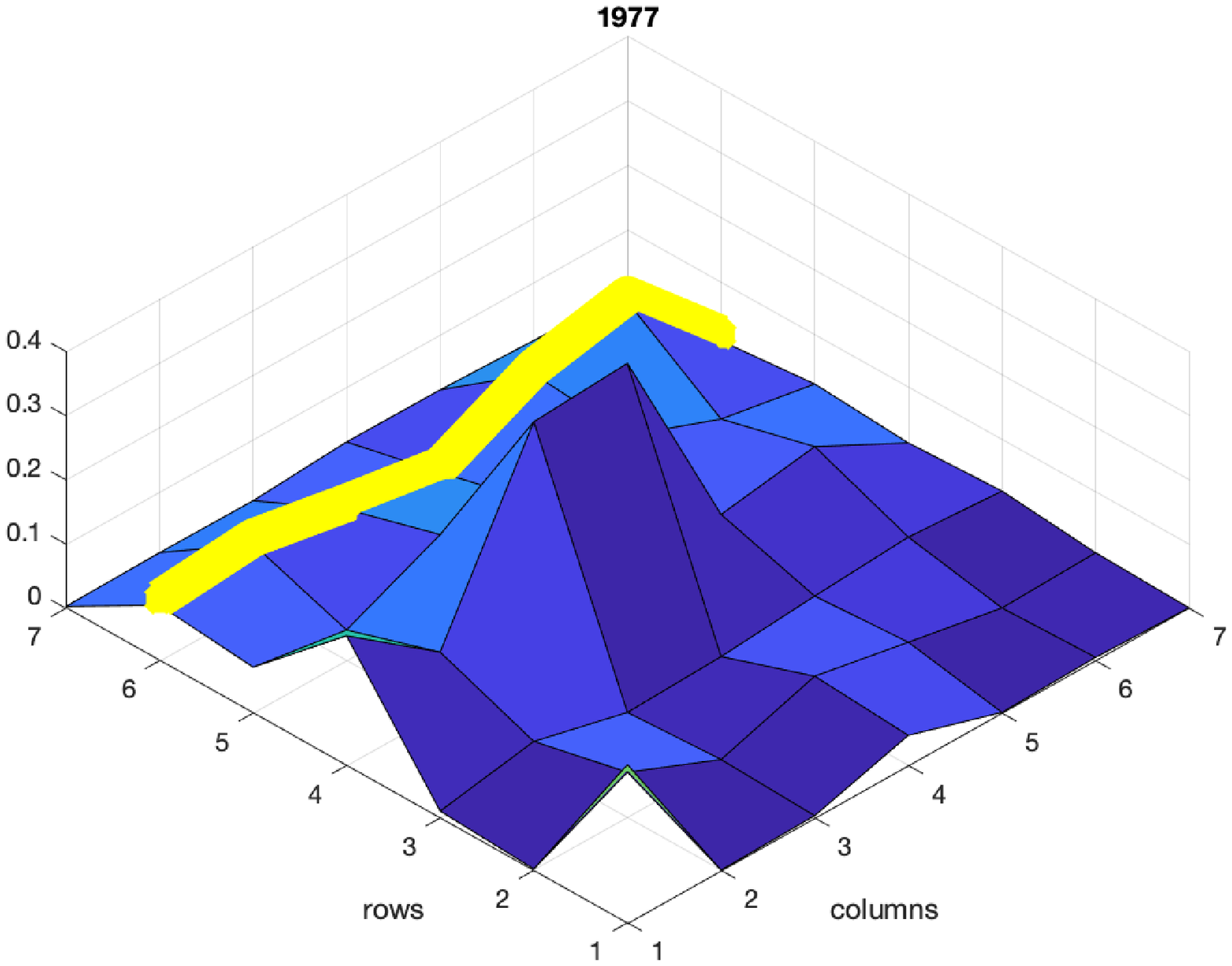} \label{fig:A_1977}}
    \subcaptionbox{1972}{\includegraphics[width=0.32\textwidth]{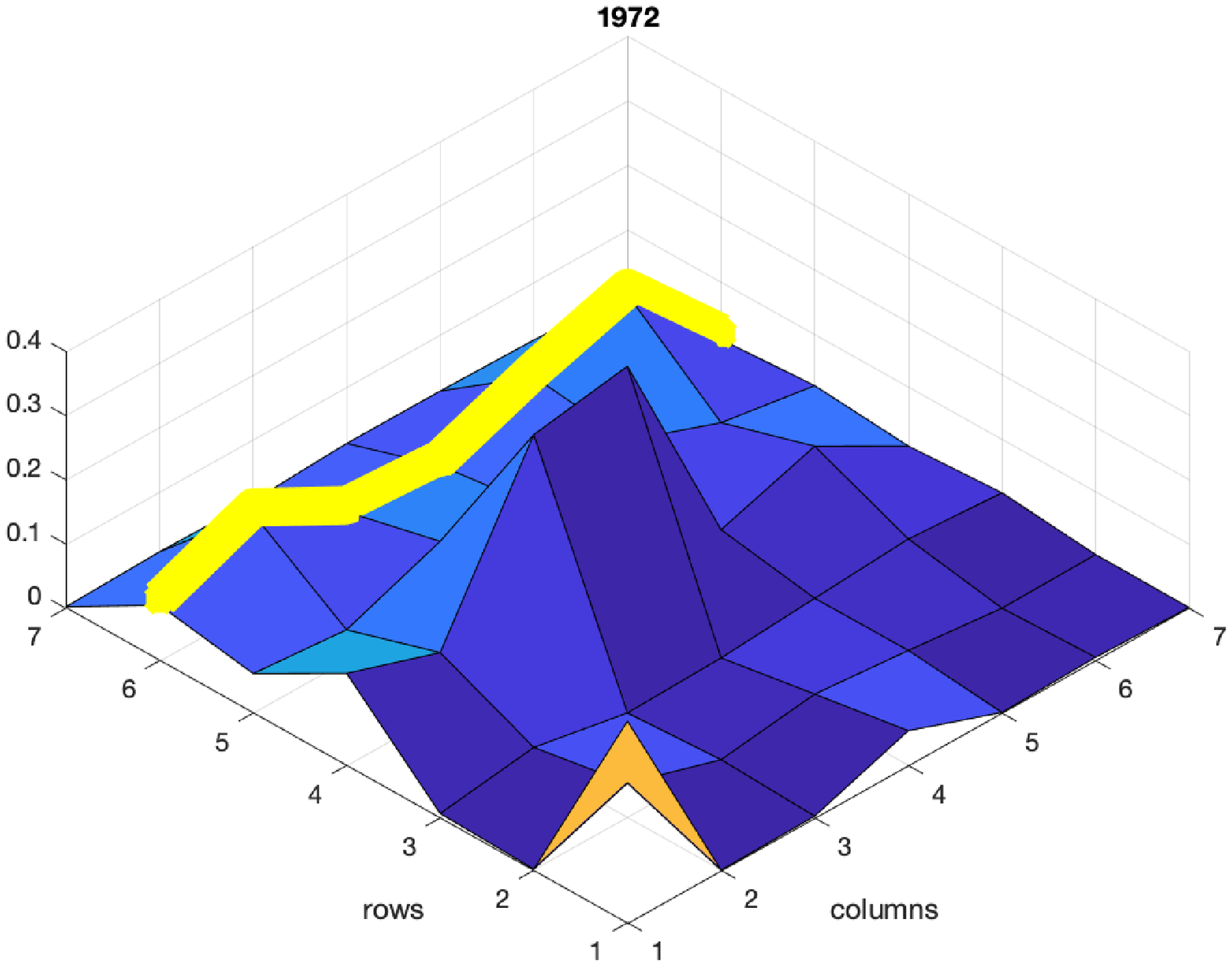} \label{fig:A_1972}}
    \subcaptionbox{1967}{\includegraphics[width=0.32\textwidth]{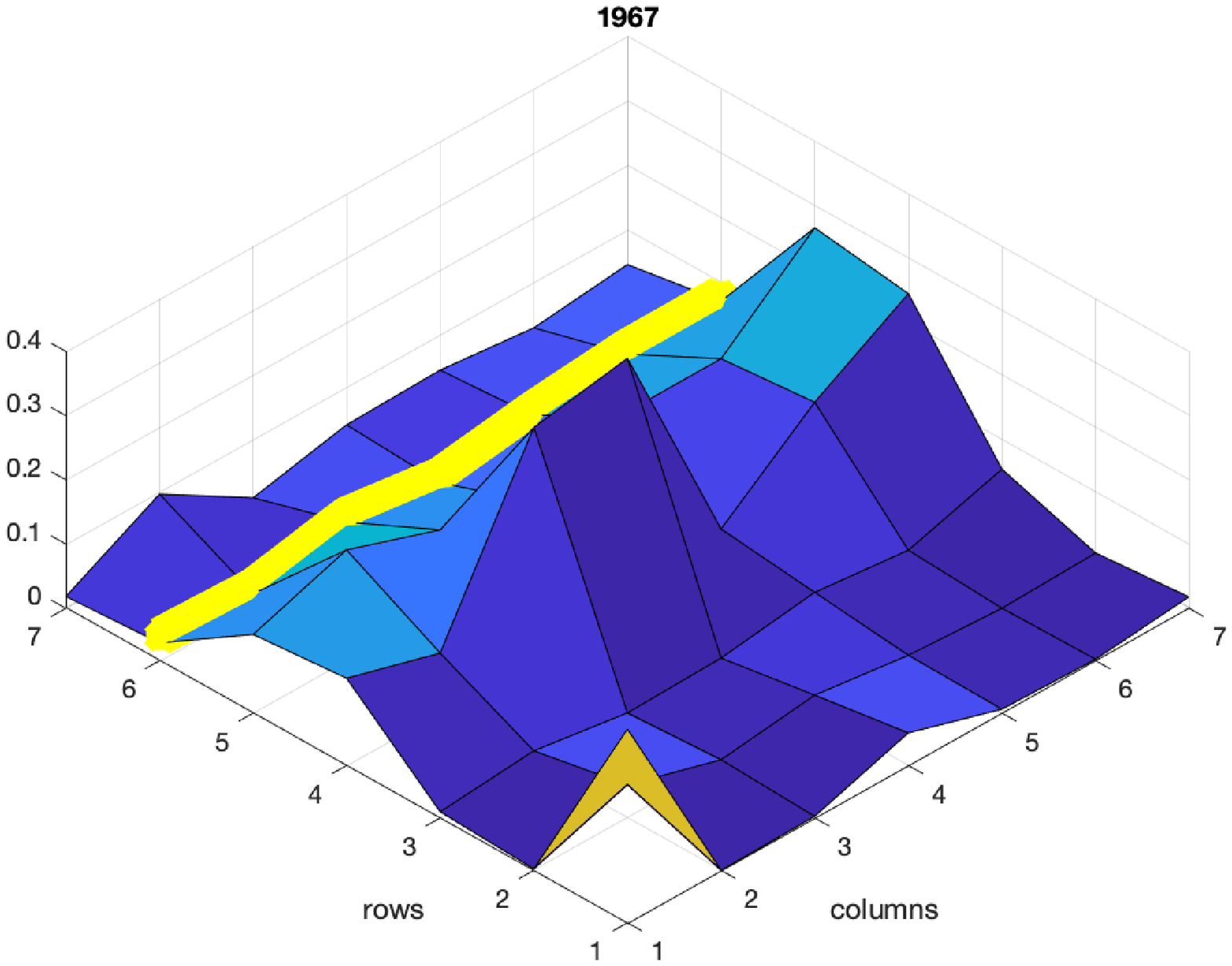} \label{fig:A_1967}}  \\
    \subcaptionbox{1963}{\includegraphics[width=0.32\textwidth]{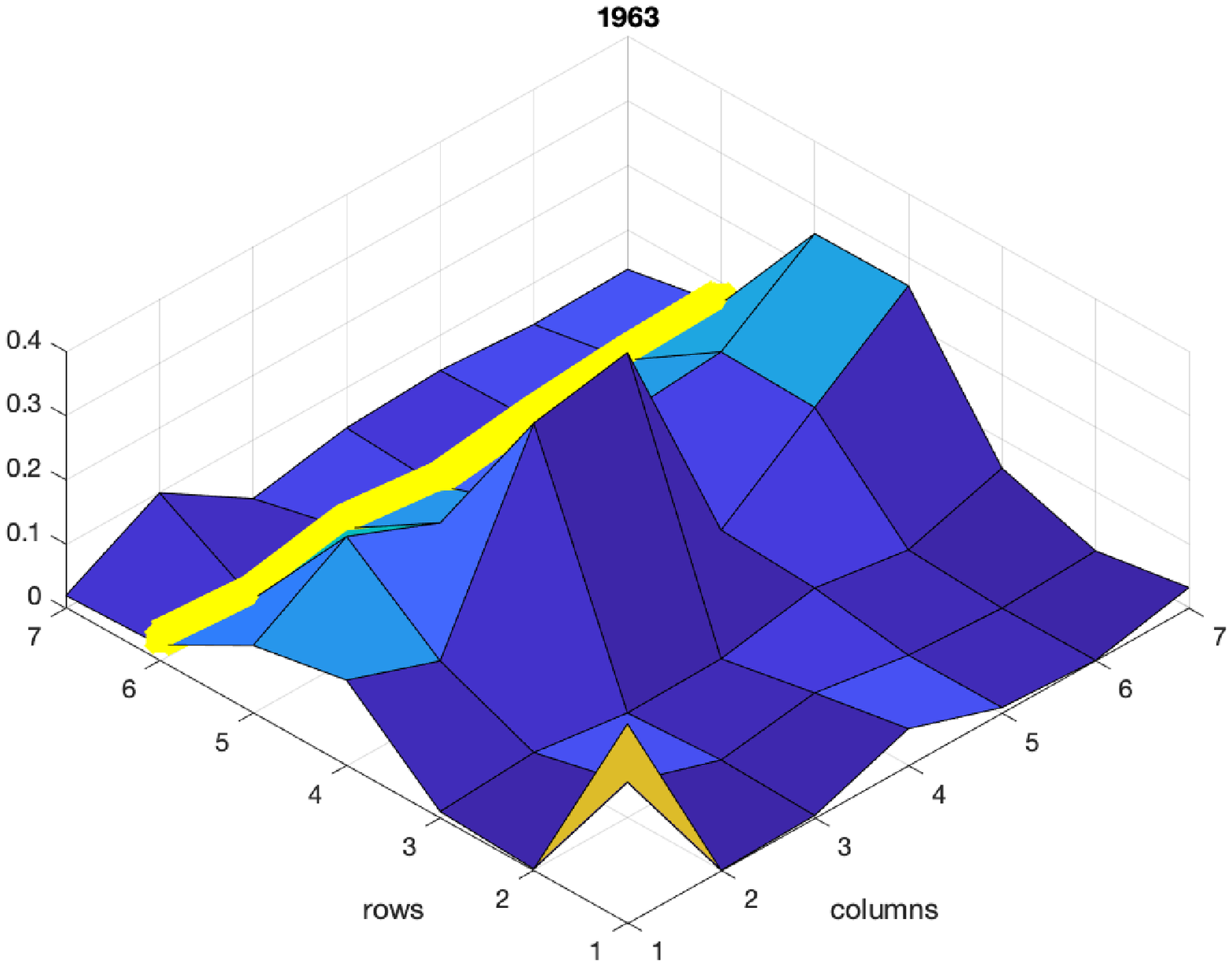} \label{fig:A_1963}}
    \subcaptionbox{1958}{\includegraphics[width=0.32\textwidth]{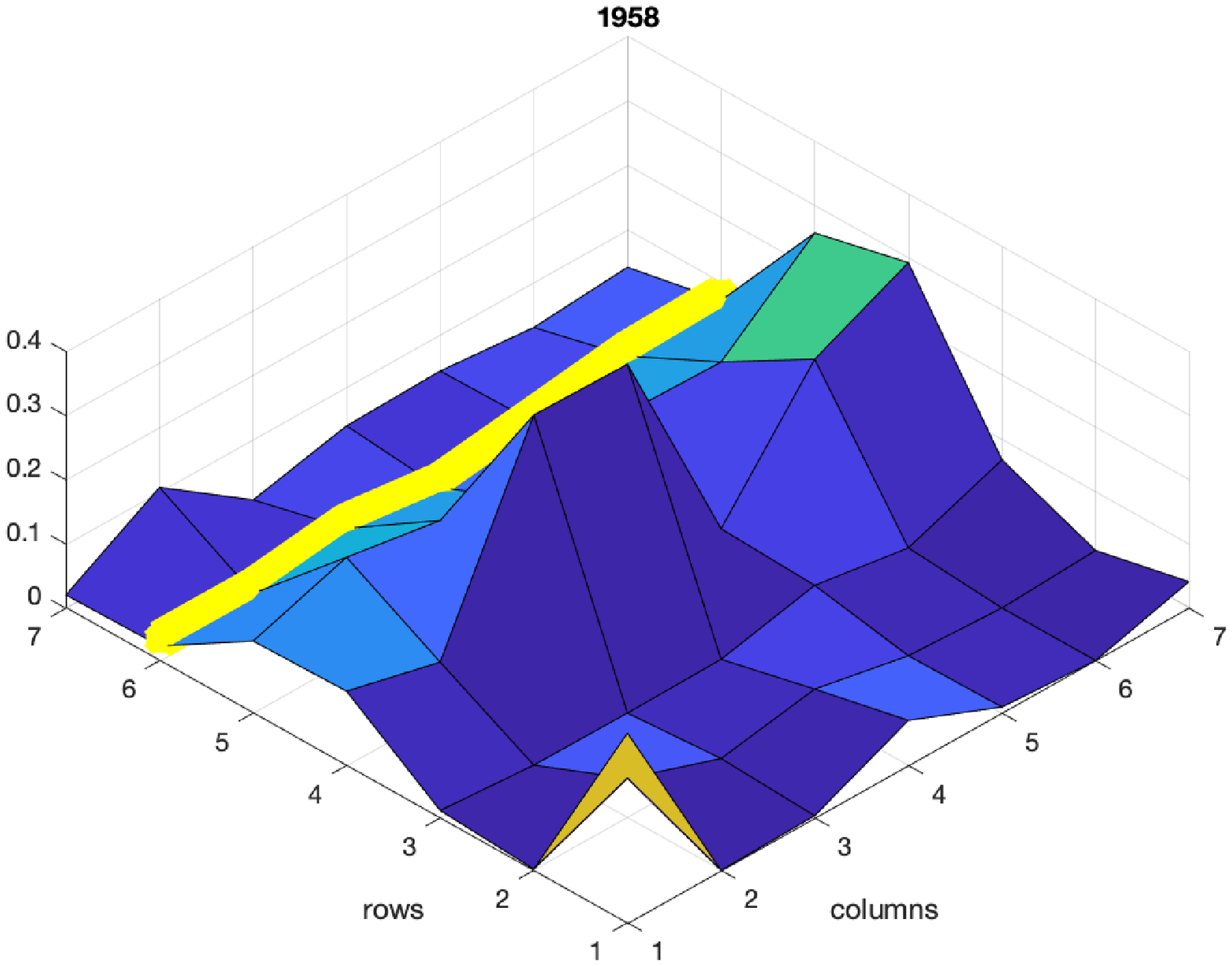} \label{fig:A_1958}}
    \subcaptionbox{1947}{\includegraphics[width=0.32\textwidth]{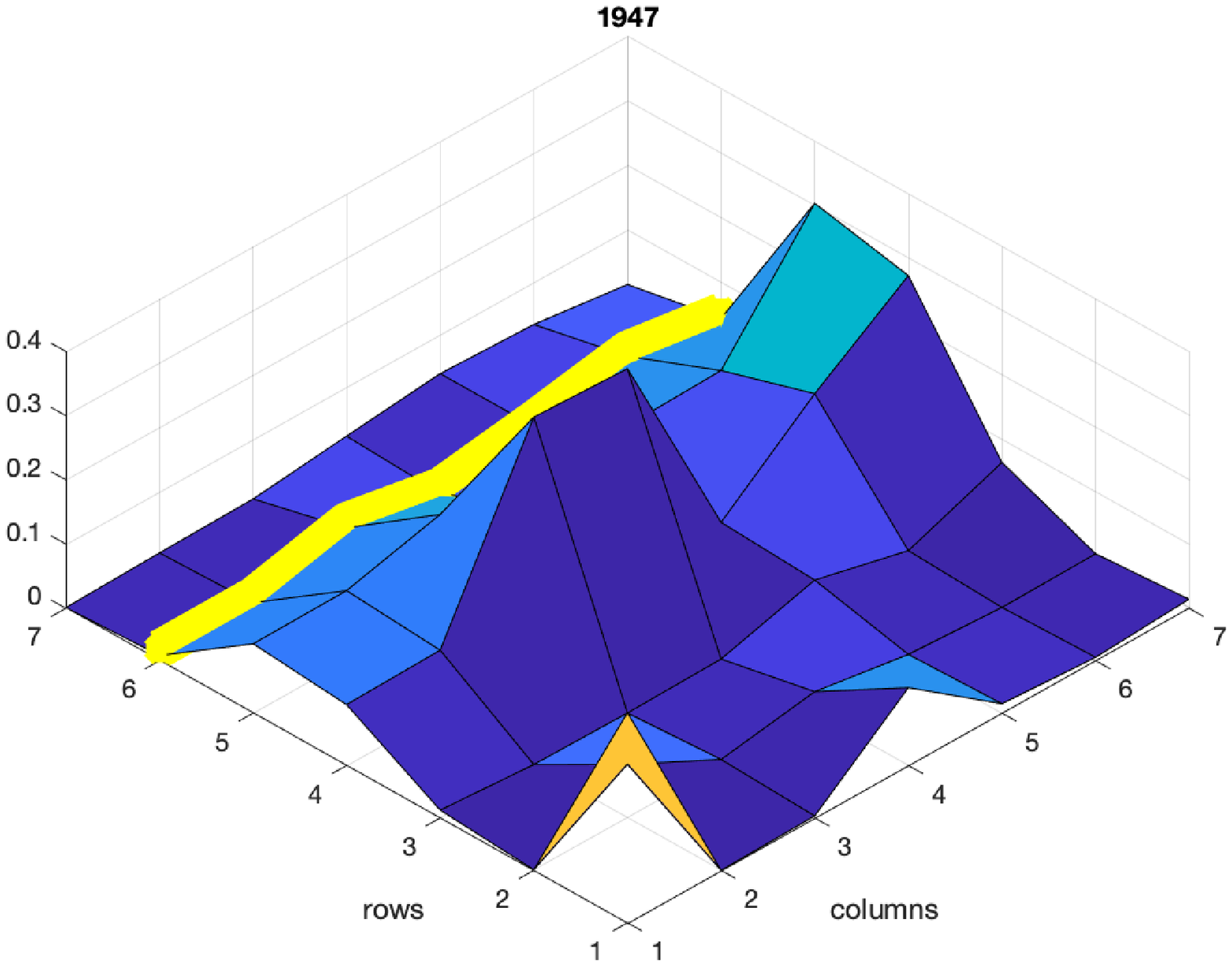} \label{fig:A_1947}}  \\
    \subcaptionbox{1939}{\includegraphics[width=0.32\textwidth]{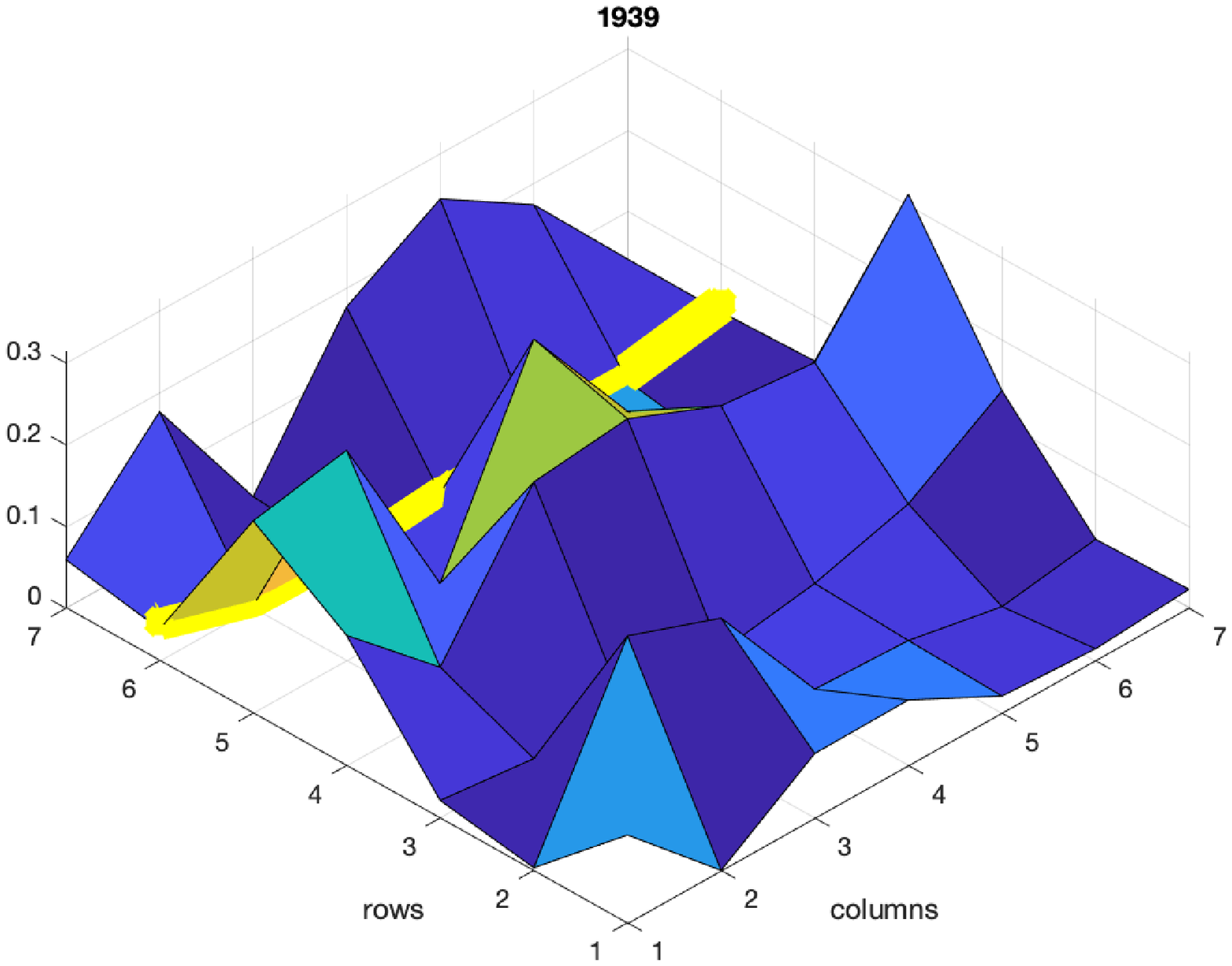} \label{fig:A_1939}}
    \subcaptionbox{1929}{\includegraphics[width=0.32\textwidth]{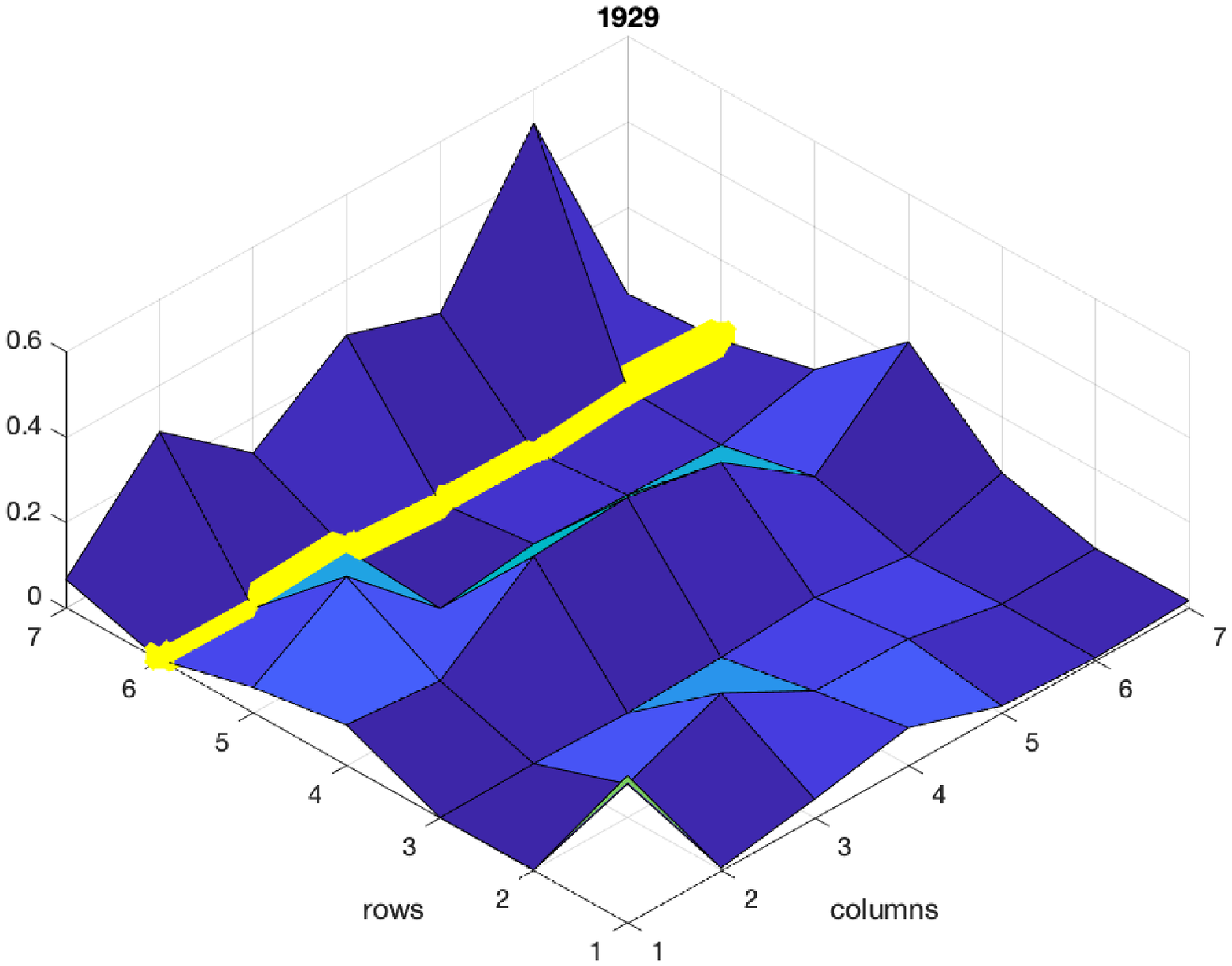} \label{fig:A_1929}}
    \subcaptionbox{1919}{\includegraphics[width=0.32\textwidth]{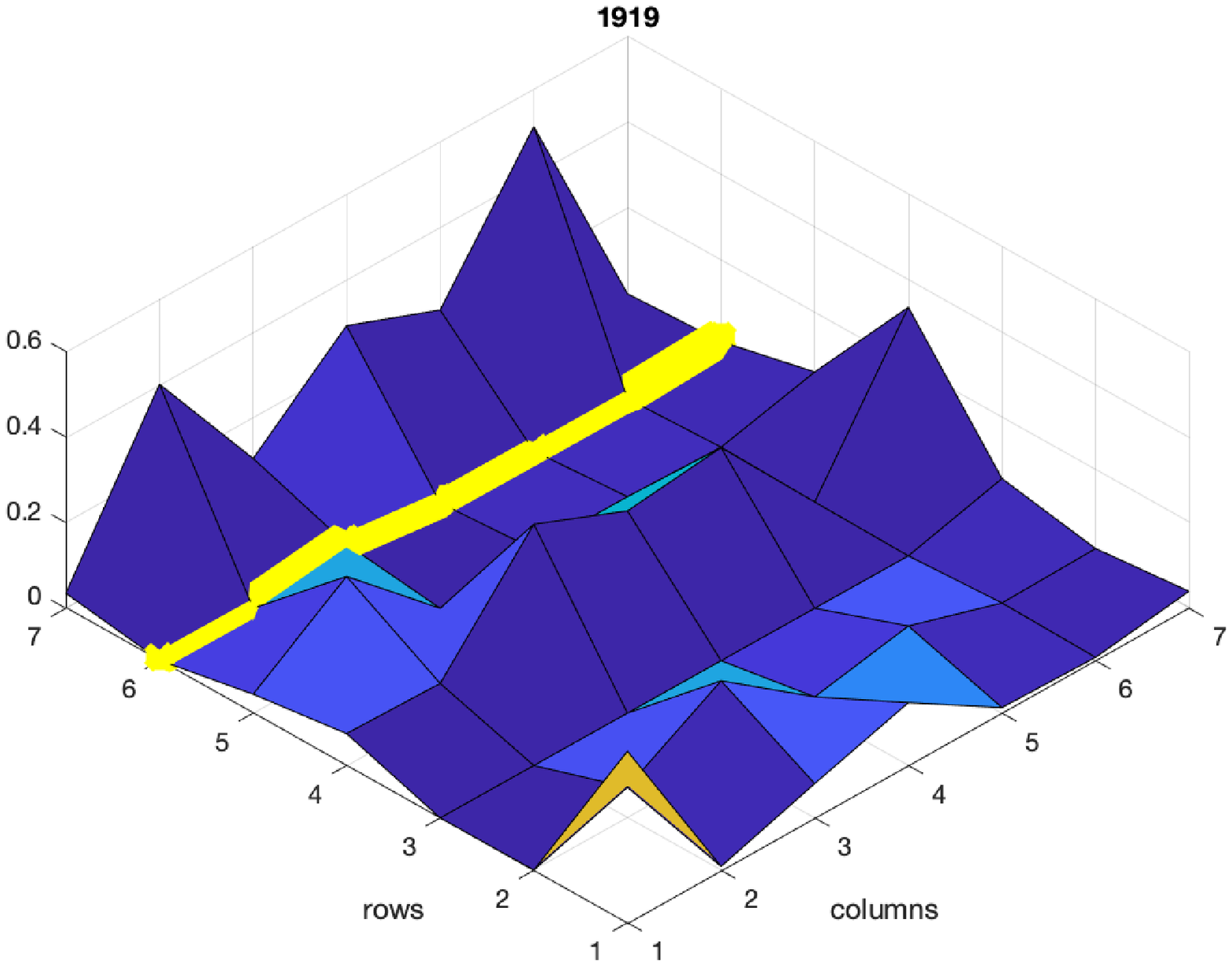} \label{fig:A_1919}}
\caption{The technical coefficients matrix ($A$) of the US economy for each year. The sectors are as follows: Agriculture (1), Mining (2), Construction (3), Manufacturing (4), Trade, Transport \& Utilities (5), Services (6), and Other (7). (Case study~\ref{sec:real}).}
\label{fig:A_dist}
\end{figure}

\begin{figure}[h]
    \centering
    \subcaptionbox{2006}{\includegraphics[width=0.32\textwidth]{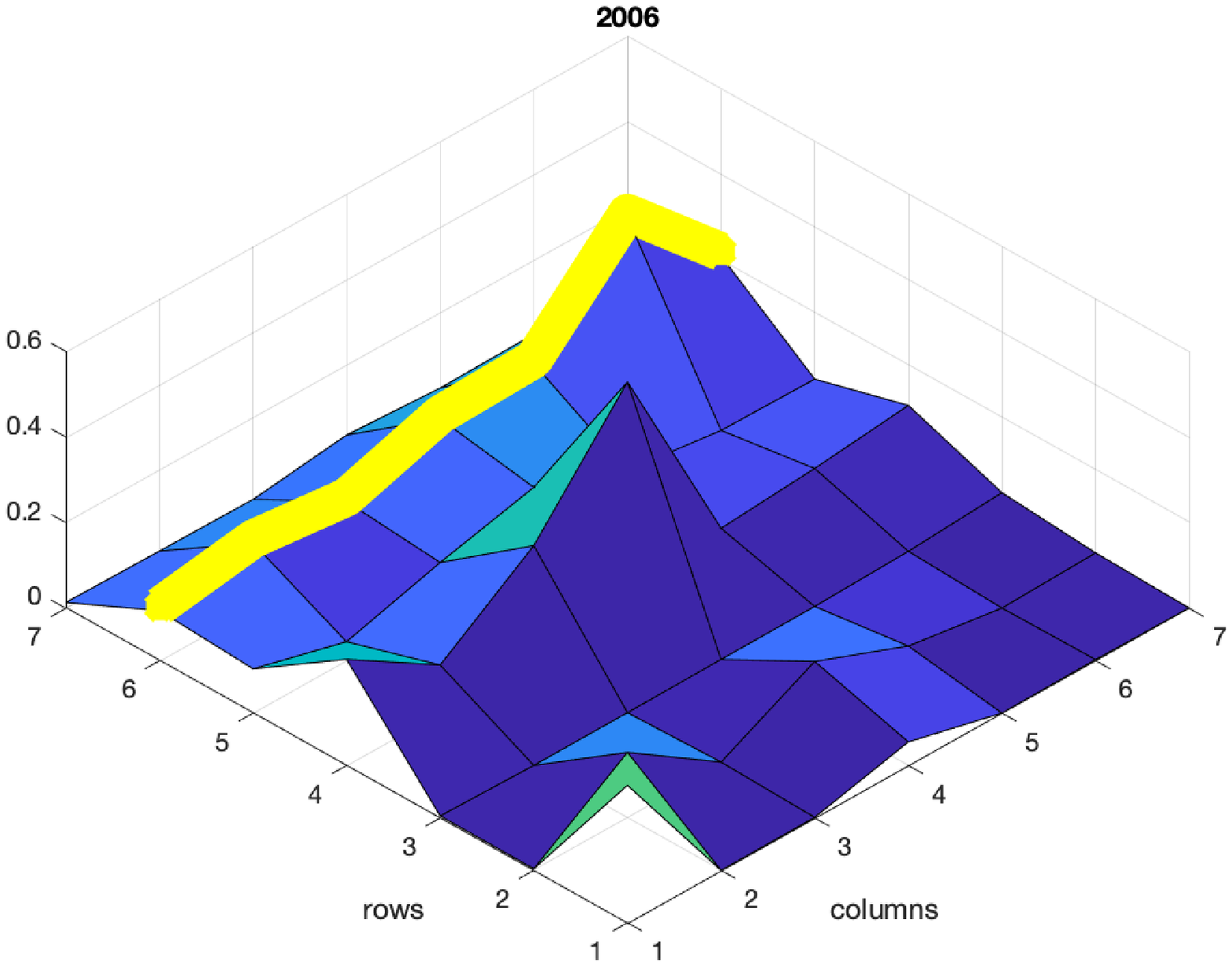} \label{fig:d_2006}}
    \subcaptionbox{2002}{\includegraphics[width=0.32\textwidth]{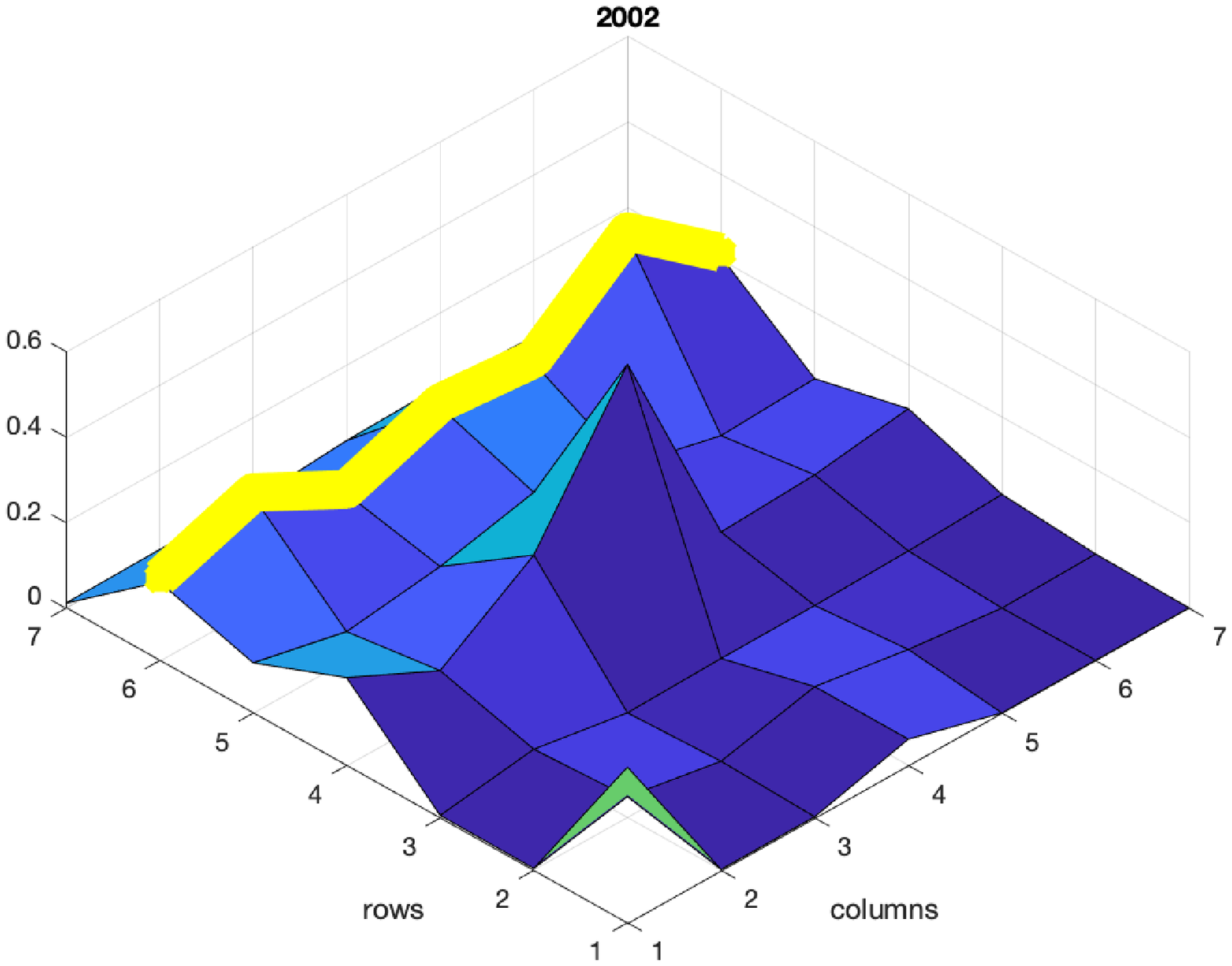} \label{fig:d_2002}}
    \subcaptionbox{1997}{\includegraphics[width=0.32\textwidth]{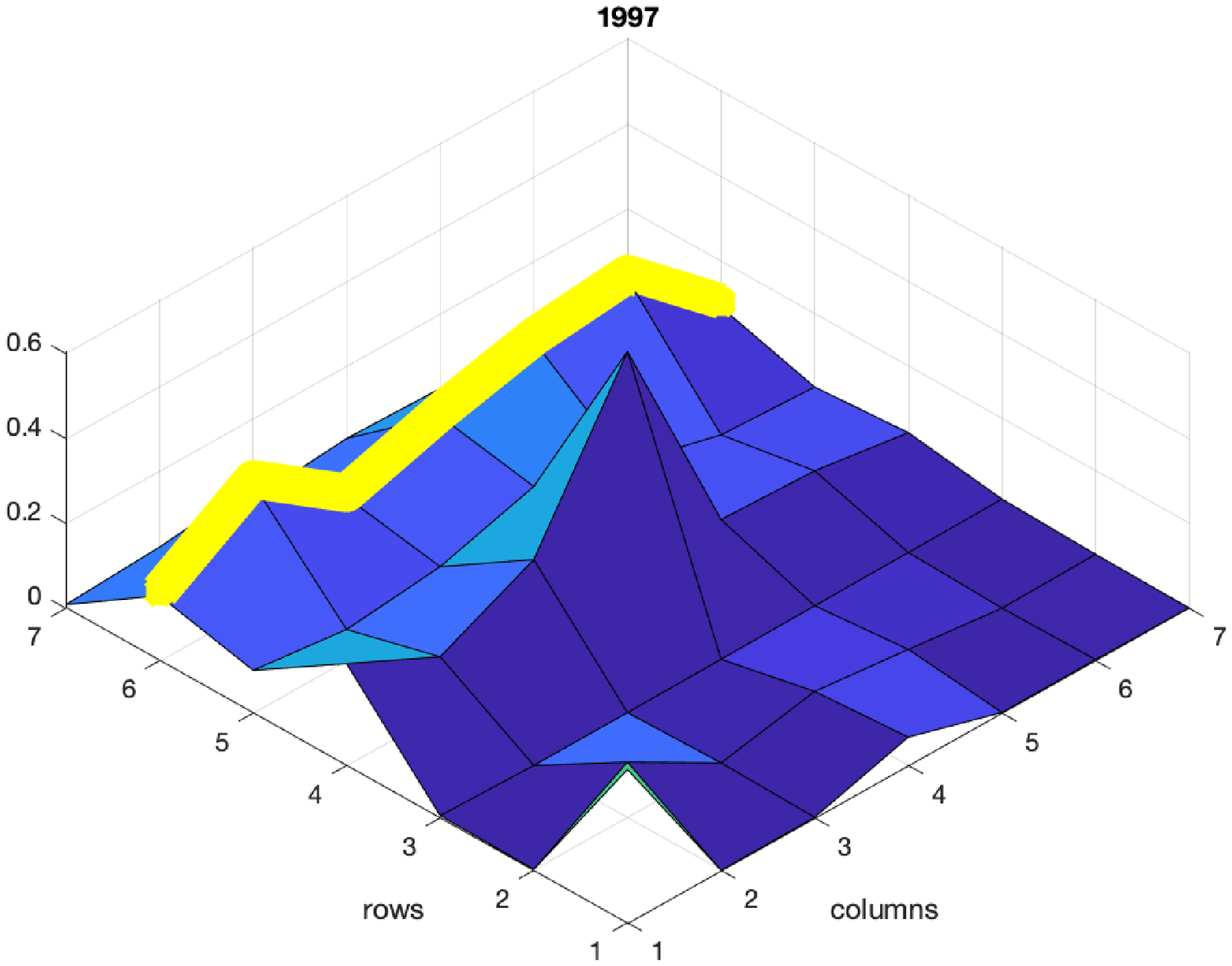} \label{fig:d_1997}}  \\
    \subcaptionbox{1992}{\includegraphics[width=0.32\textwidth]{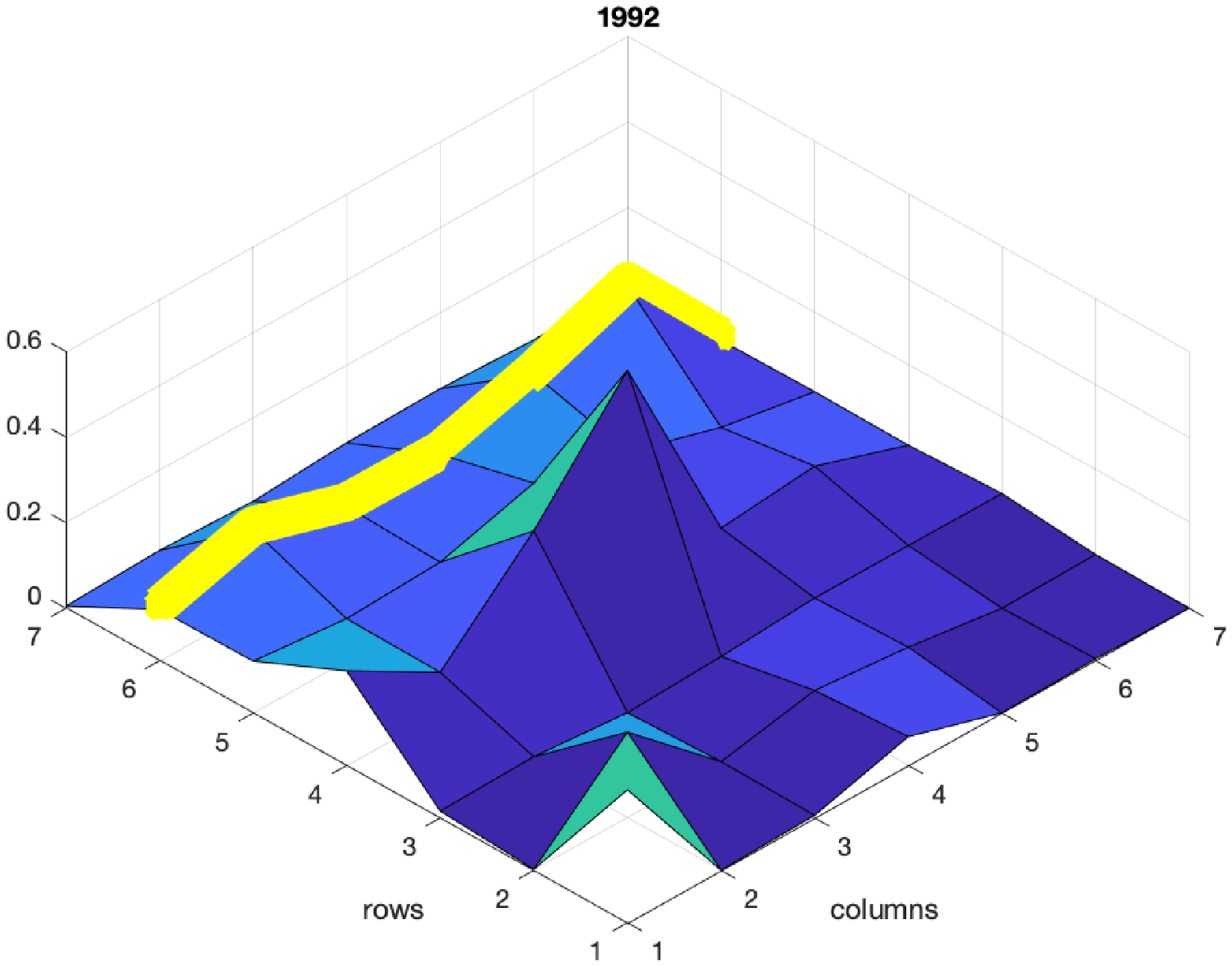} \label{fig:d_1992}}
    \subcaptionbox{1987}{\includegraphics[width=0.32\textwidth]{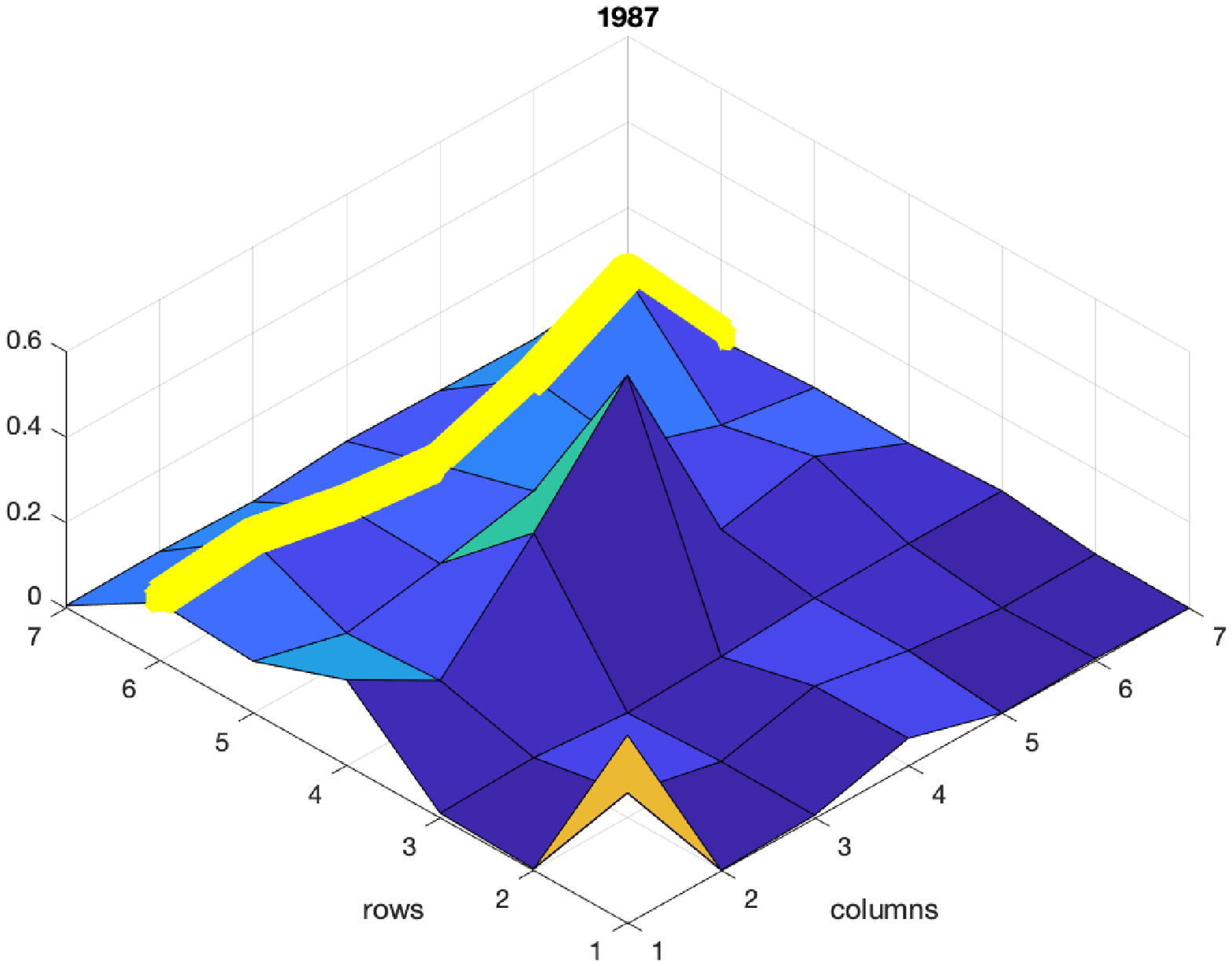} \label{fig:d_1987}}
    \subcaptionbox{1982}{\includegraphics[width=0.32\textwidth]{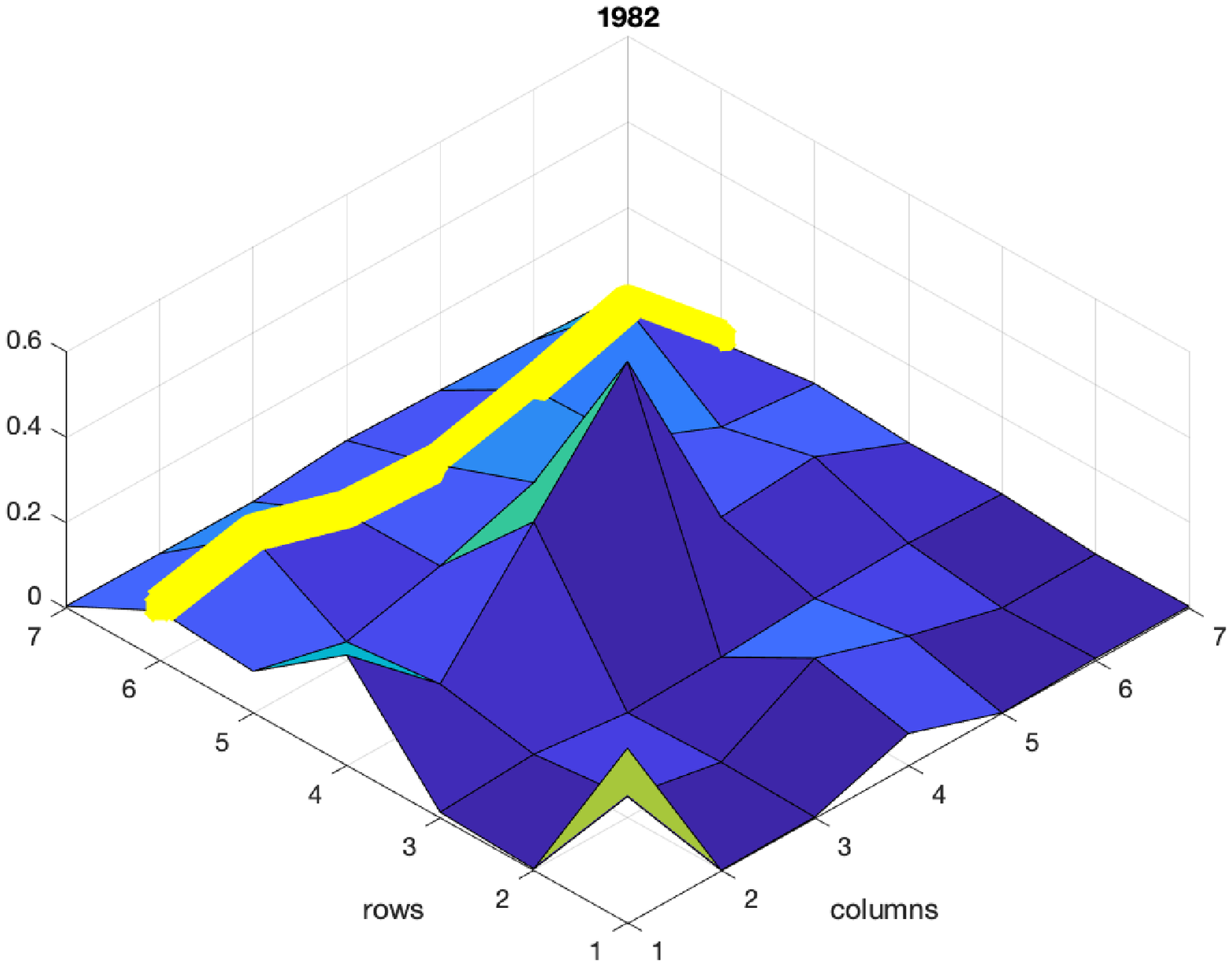} \label{fig:d_1982}}  \\
    \subcaptionbox{1977}{\includegraphics[width=0.32\textwidth]{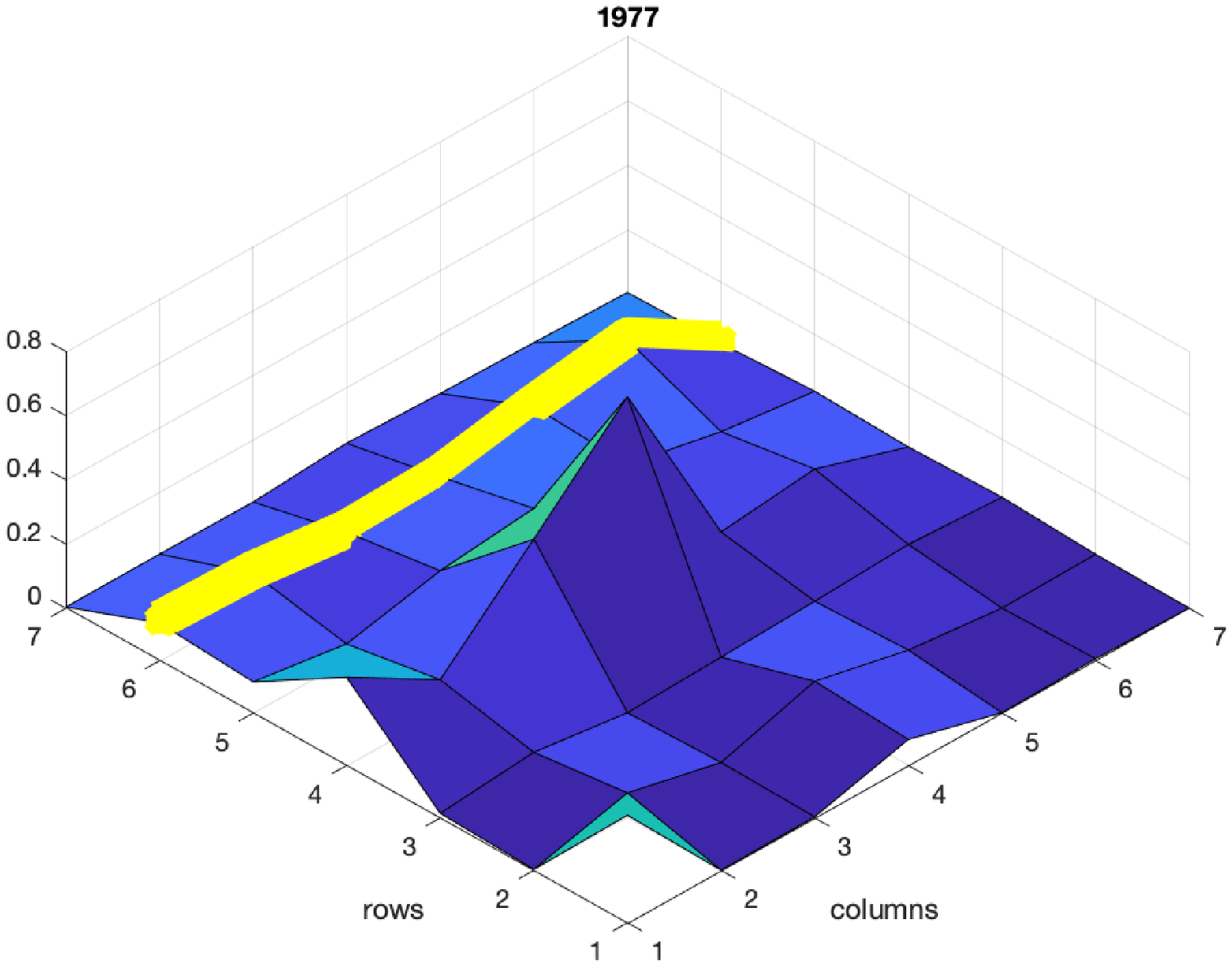} \label{fig:d_1977}}
    \subcaptionbox{1972}{\includegraphics[width=0.32\textwidth]{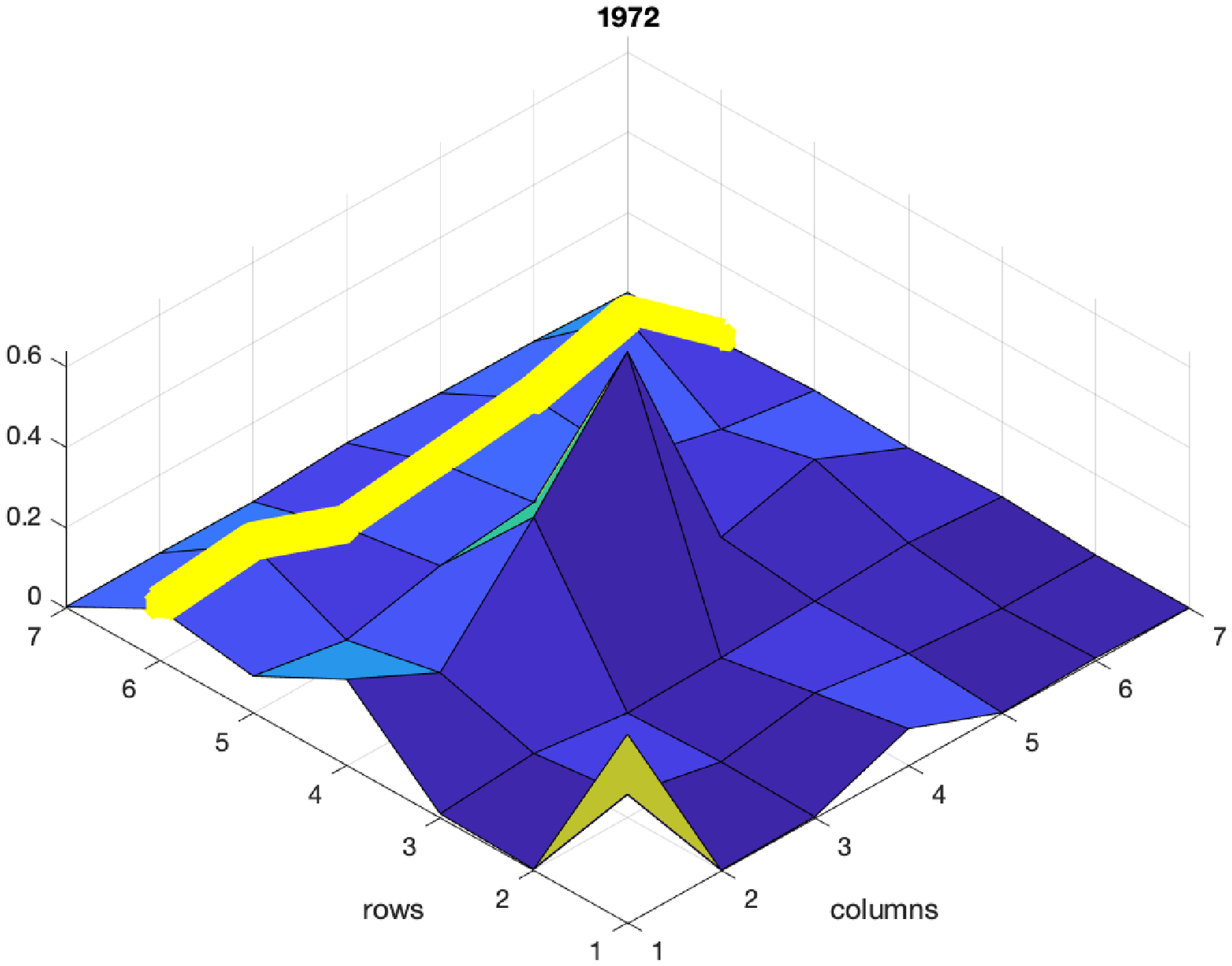} \label{fig:d_1972}}
    \subcaptionbox{1967}{\includegraphics[width=0.32\textwidth]{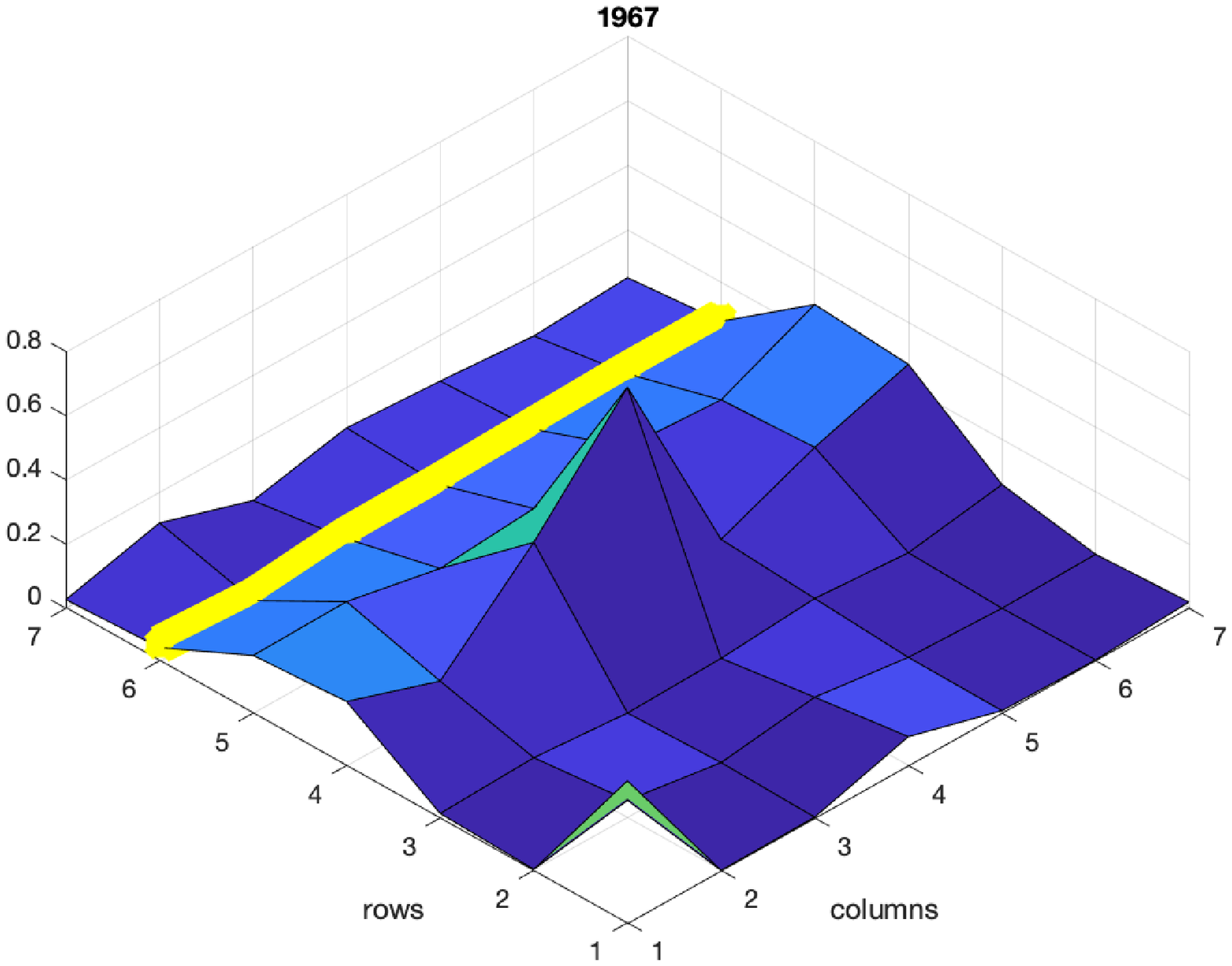} \label{fig:d_1967}}  \\
    \subcaptionbox{1963}{\includegraphics[width=0.32\textwidth]{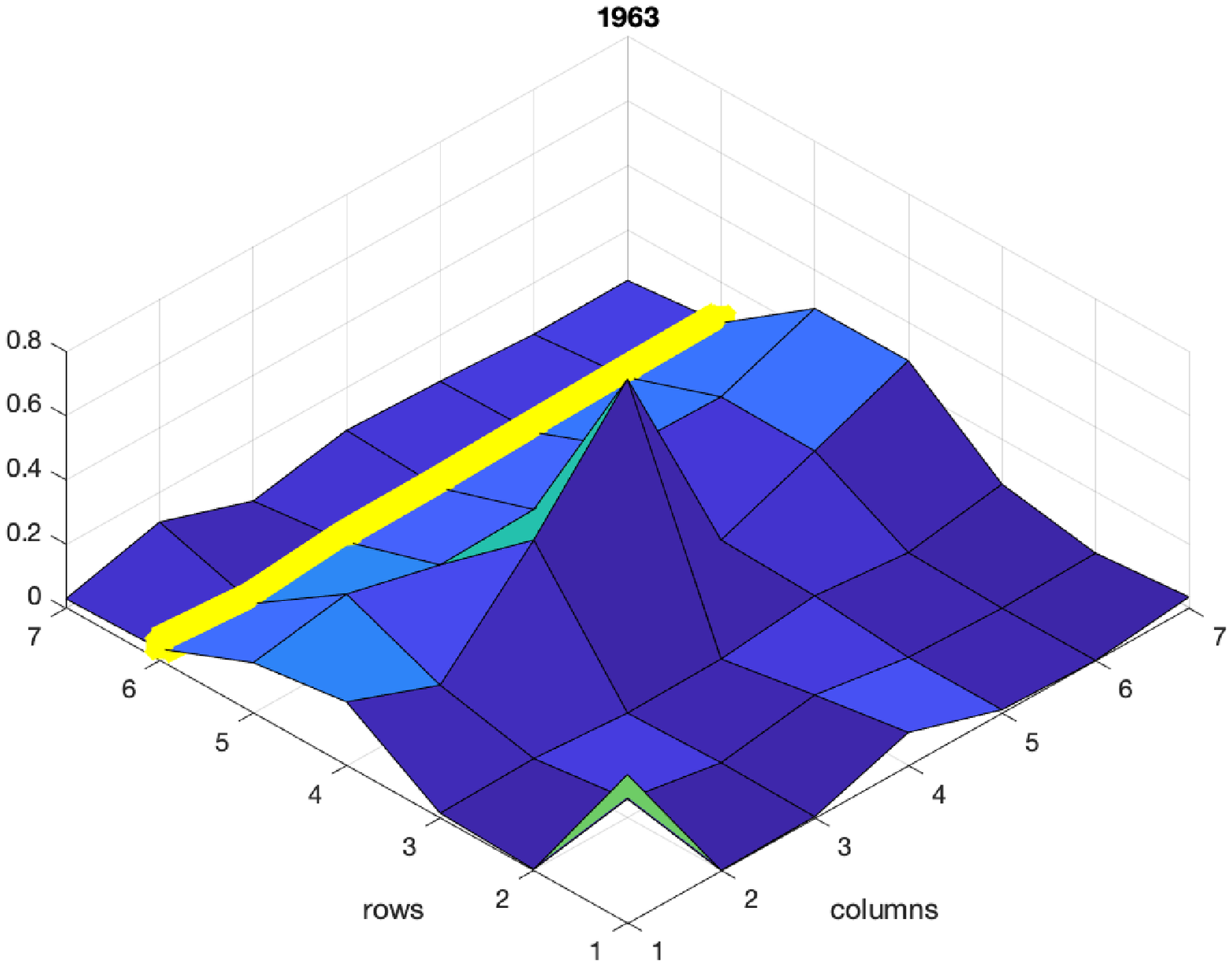} \label{fig:d_1963}}
    \subcaptionbox{1958}{\includegraphics[width=0.32\textwidth]{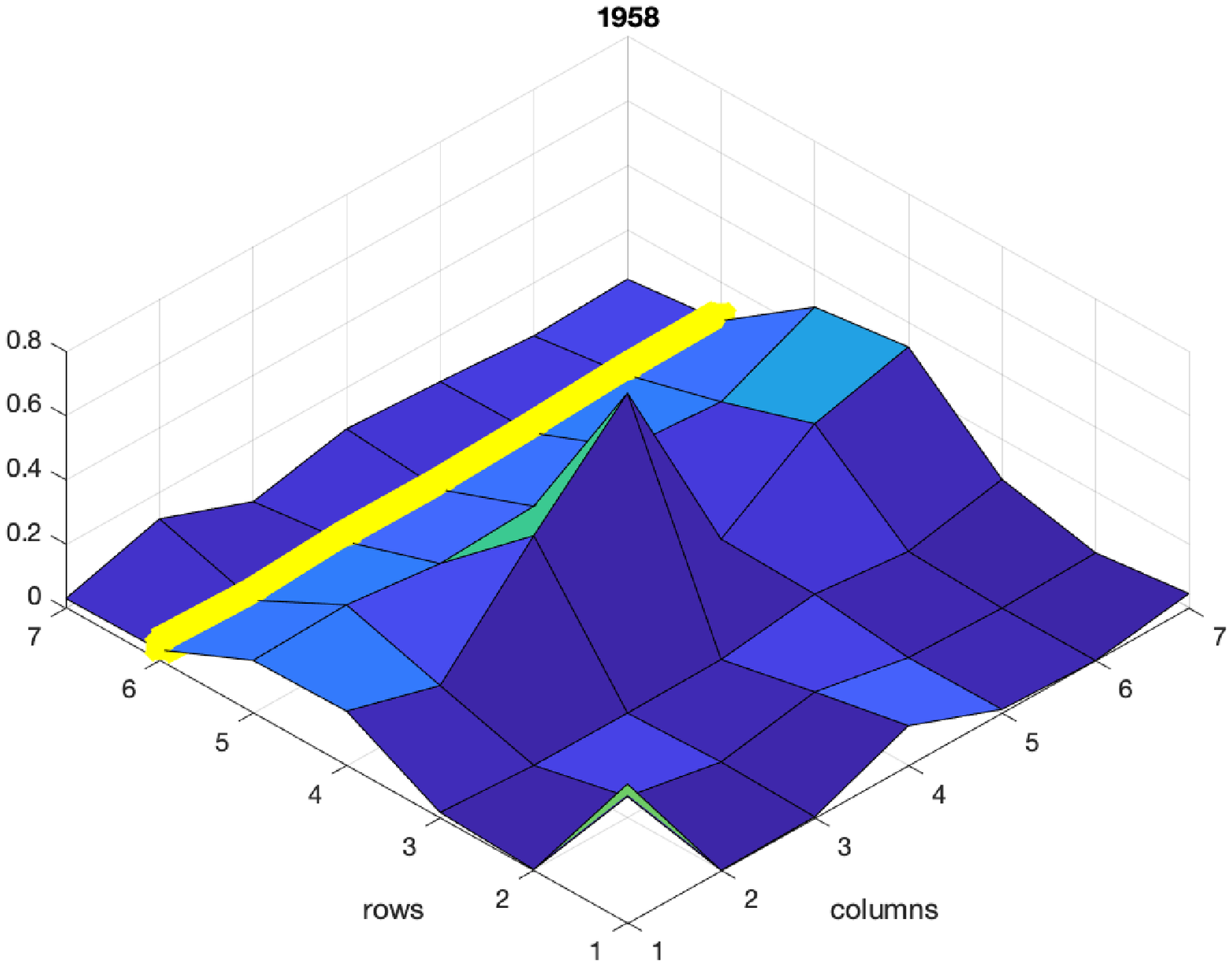} \label{fig:d_1958}}
    \subcaptionbox{1947}{\includegraphics[width=0.32\textwidth]{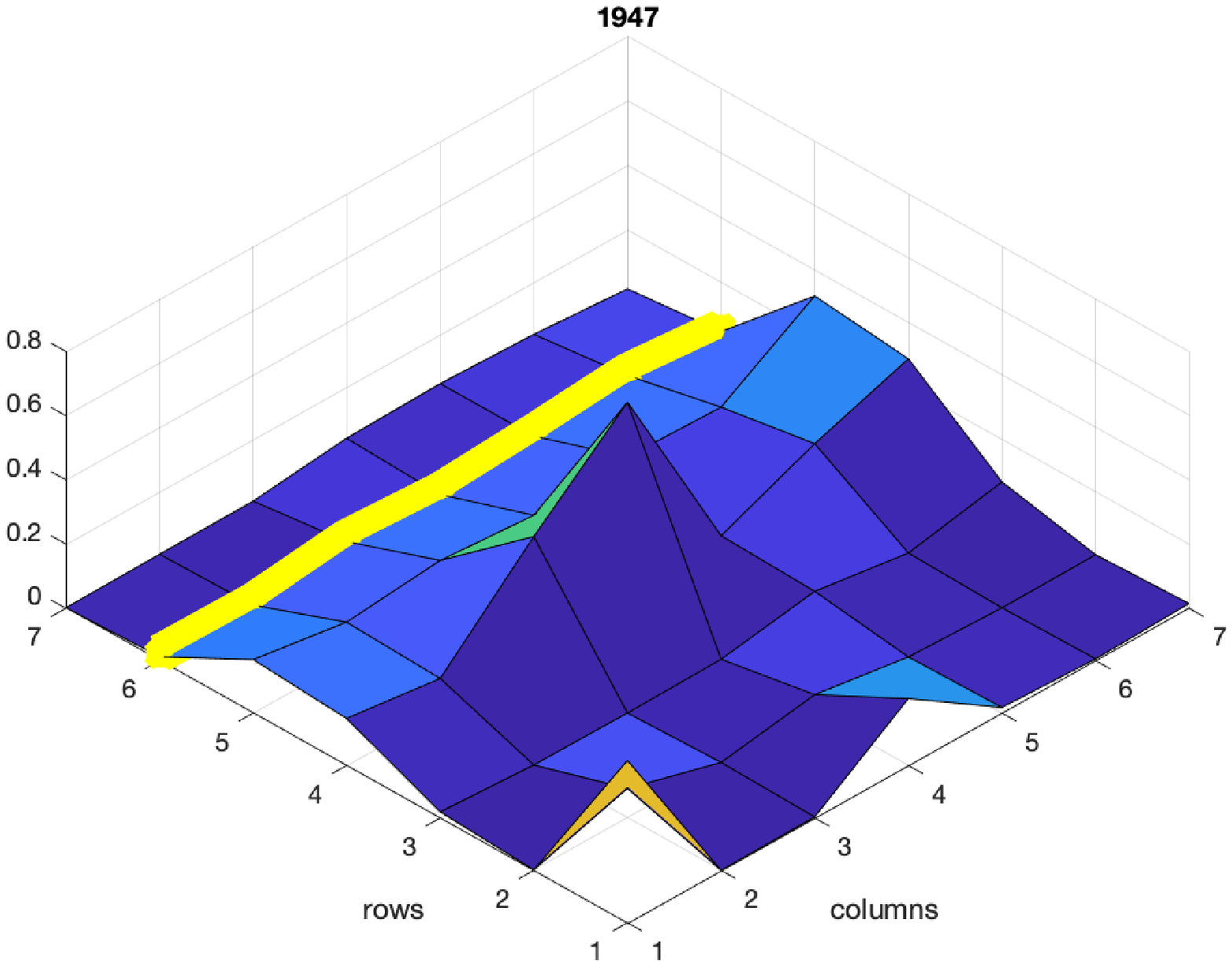} \label{fig:d_1947}}  \\
    \subcaptionbox{1939}{\includegraphics[width=0.32\textwidth]{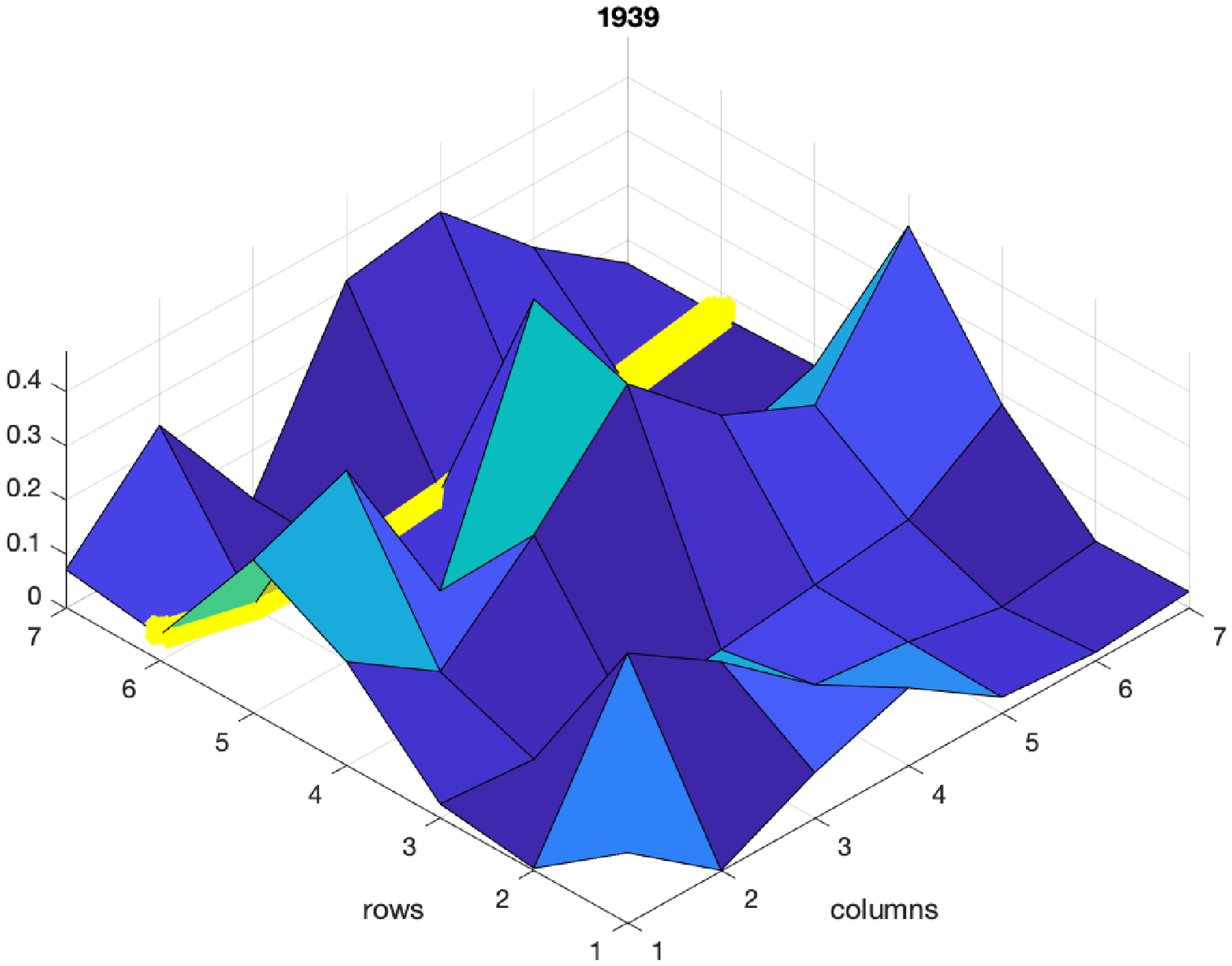} \label{fig:d_1939}}
    \subcaptionbox{1929}{\includegraphics[width=0.32\textwidth]{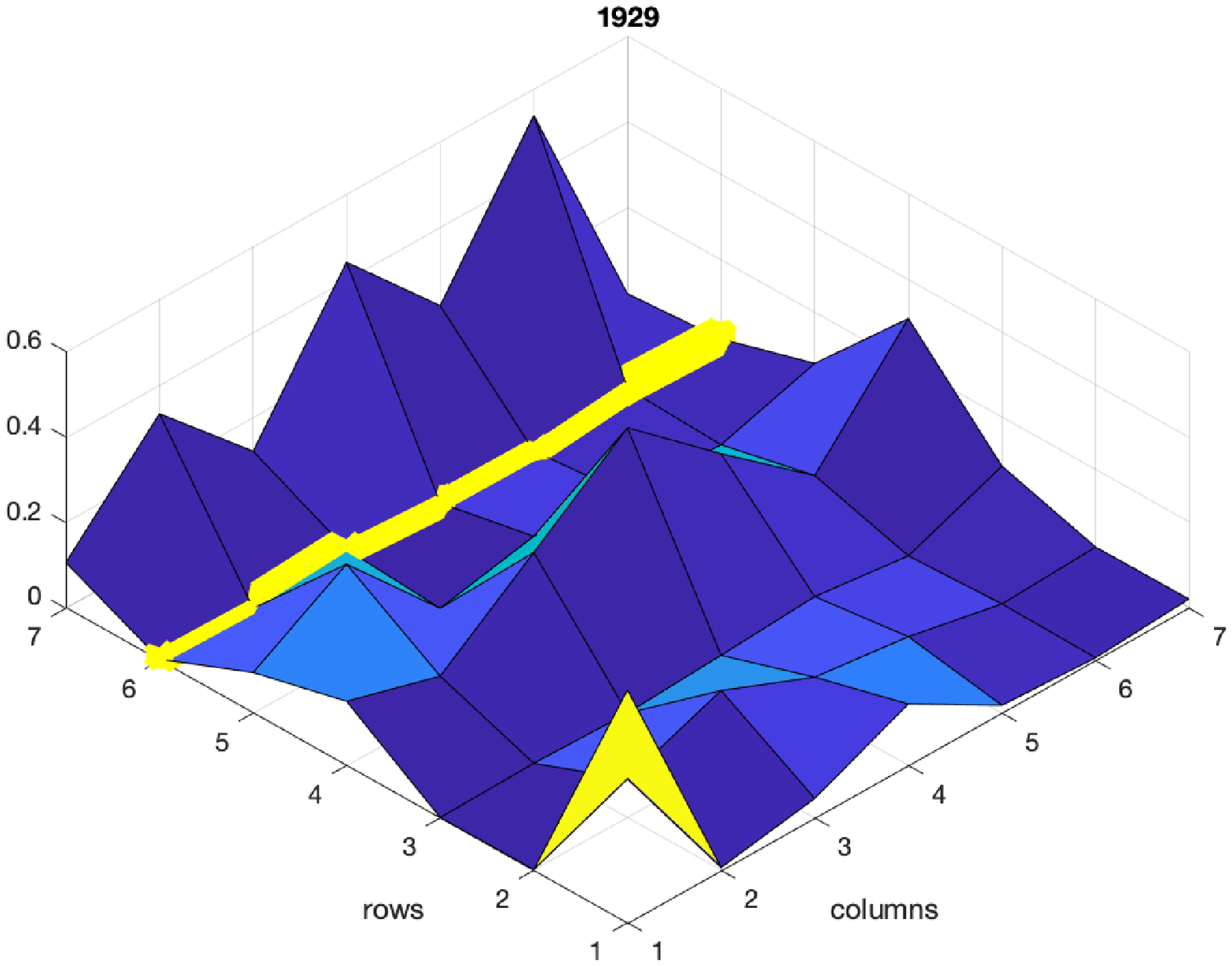} \label{fig:d_1929}}
    \subcaptionbox{1919}{\includegraphics[width=0.32\textwidth]{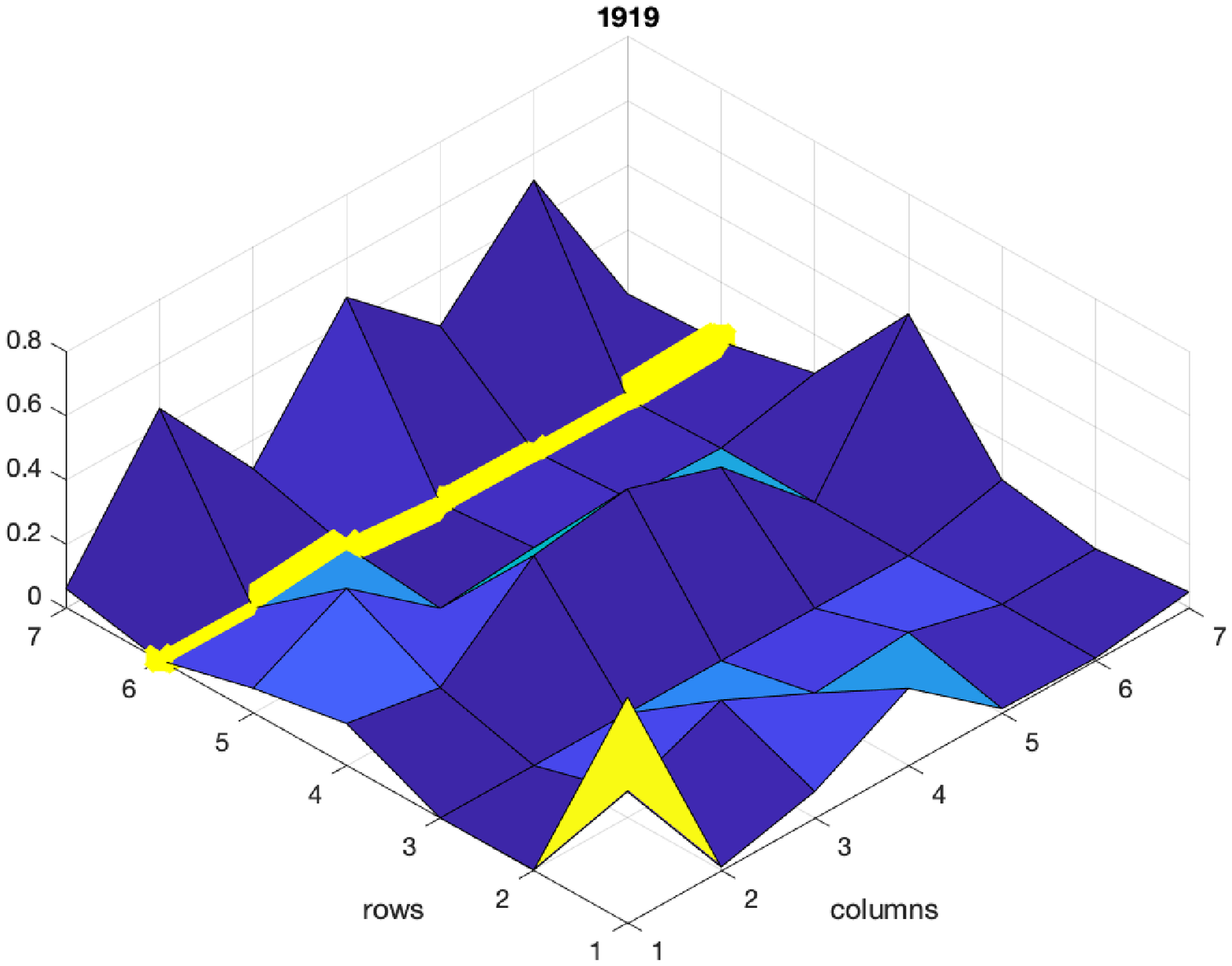} \label{fig:d_1919}}
\caption{The simple direct requirements matrices ($N^\texttt{d}$) of the US economy for each year. The sectors are as follows: Agriculture (1), Mining (2), Construction (3), Manufacturing (4), Trade, Transport \& Utilities (5), Services (6), and Other (7). (Case study~\ref{sec:real}).}
\label{fig:d_dist}
\end{figure}

\begin{figure}[h]
    \centering
    \subcaptionbox{2006}{\includegraphics[width=0.32\textwidth]{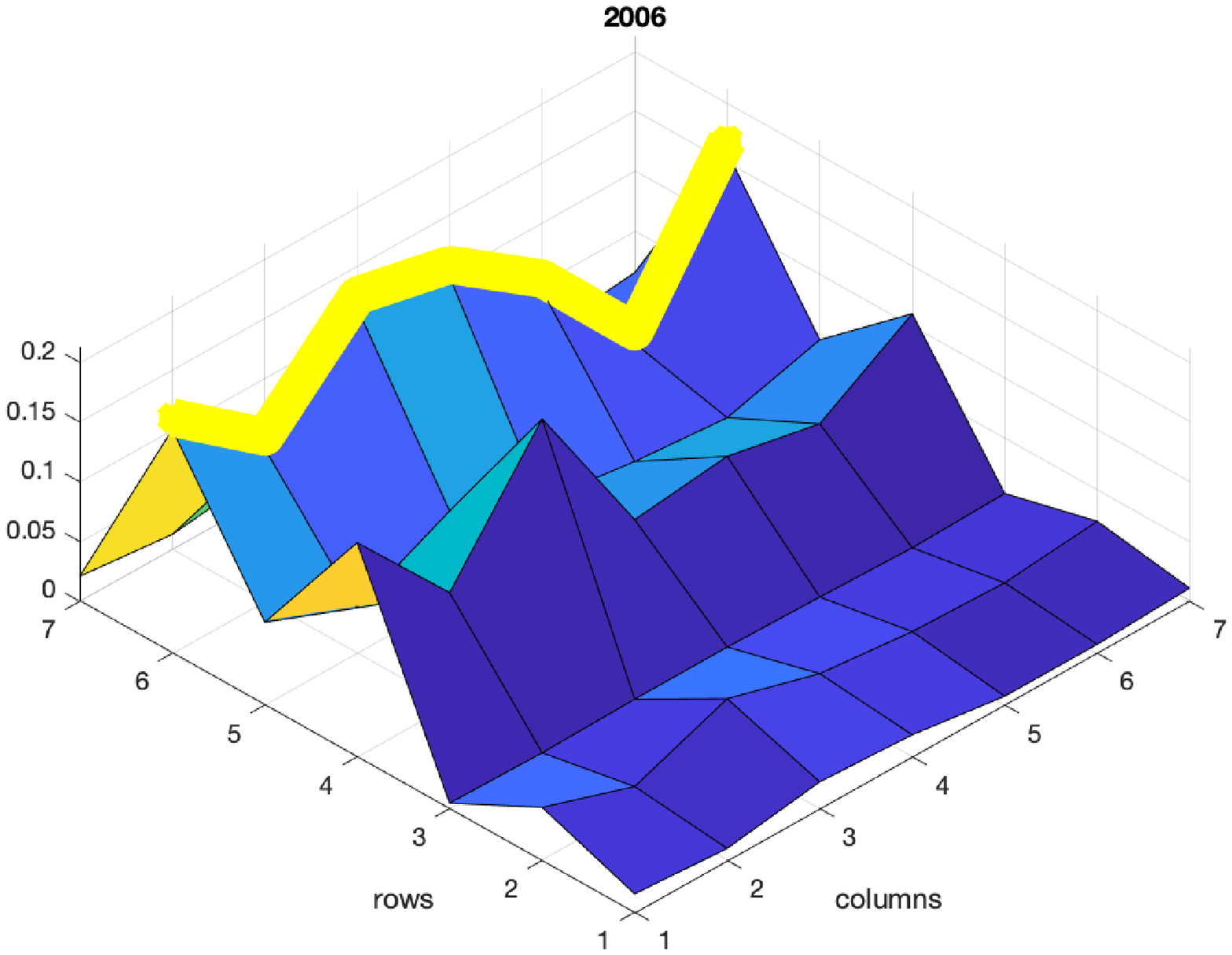} \label{fig:i_2006}}
    \subcaptionbox{2002}{\includegraphics[width=0.32\textwidth]{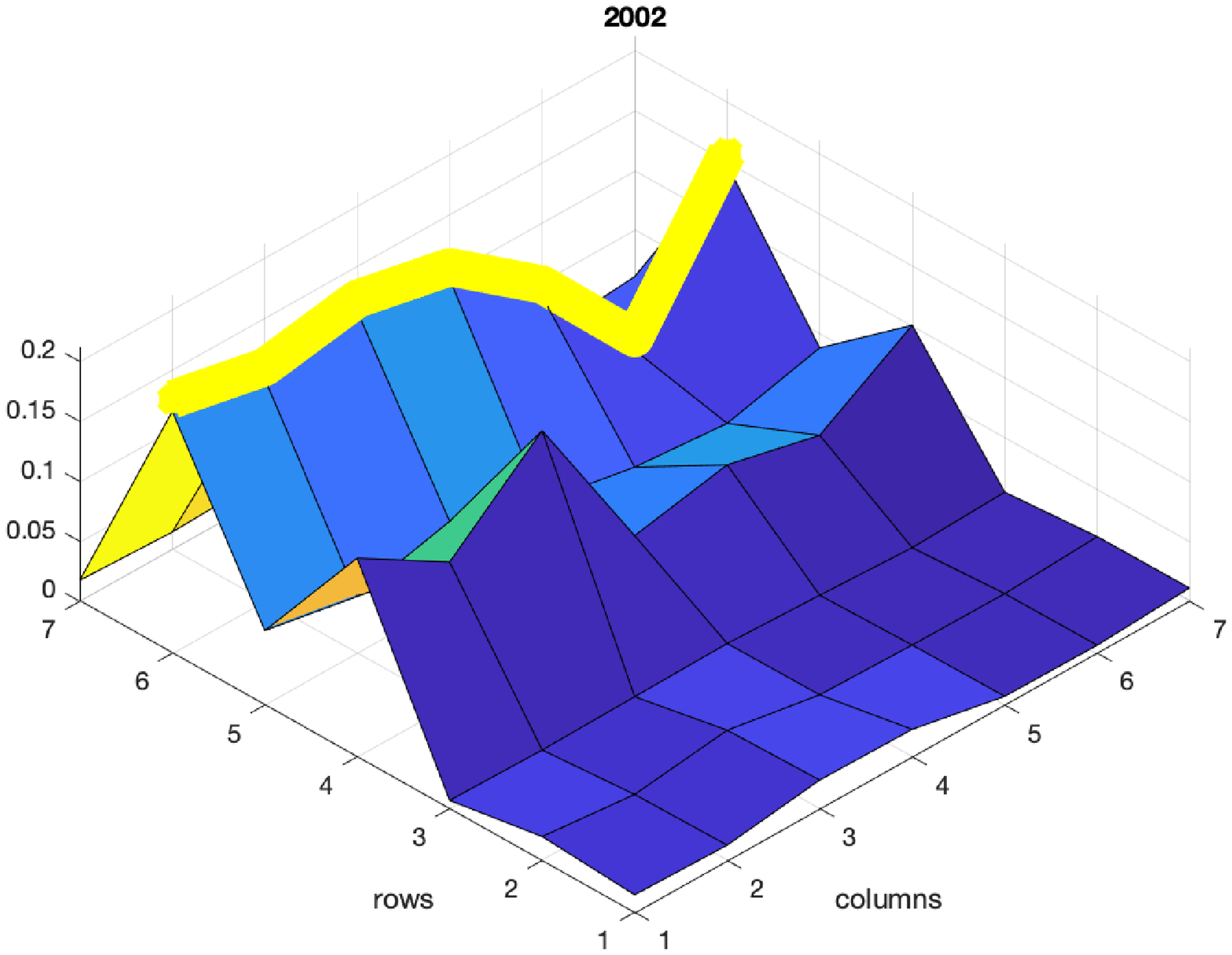} \label{fig:i_2002}}
    \subcaptionbox{1997}{\includegraphics[width=0.32\textwidth]{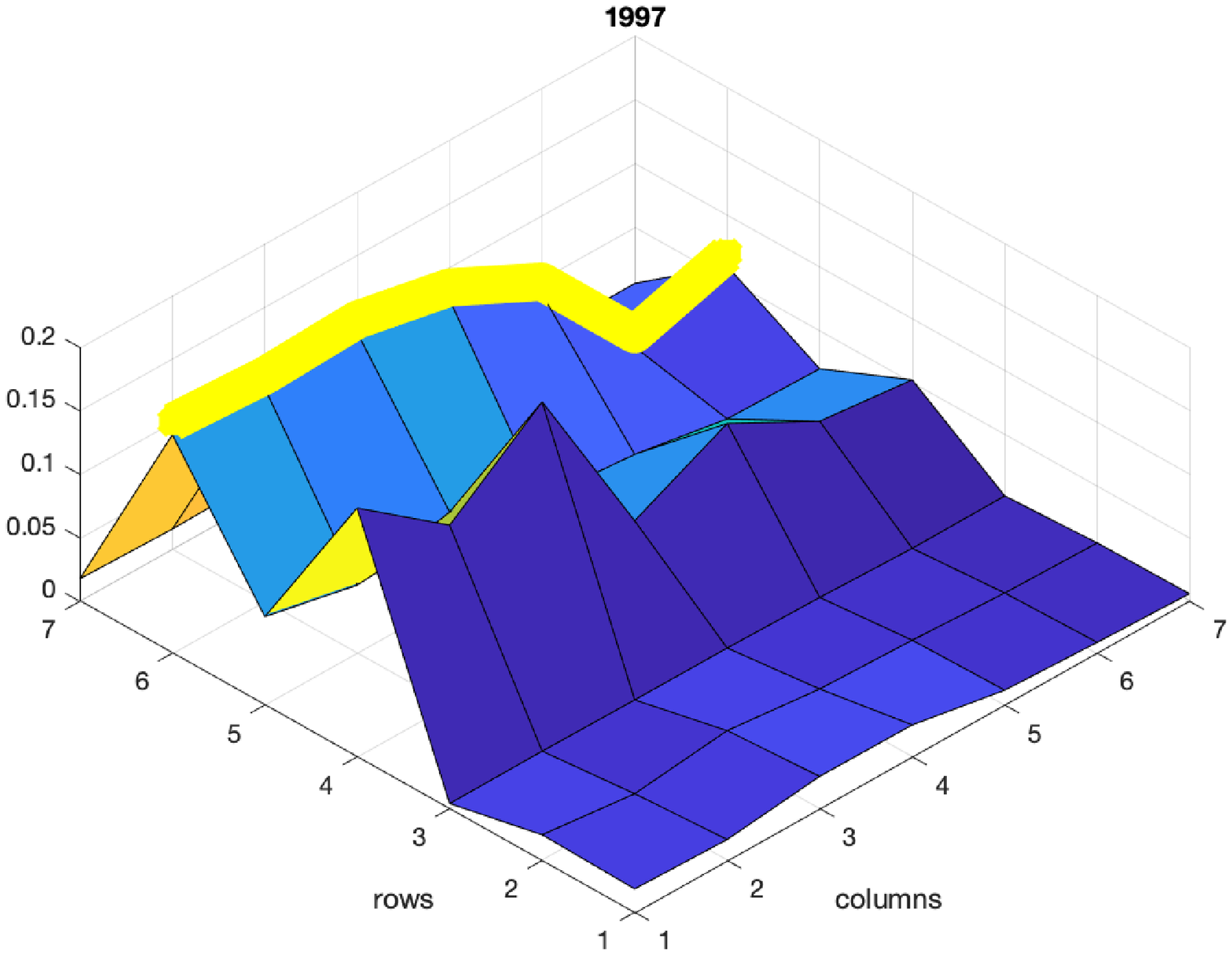} \label{fig:i_1997}}  \\
    \subcaptionbox{1992}{\includegraphics[width=0.32\textwidth]{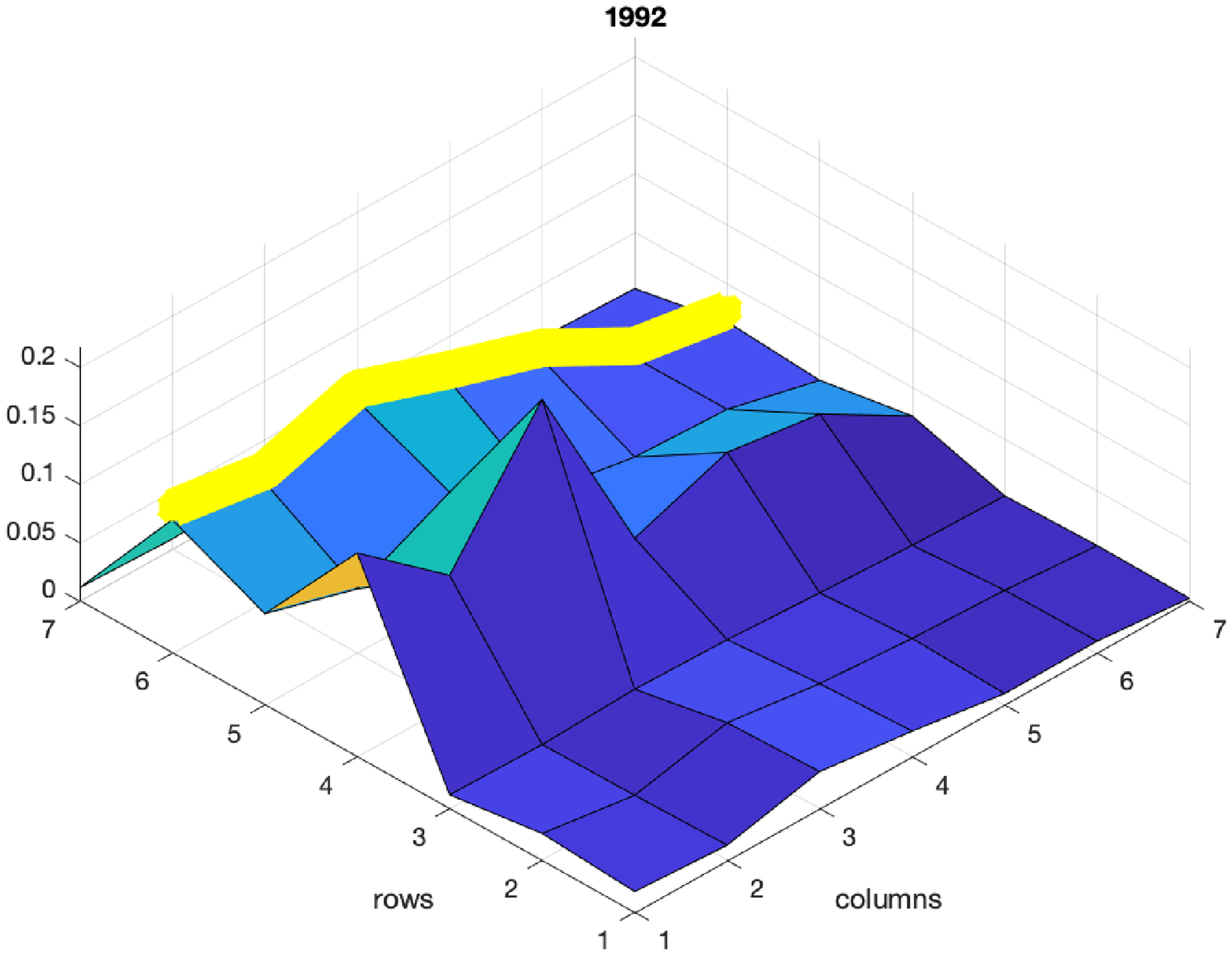} \label{fig:i_1992}}
    \subcaptionbox{1987}{\includegraphics[width=0.32\textwidth]{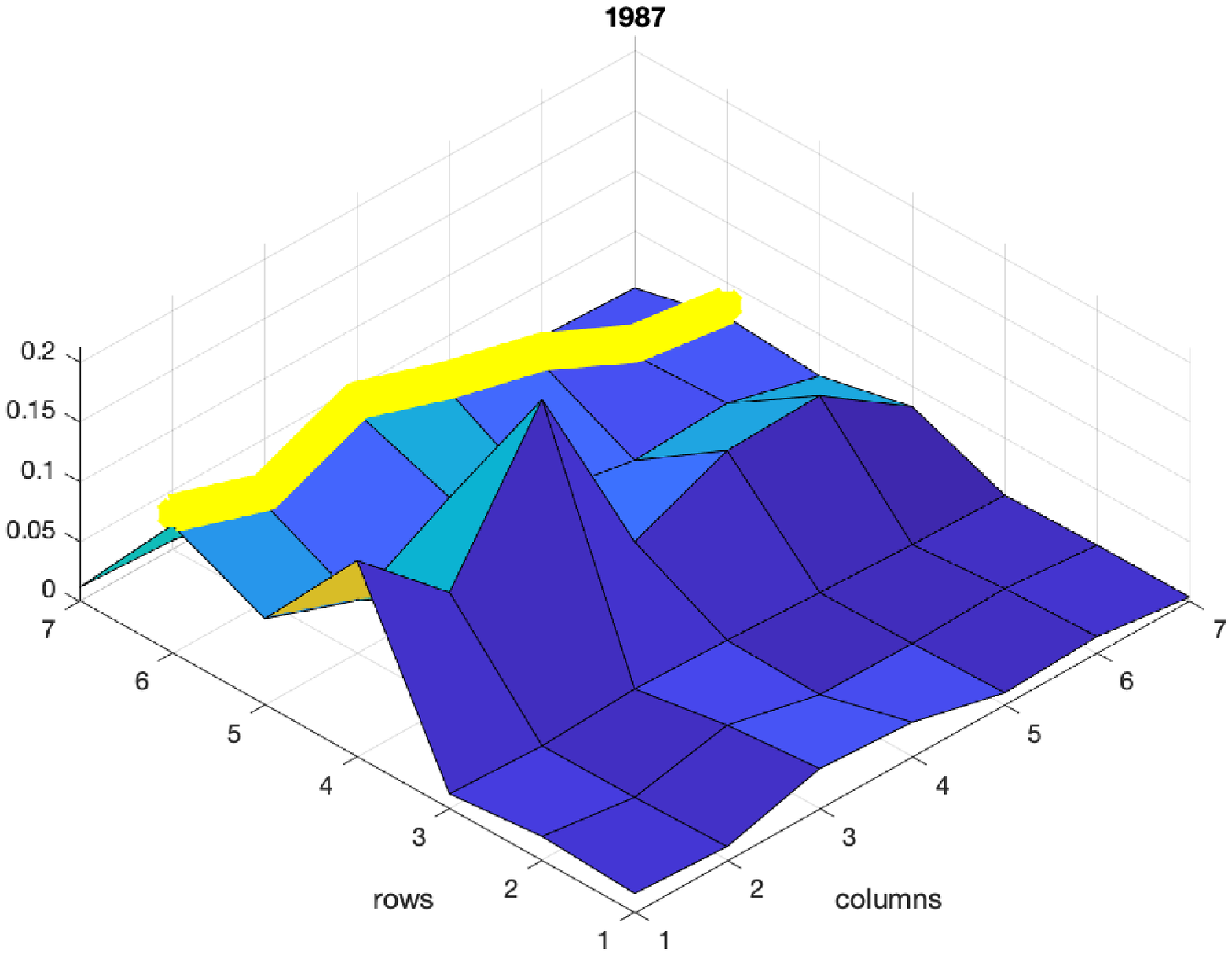} \label{fig:i_1987}}
    \subcaptionbox{1982}{\includegraphics[width=0.32\textwidth]{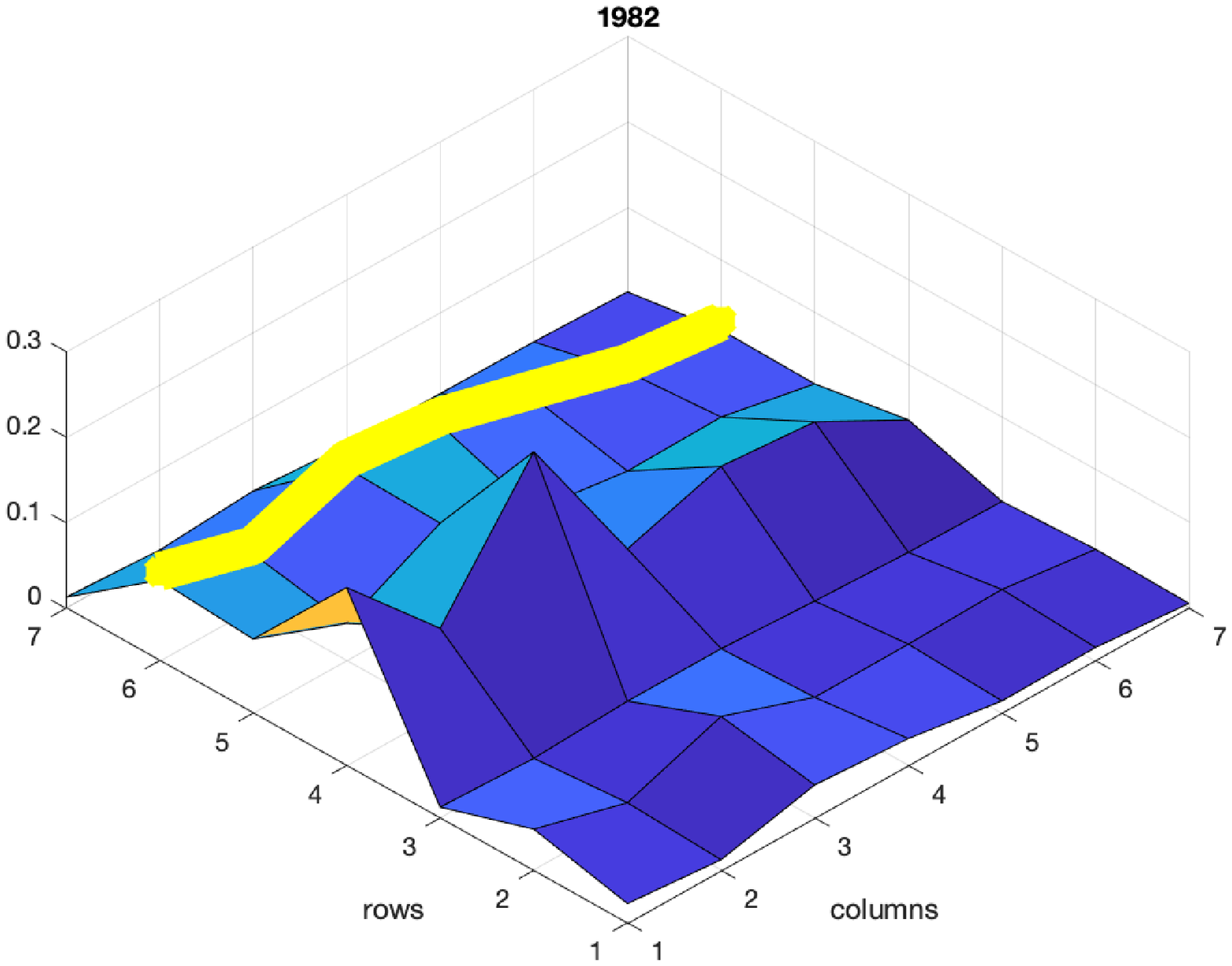} \label{fig:i_1982}}  \\
    \subcaptionbox{1977}{\includegraphics[width=0.32\textwidth]{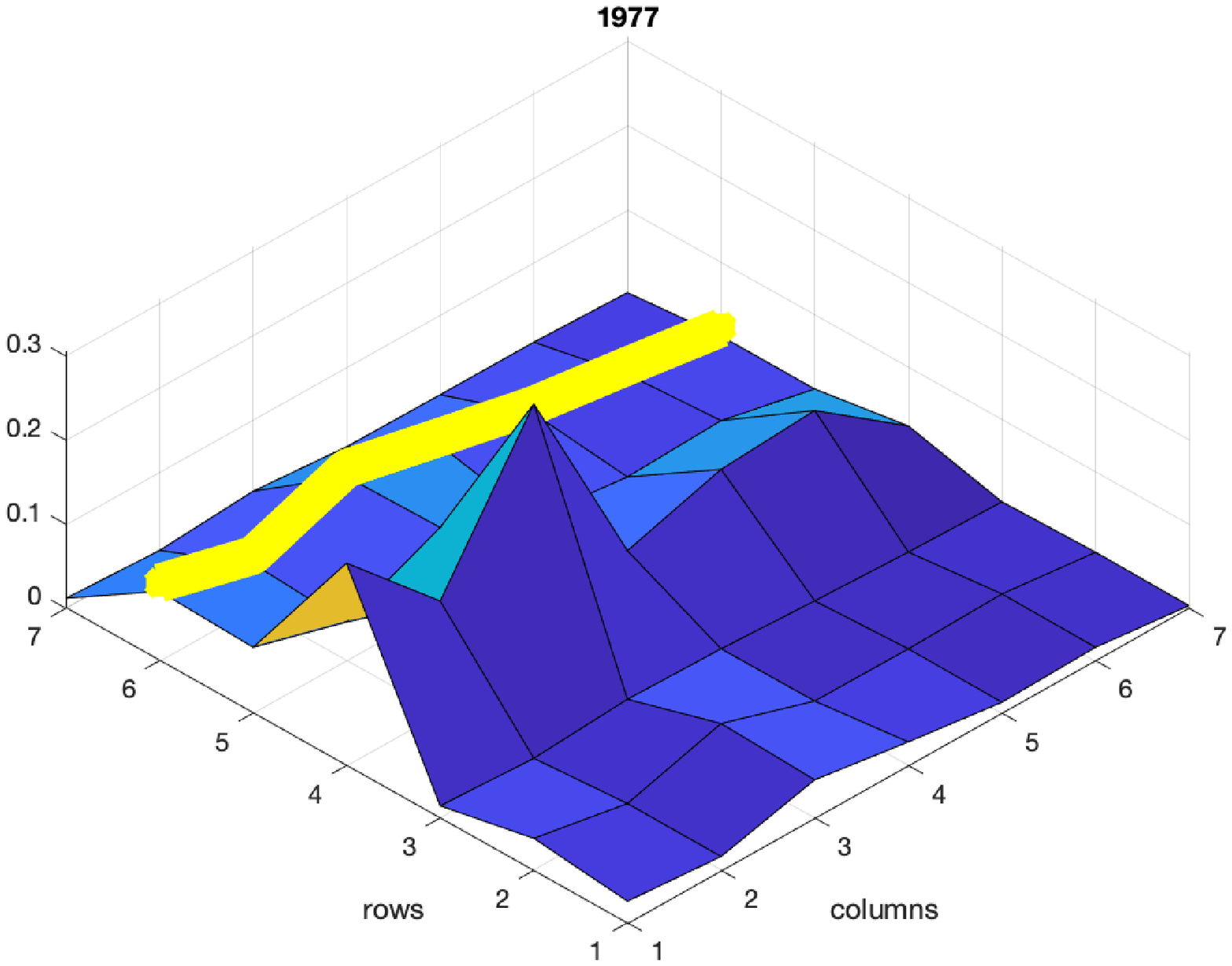} \label{fig:i_1977}}
    \subcaptionbox{1972}{\includegraphics[width=0.32\textwidth]{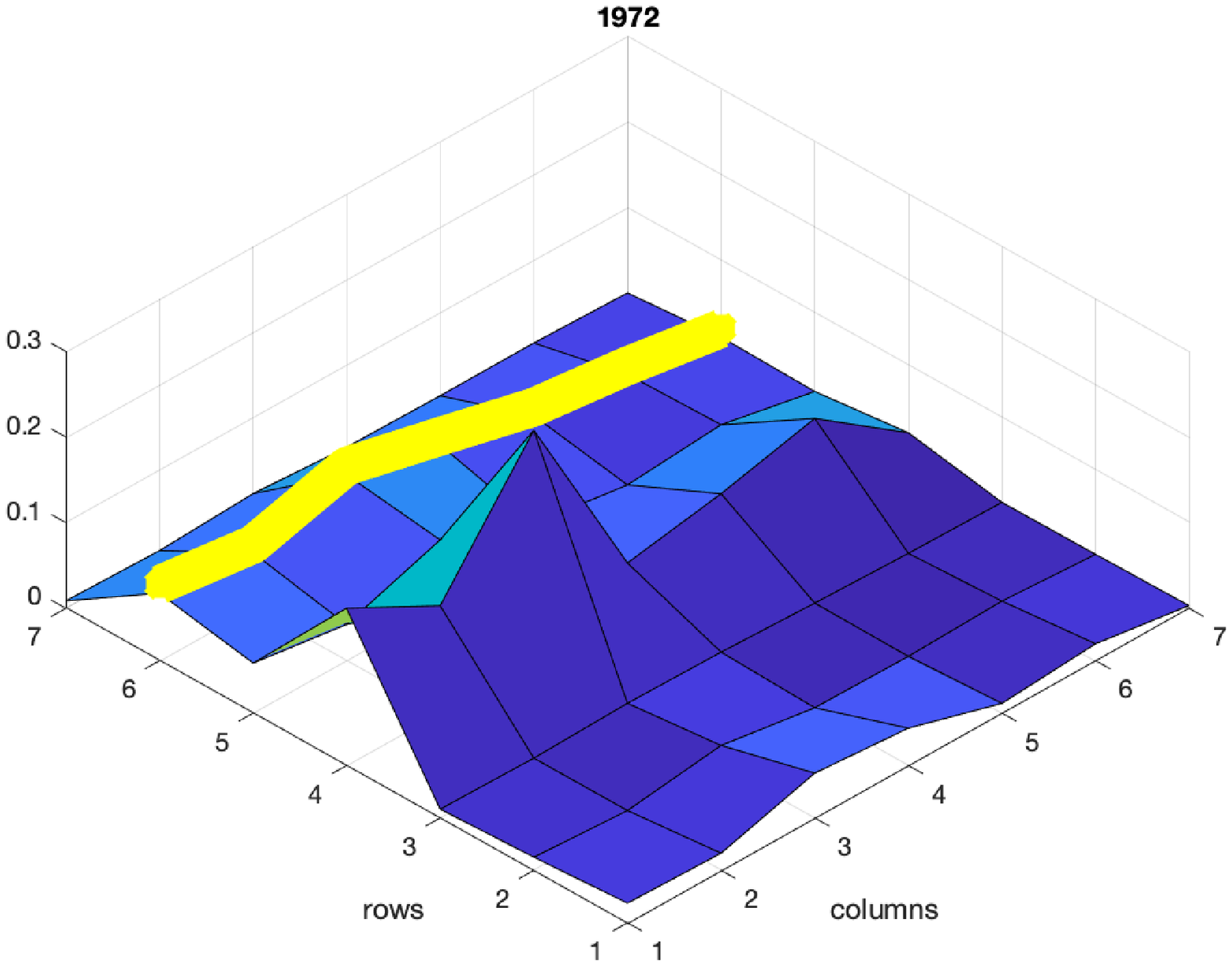} \label{fig:i_1972}}
    \subcaptionbox{1967}{\includegraphics[width=0.32\textwidth]{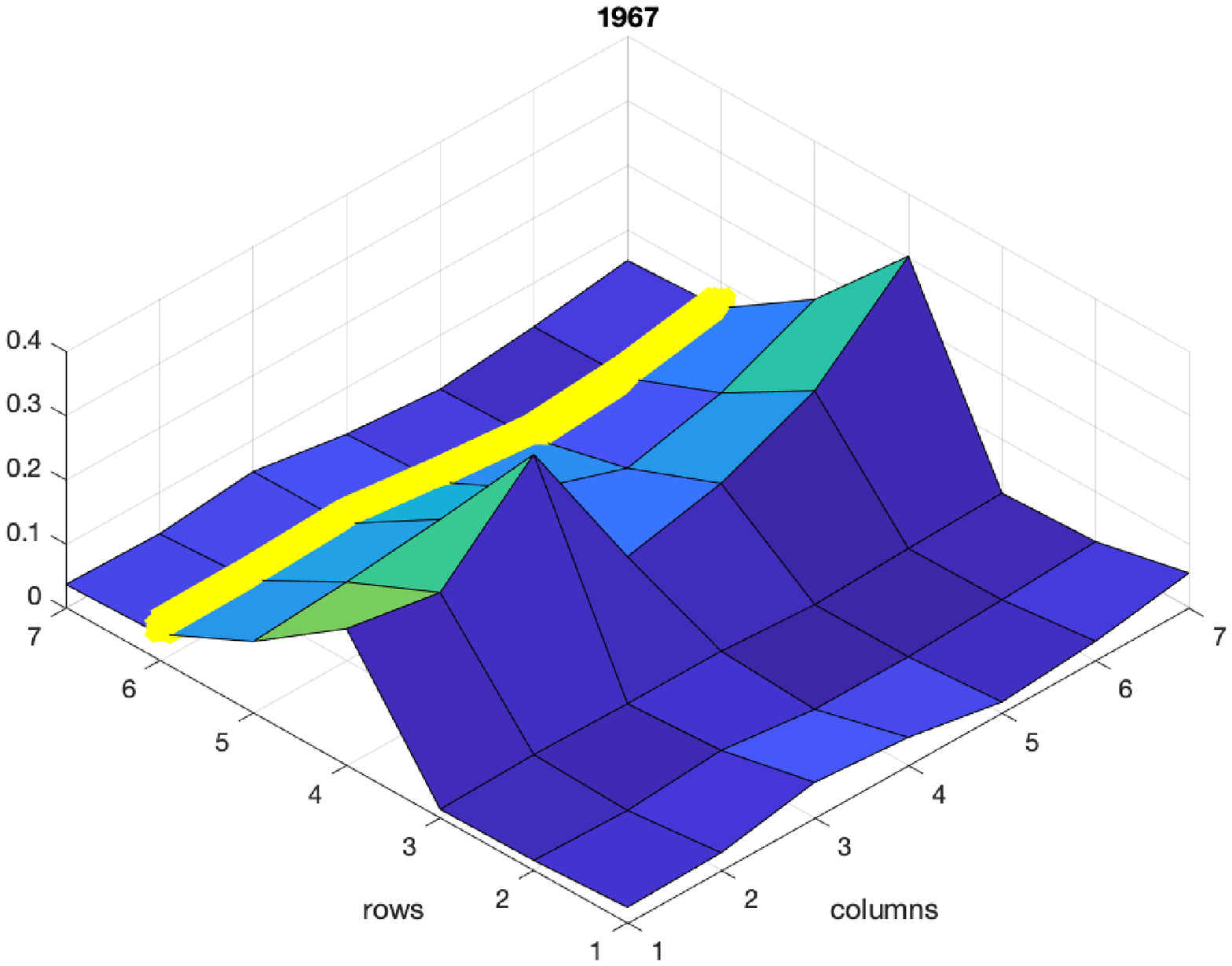} \label{fig:i_1967}}  \\
    \subcaptionbox{1963}{\includegraphics[width=0.32\textwidth]{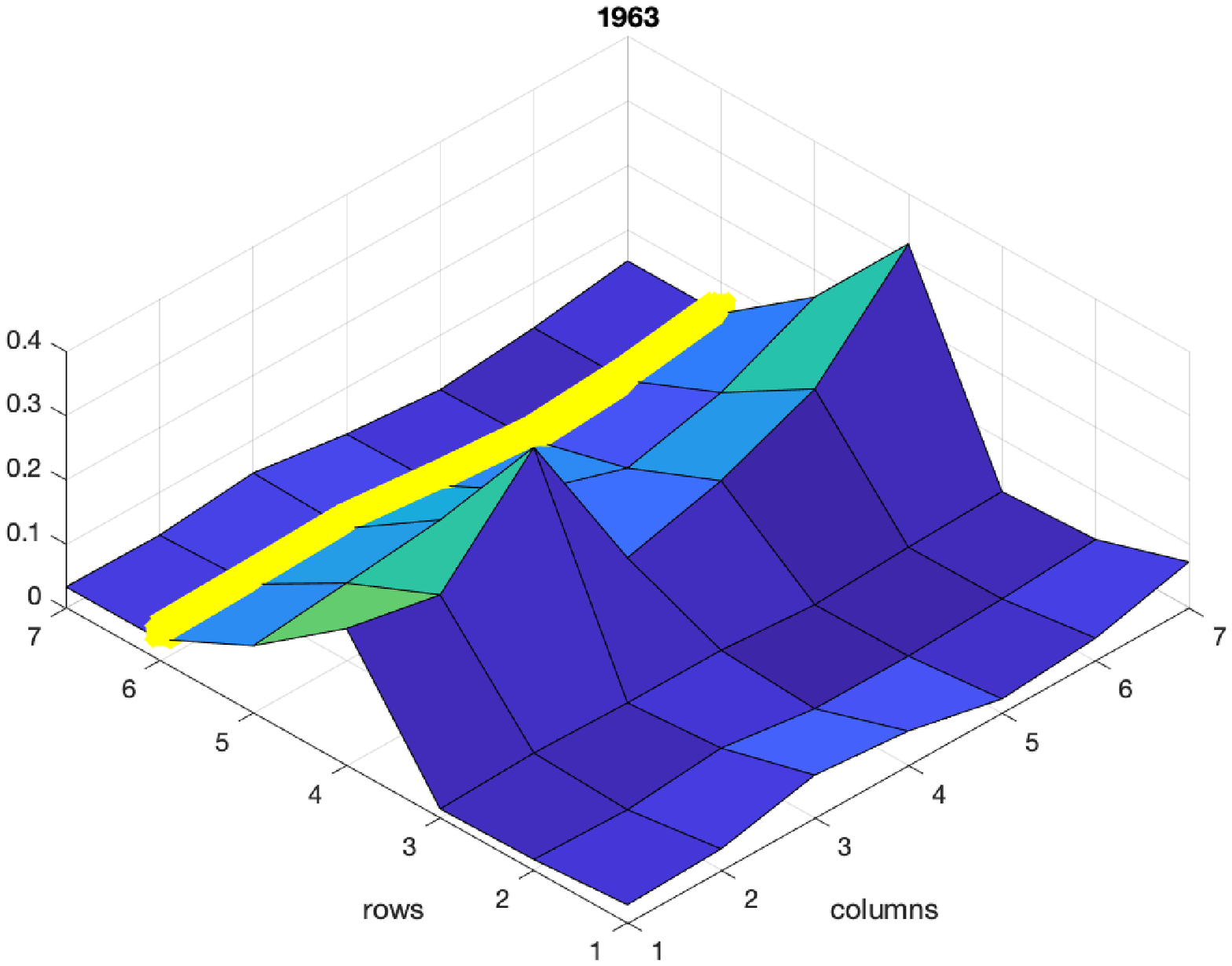} \label{fig:i_1963}}
    \subcaptionbox{1958}{\includegraphics[width=0.32\textwidth]{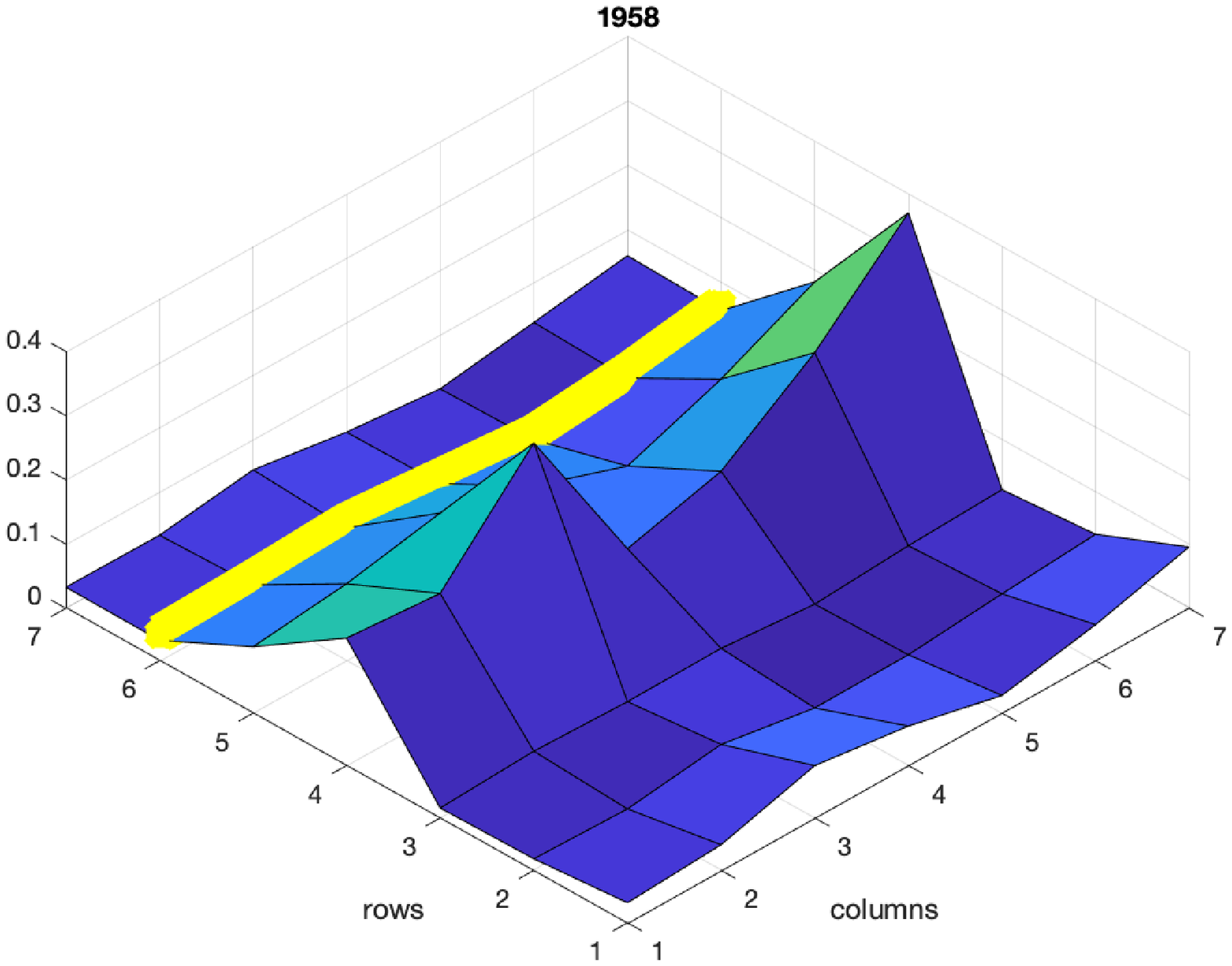} \label{fig:i_1958}}
    \subcaptionbox{1947}{\includegraphics[width=0.32\textwidth]{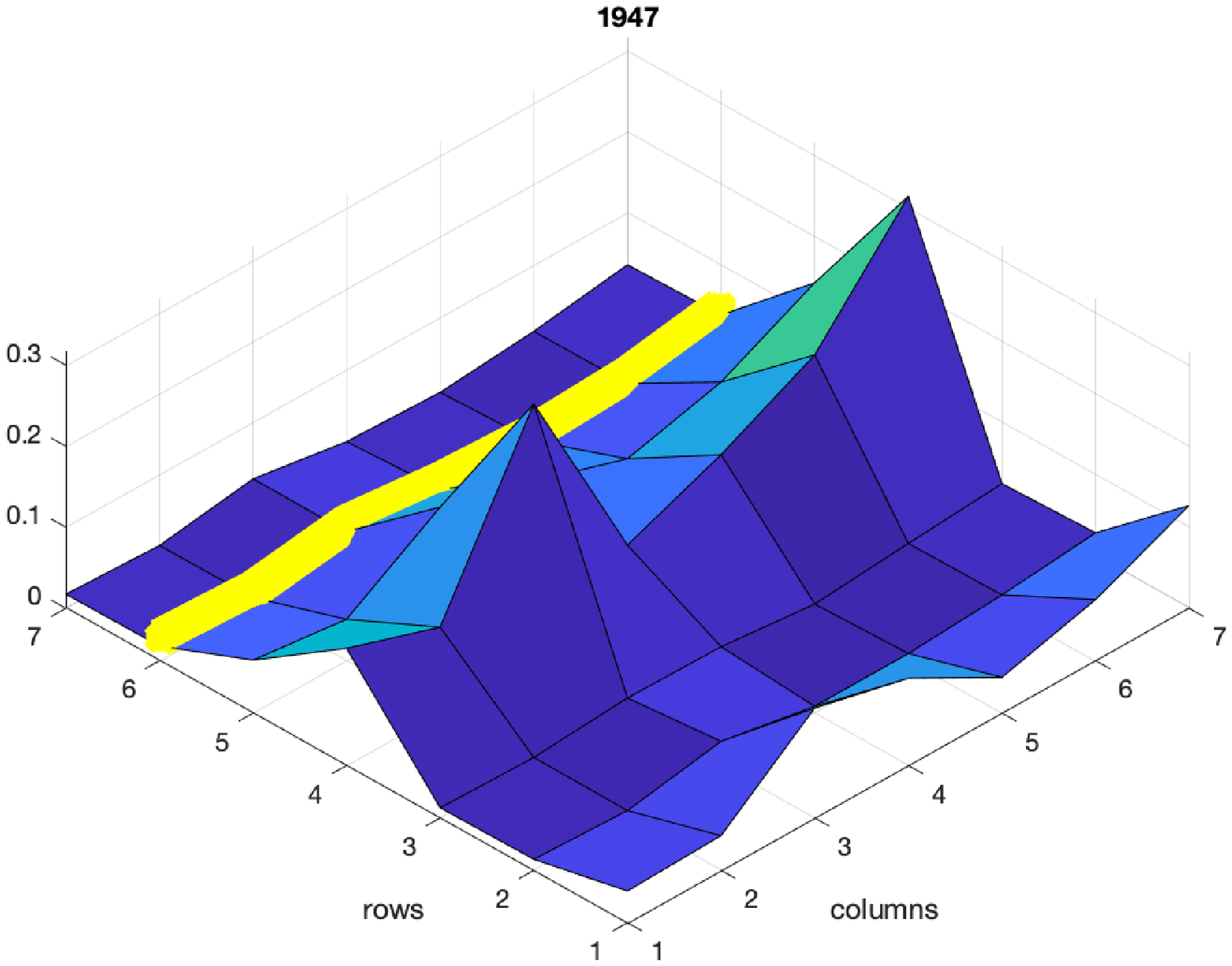} \label{fig:i_1947}}  \\
    \subcaptionbox{1939}{\includegraphics[width=0.32\textwidth]{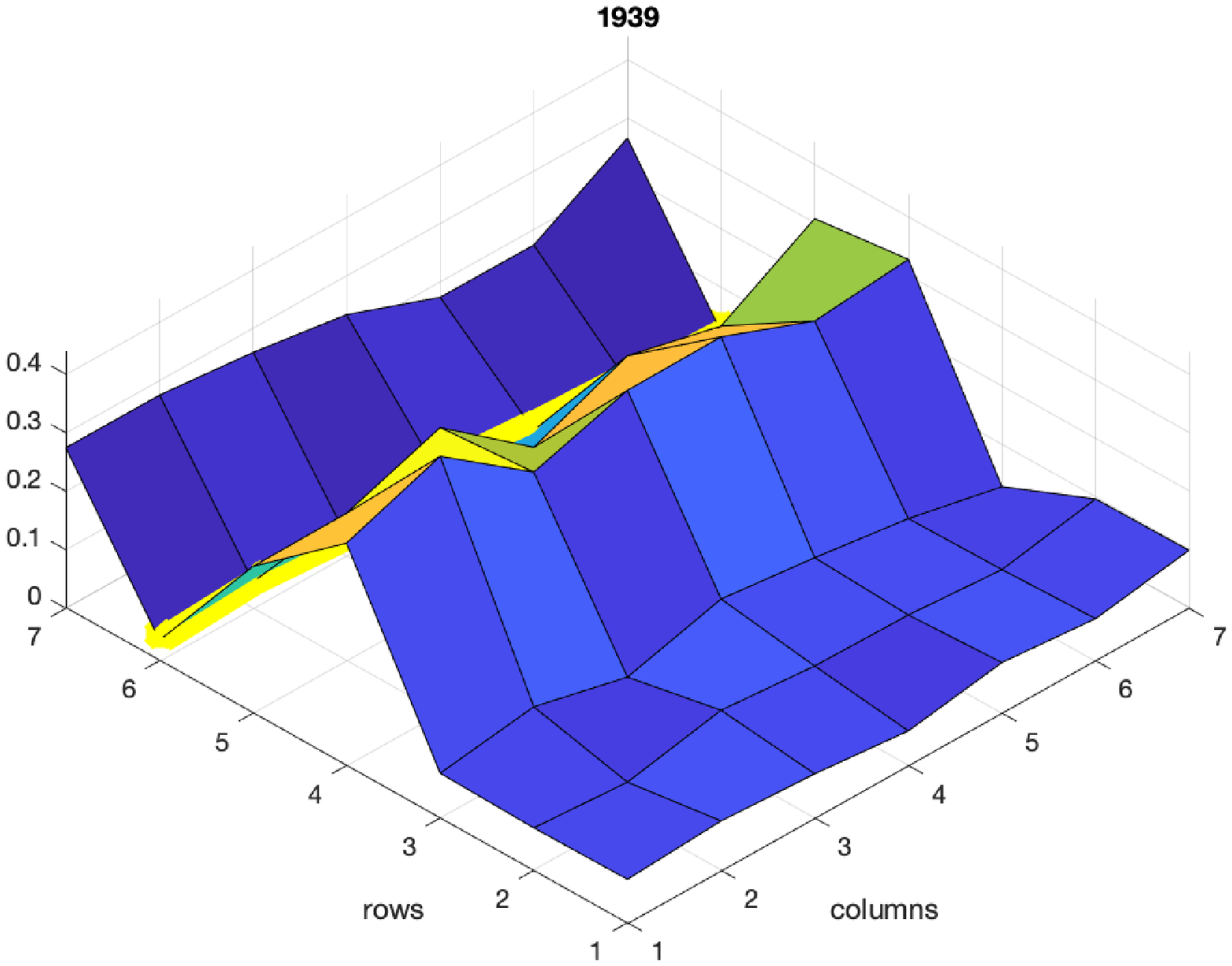} \label{fig:i_1939}}
    \subcaptionbox{1929}{\includegraphics[width=0.32\textwidth]{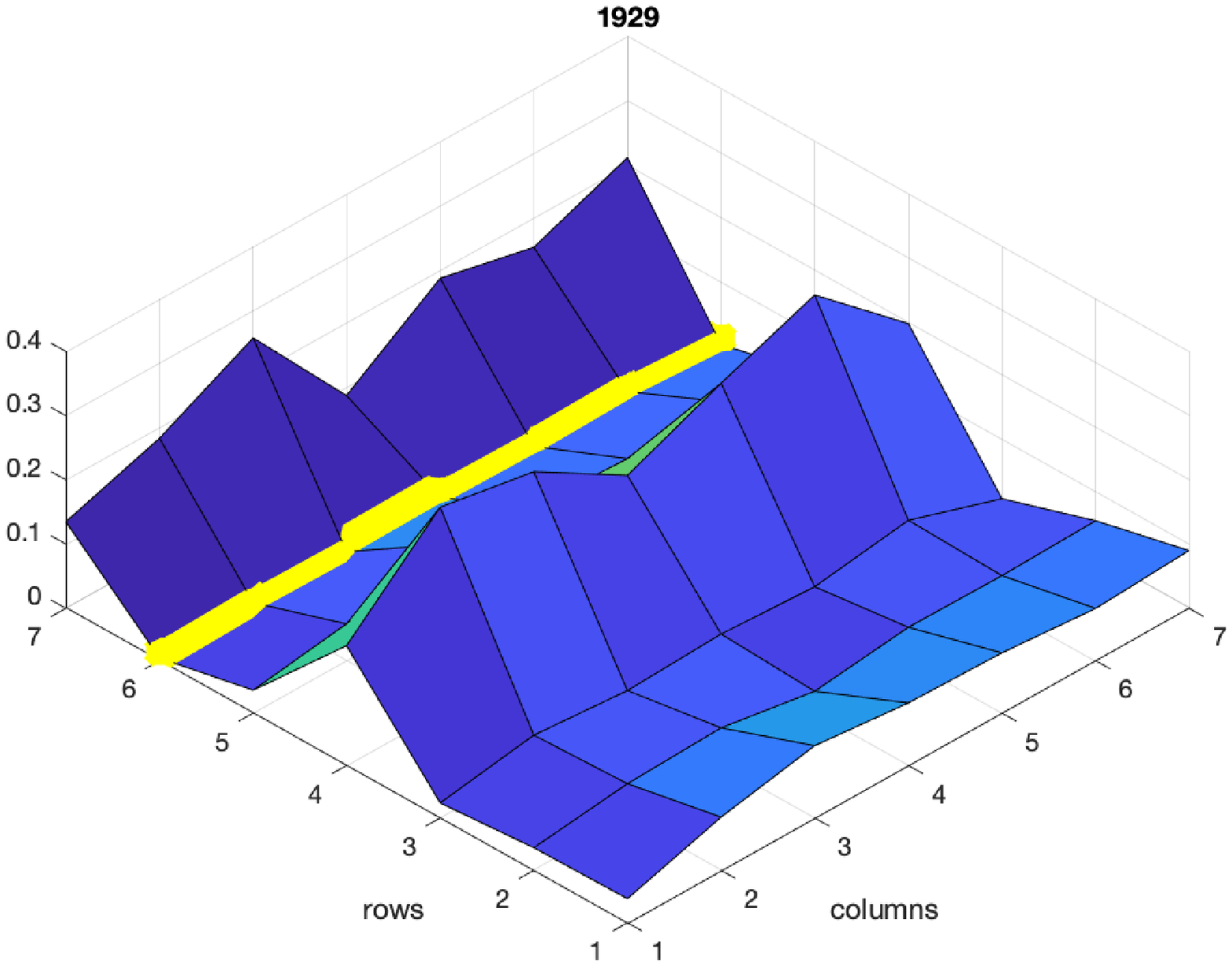} \label{fig:i_1929}}
    \subcaptionbox{1919}{\includegraphics[width=0.32\textwidth]{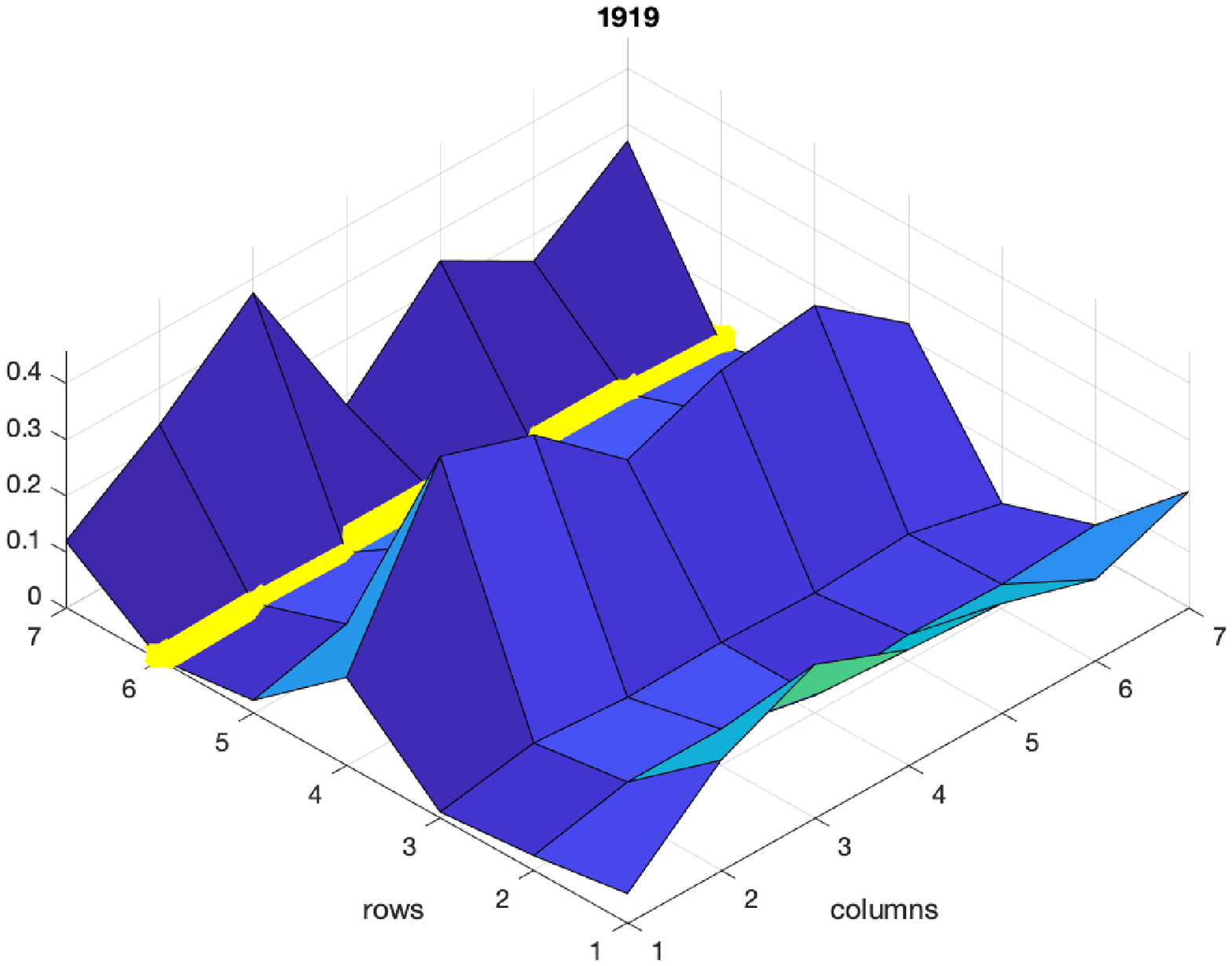} \label{fig:i_1919}}
\caption{The simple indirect requirements matrices ($N^\texttt{i}$) of the US economy for each year. The sectors are as follows: Agriculture (1), Mining (2), Construction (3), Manufacturing (4), Trade, Transport \& Utilities (5), Services (6), and Other (7). (Case study~\ref{sec:real}).}
\label{fig:i_dist}
\end{figure}

\begin{figure}[h]
    \centering
    \subcaptionbox{2006}{\includegraphics[width=0.32\textwidth]{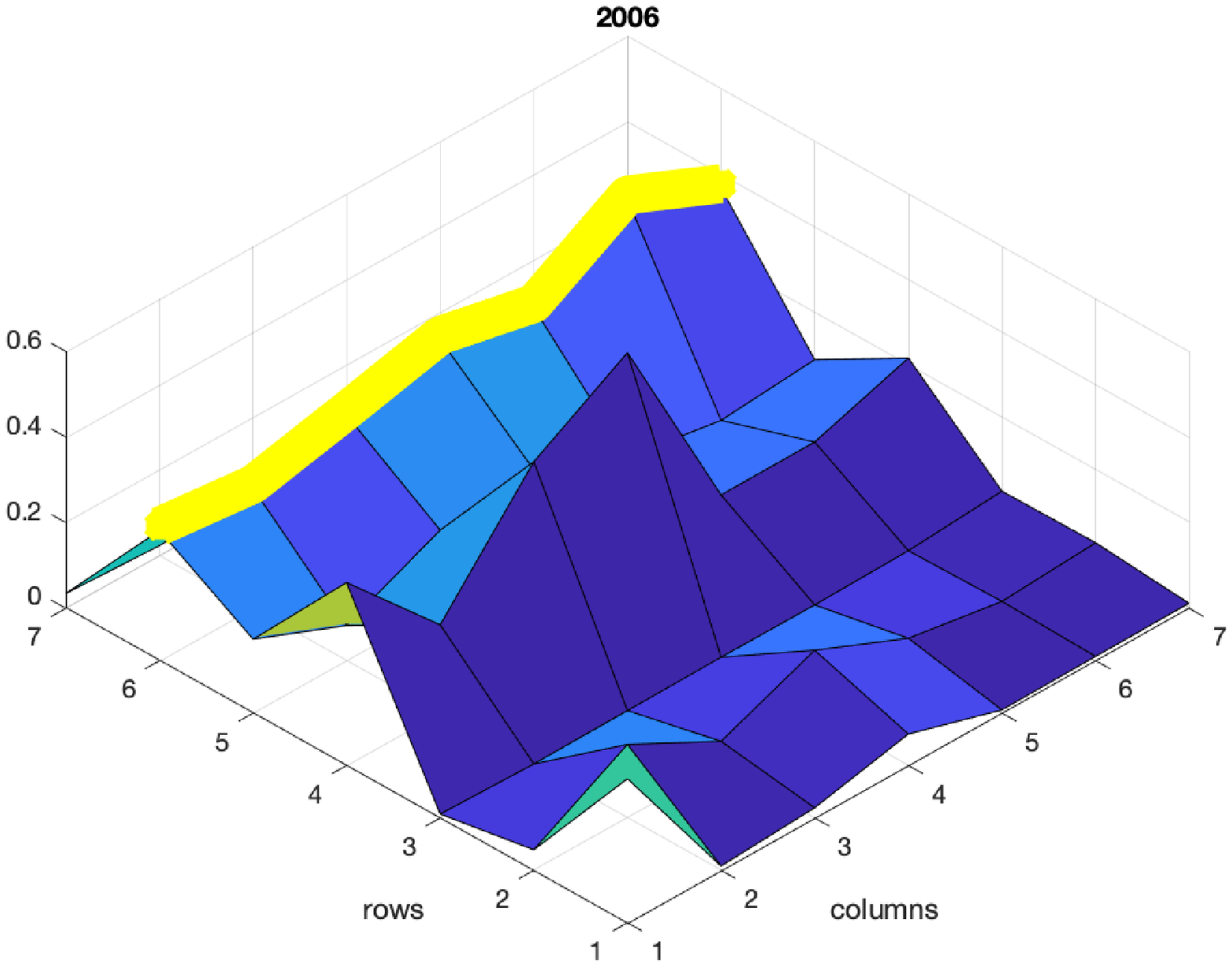} \label{fig:t_2006}}
    \subcaptionbox{2002}{\includegraphics[width=0.32\textwidth]{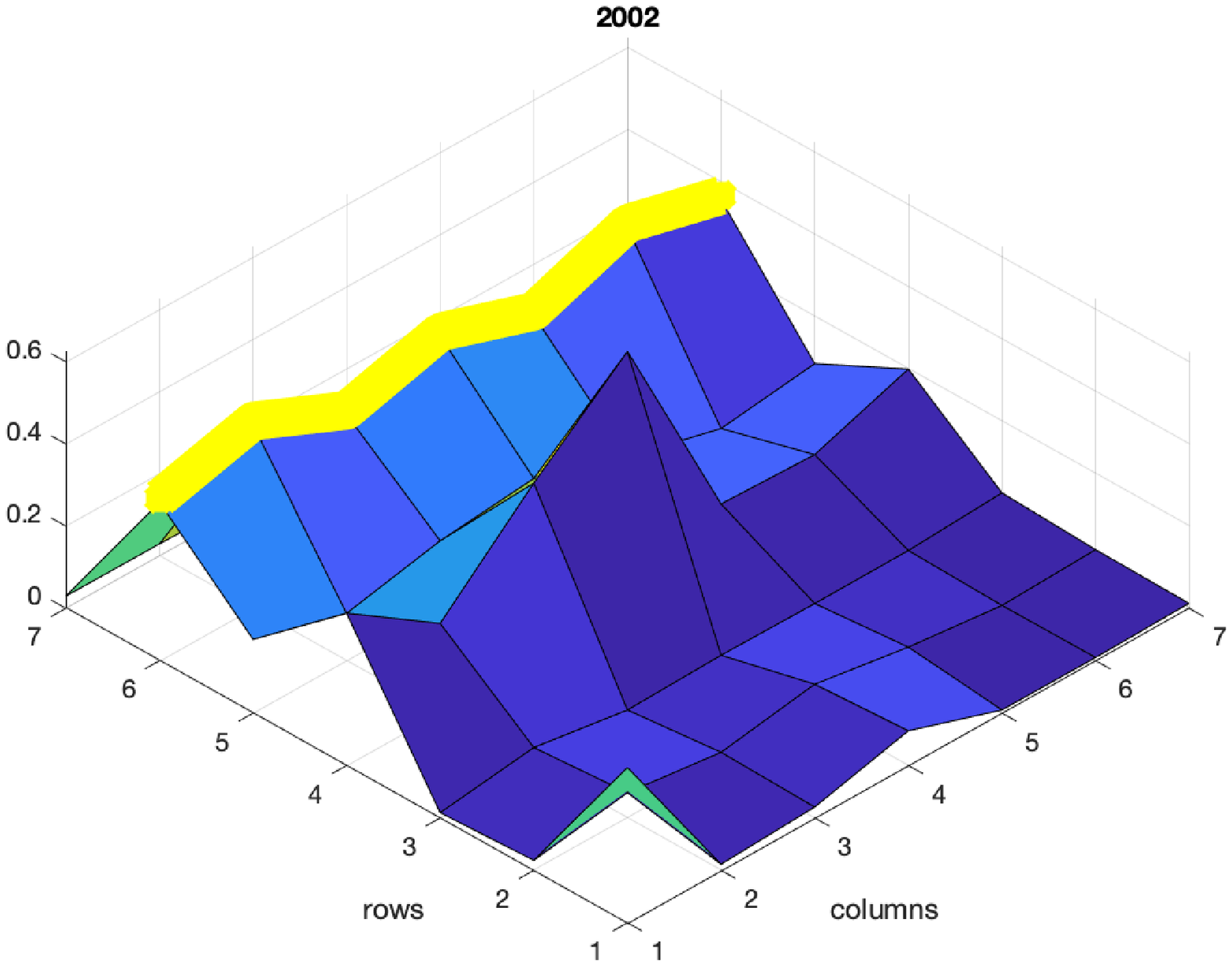} \label{fig:t_2002}}
    \subcaptionbox{1997}{\includegraphics[width=0.32\textwidth]{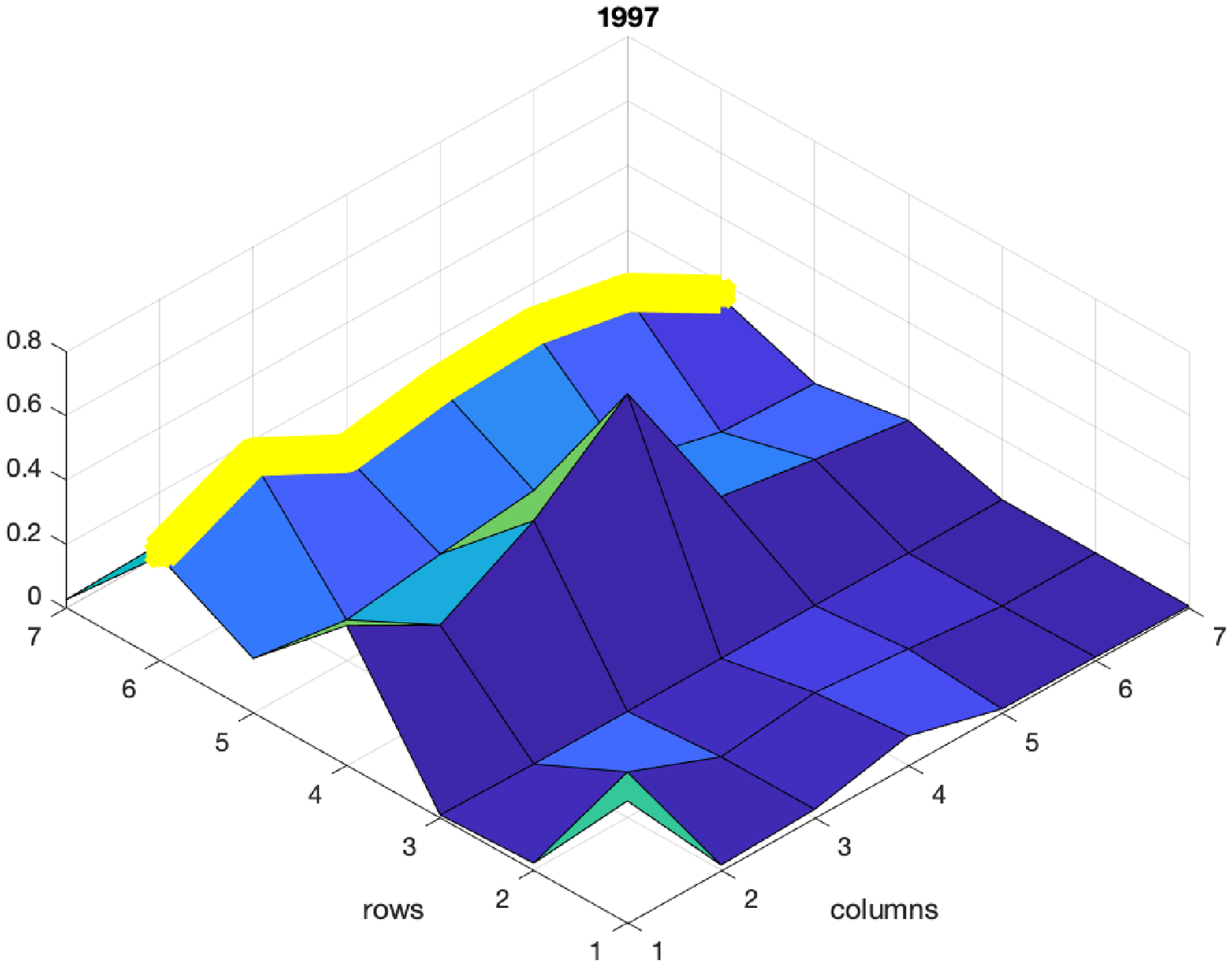} \label{fig:t_1997}}  \\
    \subcaptionbox{1992}{\includegraphics[width=0.32\textwidth]{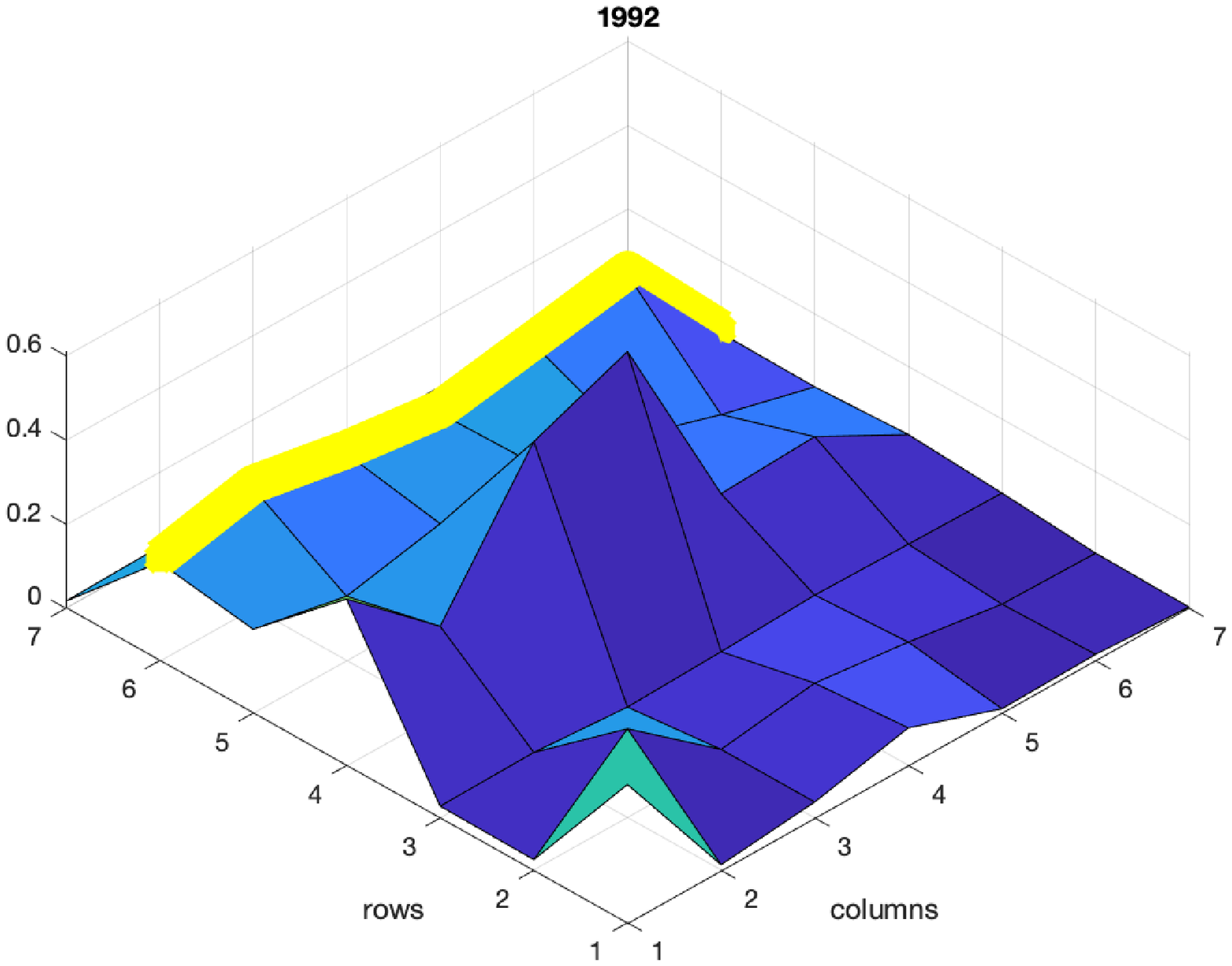} \label{fig:t_1992}}
    \subcaptionbox{1987}{\includegraphics[width=0.32\textwidth]{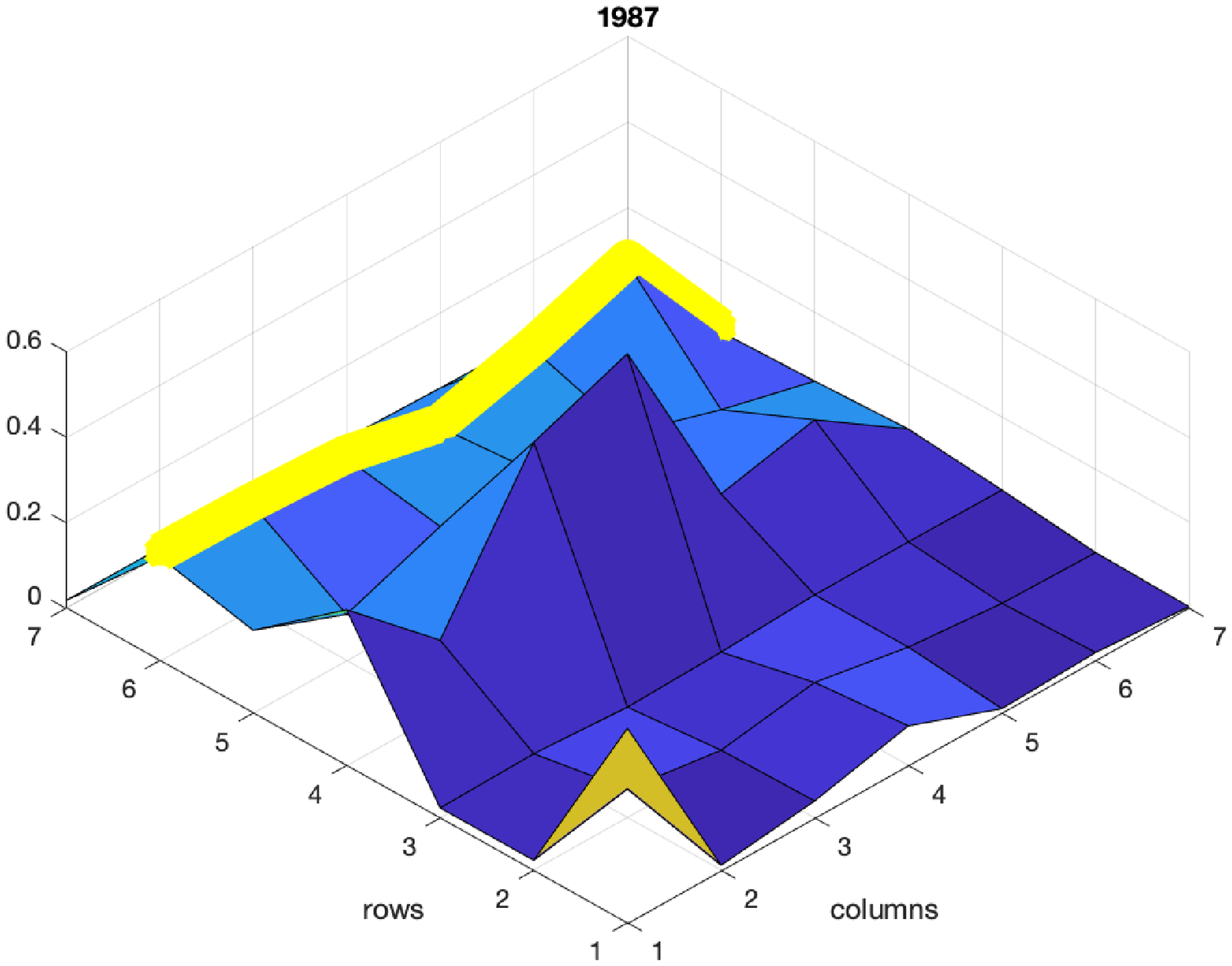} \label{fig:t_1987}}
    \subcaptionbox{1982}{\includegraphics[width=0.32\textwidth]{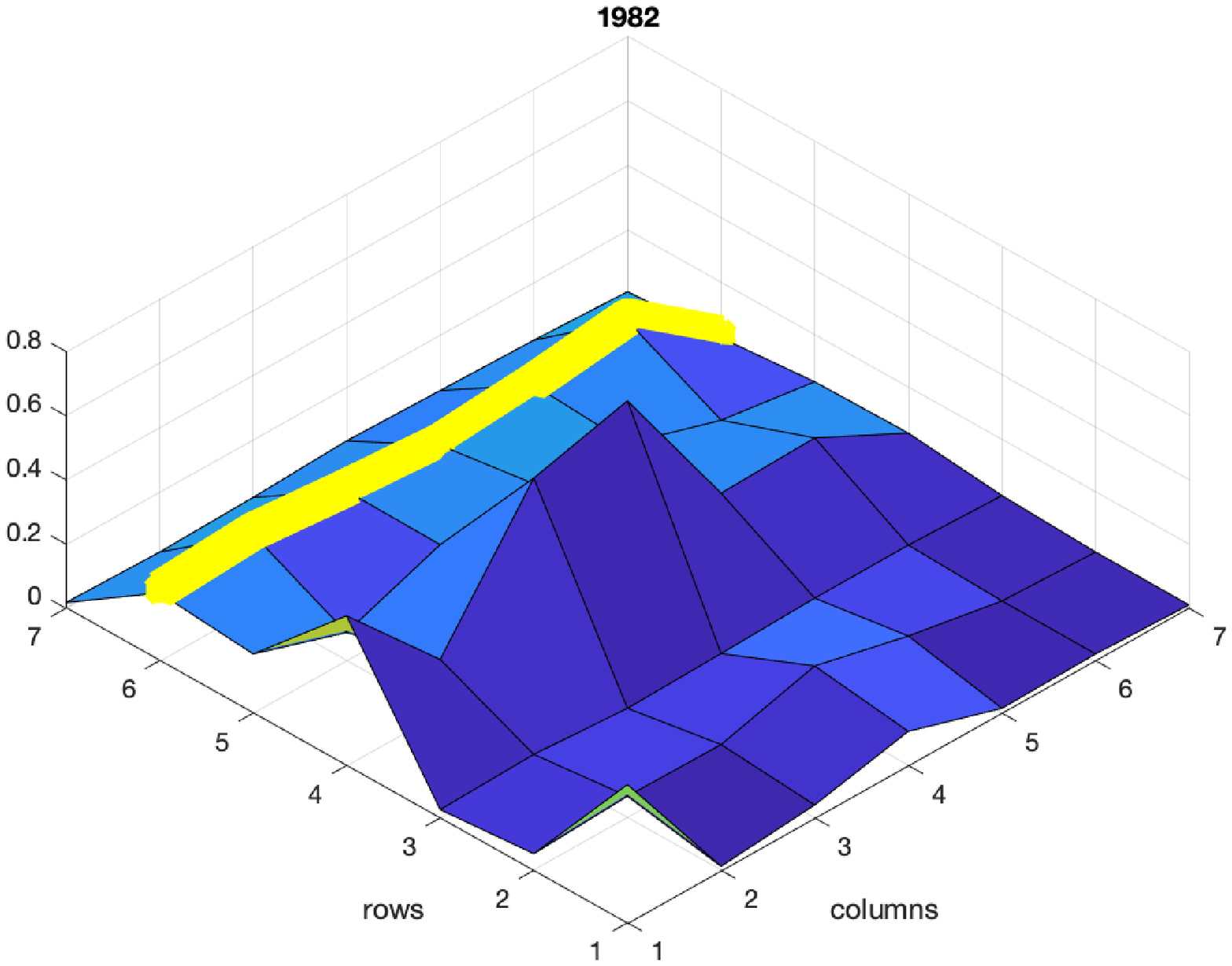} \label{fig:t_1982}}  \\
    \subcaptionbox{1977}{\includegraphics[width=0.32\textwidth]{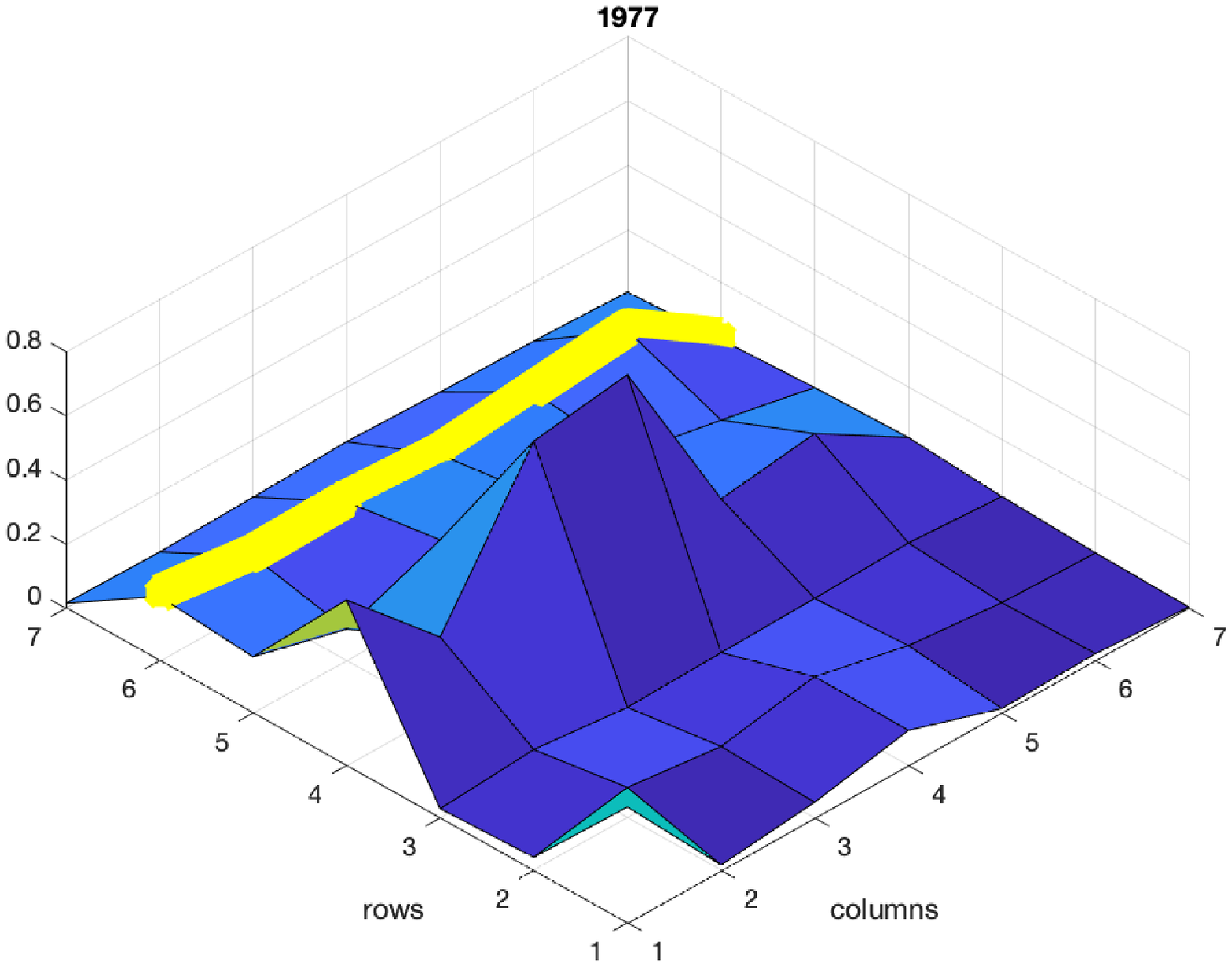} \label{fig:t_1977}}
    \subcaptionbox{1972}{\includegraphics[width=0.32\textwidth]{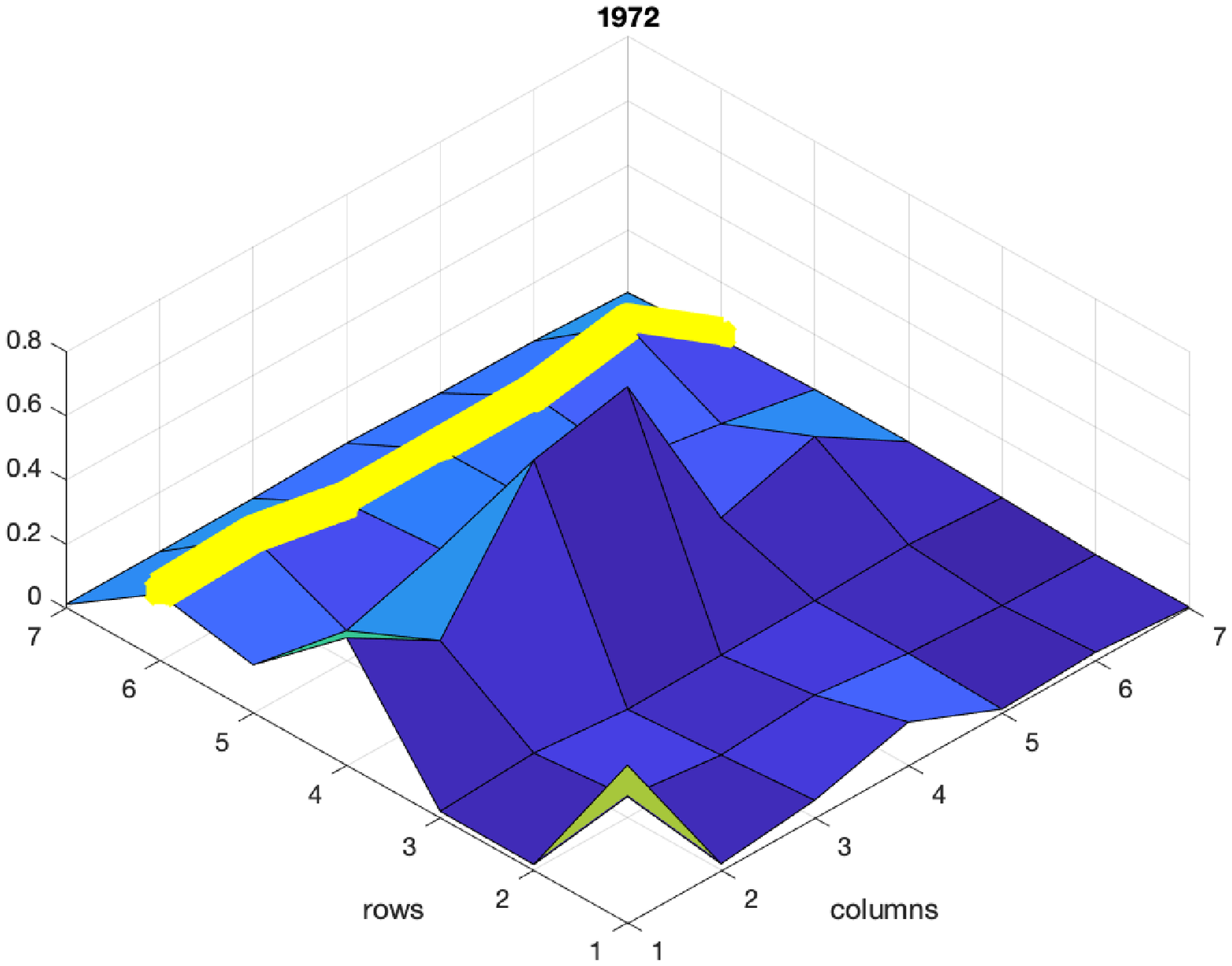} \label{fig:t_1972}}
    \subcaptionbox{1967}{\includegraphics[width=0.32\textwidth]{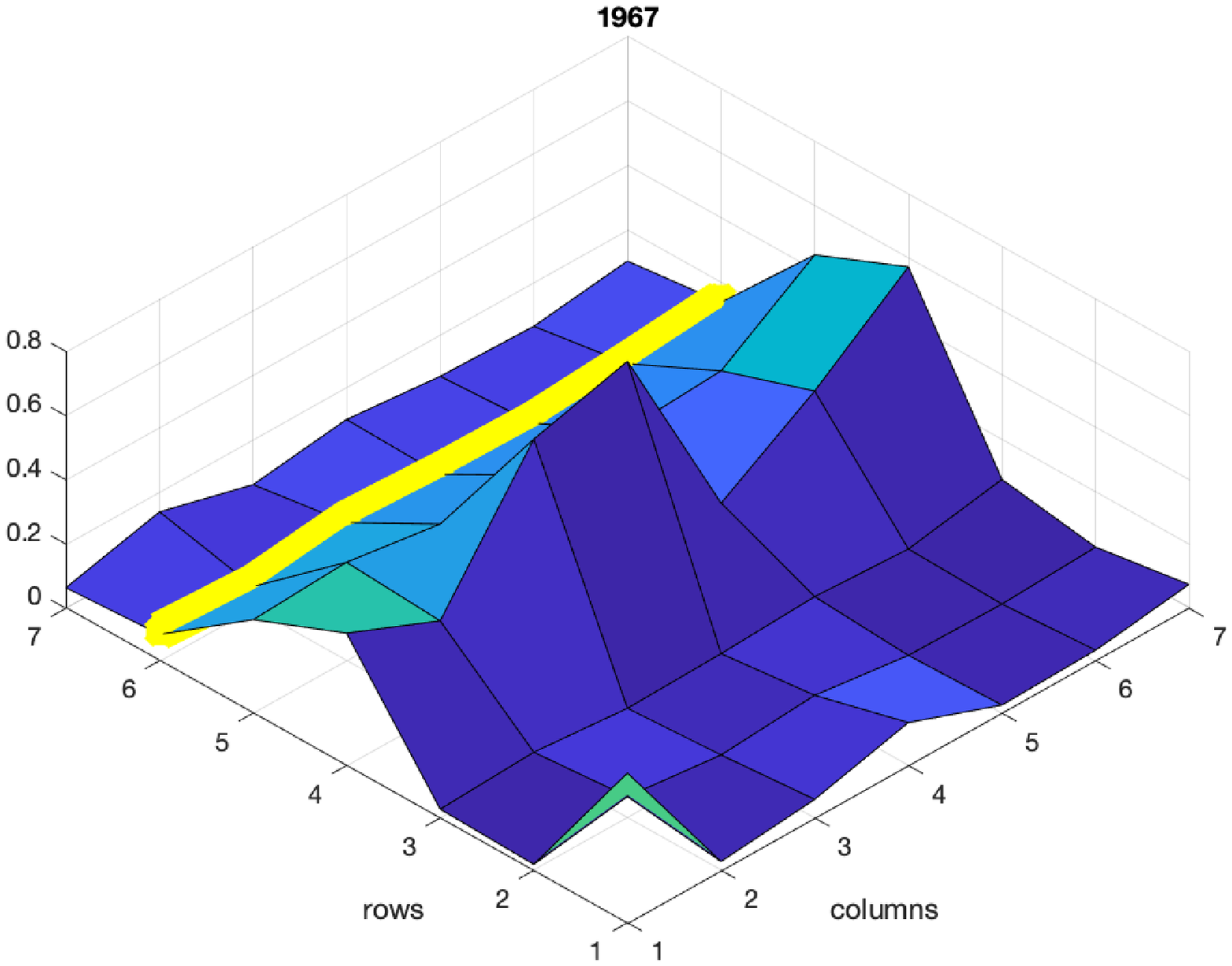} \label{fig:t_1967}}  \\
    \subcaptionbox{1963}{\includegraphics[width=0.32\textwidth]{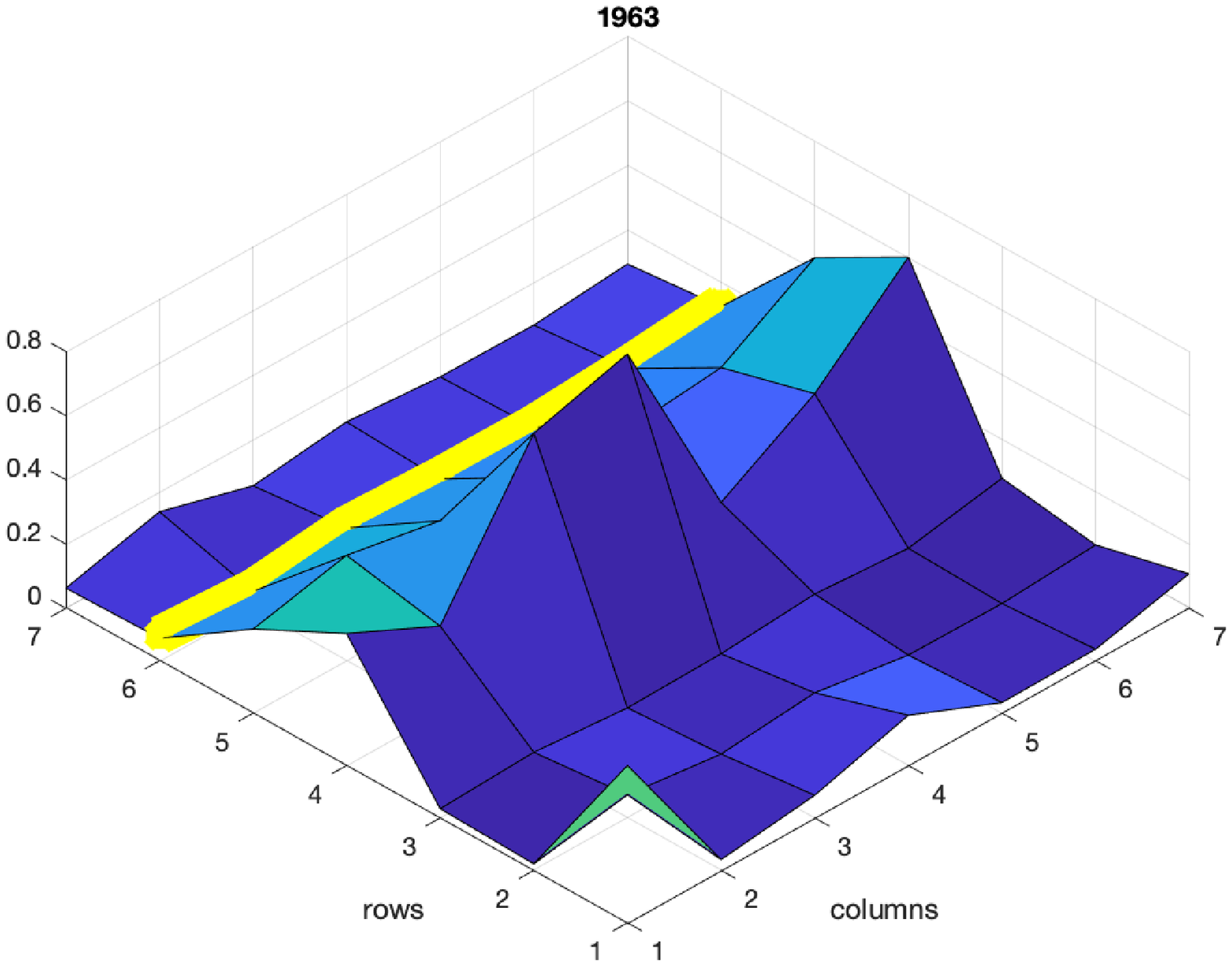} \label{fig:t_1963}}
    \subcaptionbox{1958}{\includegraphics[width=0.32\textwidth]{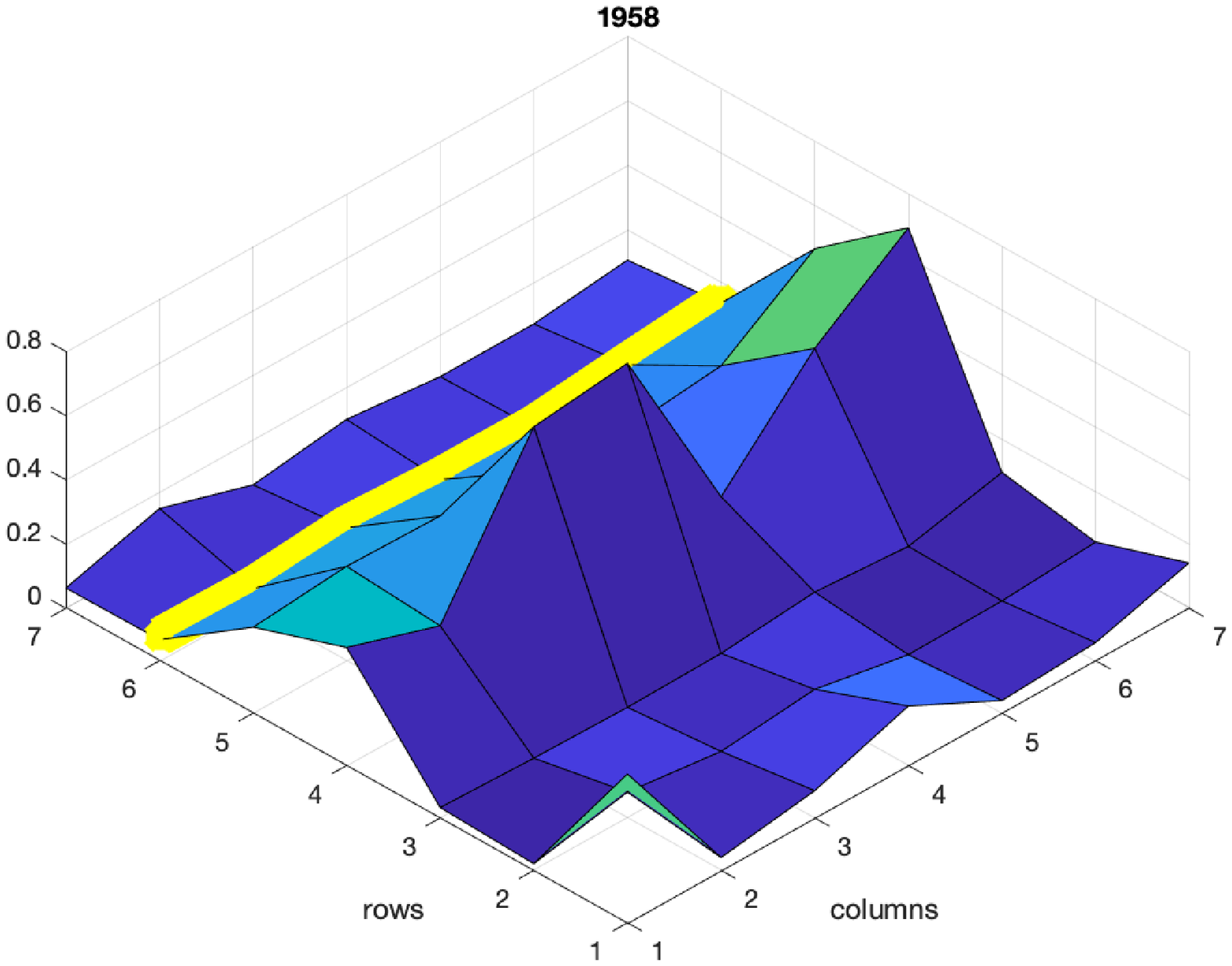} \label{fig:t_1958}}
    \subcaptionbox{1947}{\includegraphics[width=0.32\textwidth]{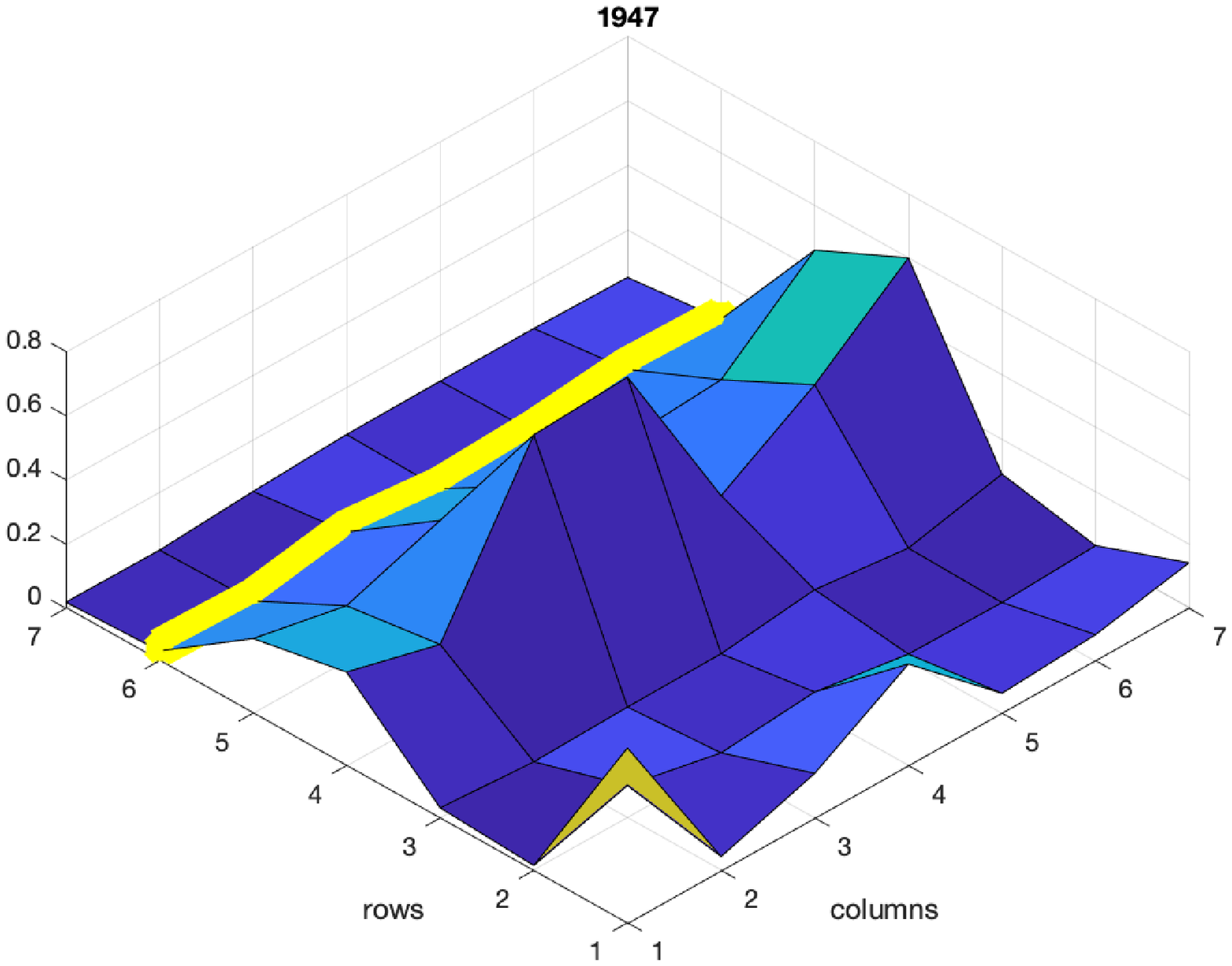} \label{fig:t_1947}}  \\
    \subcaptionbox{1939}{\includegraphics[width=0.32\textwidth]{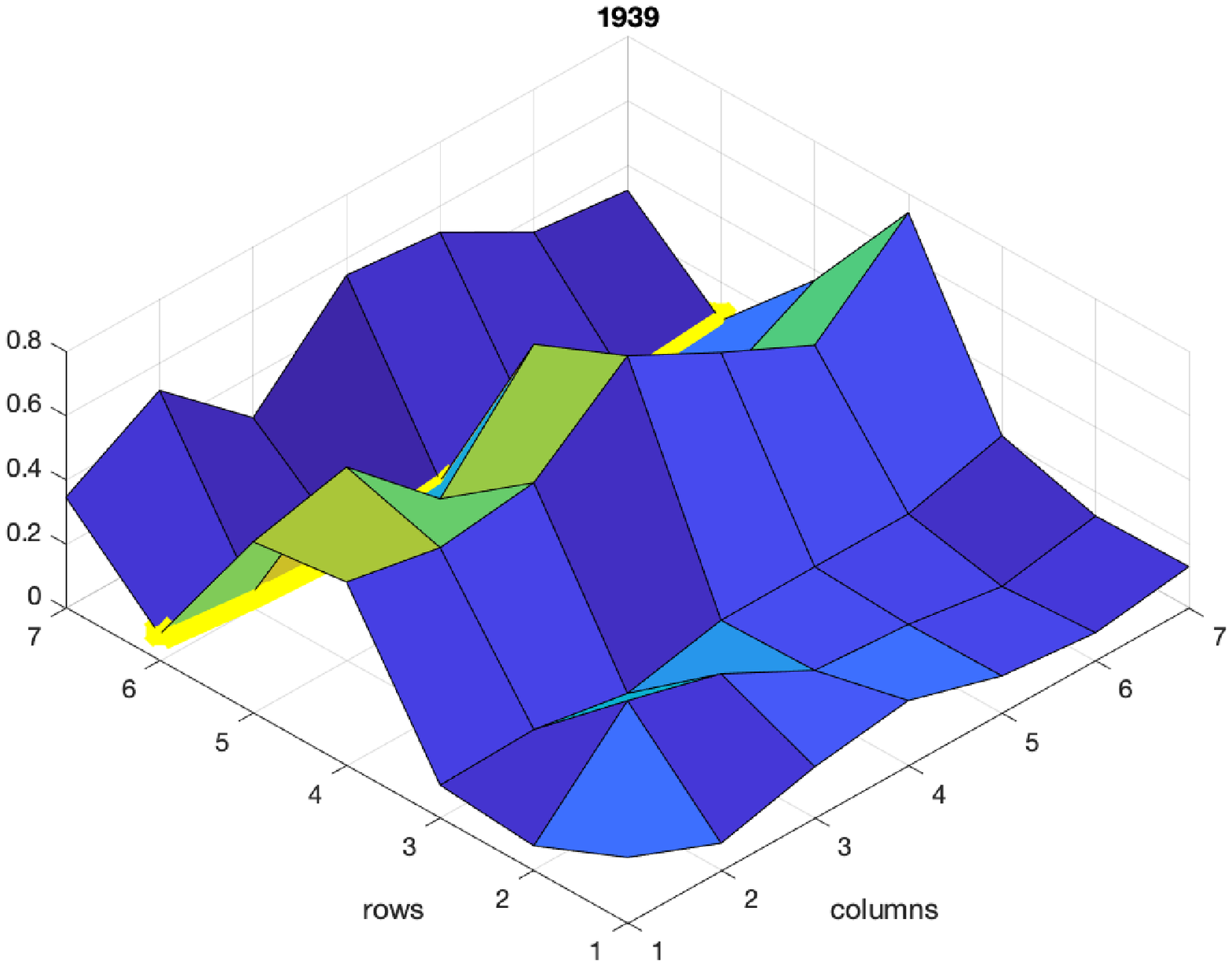} \label{fig:t_1939}}
    \subcaptionbox{1929}{\includegraphics[width=0.32\textwidth]{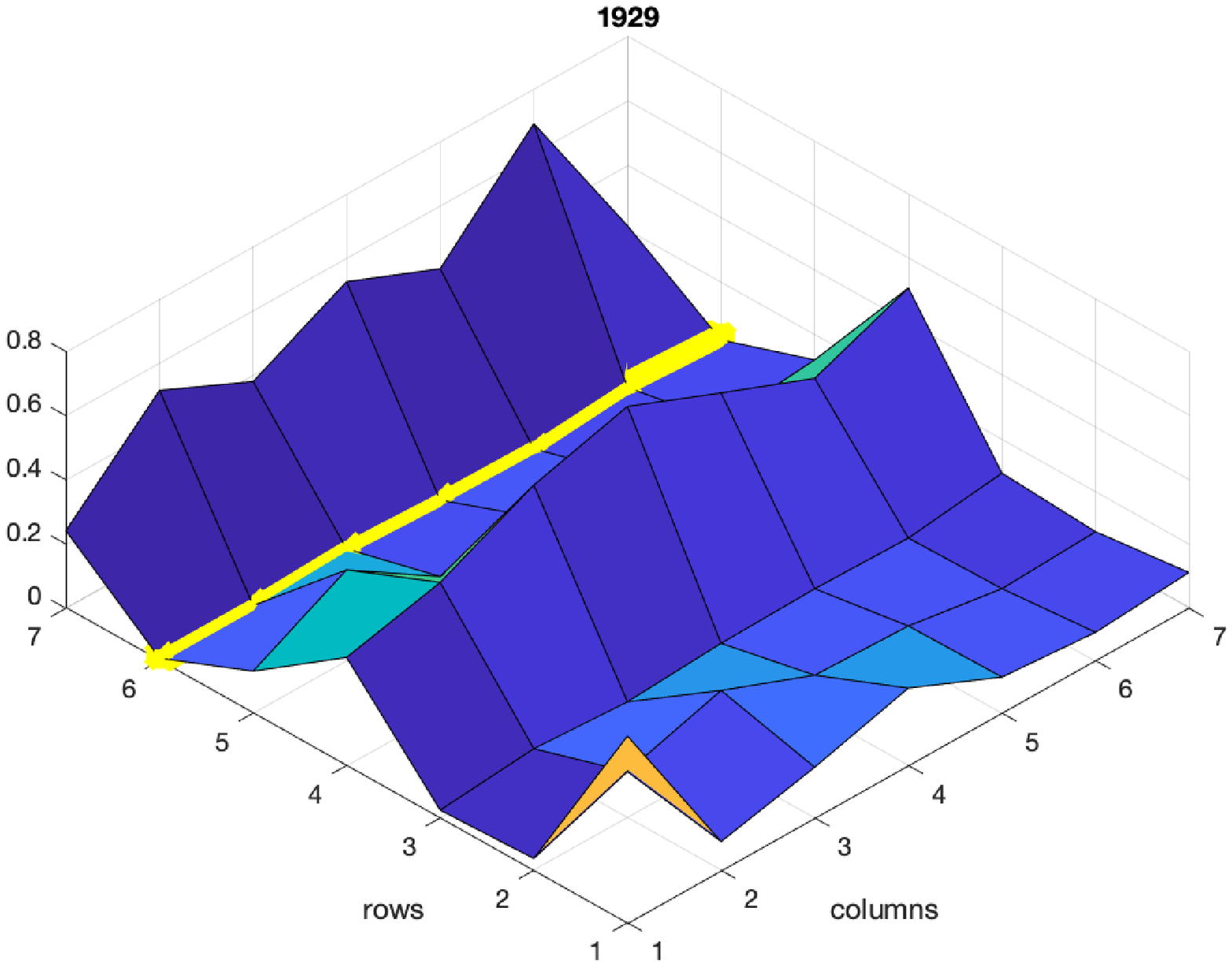} \label{fig:t_1929}}
    \subcaptionbox{1919}{\includegraphics[width=0.32\textwidth]{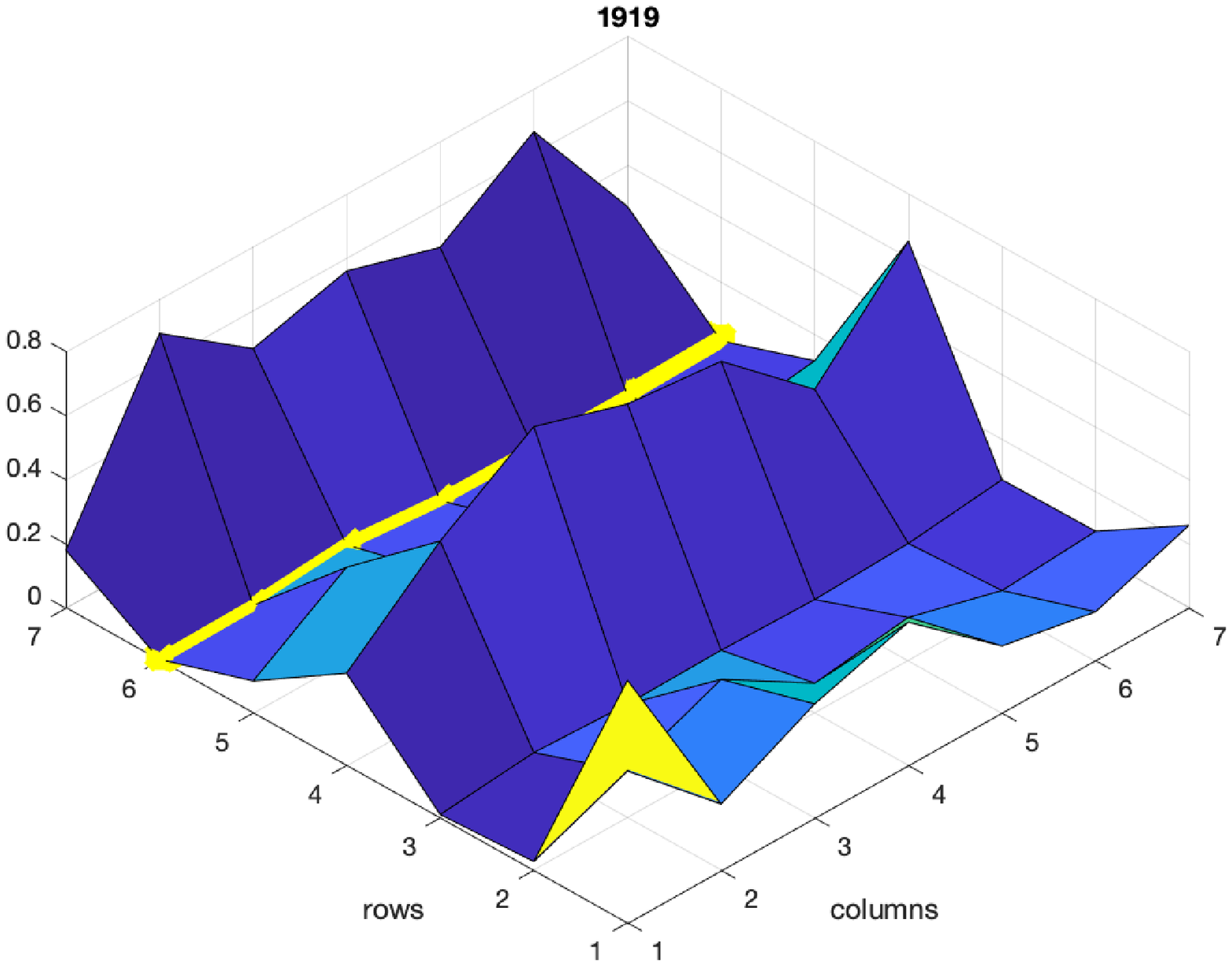} \label{fig:t_1919}}
\caption{The simple transfer (total) requirements matrices ($N^\texttt{t}$) of the US economy for each year. The sectors are as follows: Agriculture (1), Mining (2), Construction (3), Manufacturing (4), Trade, Transport \& Utilities (5), Services (6), and Other (7). (Case study~\ref{sec:real}).}
\label{fig:t_dist}
\end{figure}

The numerical results for the proposed simple direct, indirect, and transfer (total) requirements tables, $N^\texttt{d}$, $N^\texttt{i}$, and $N^\texttt{t}$ are presented in Tables~\ref{tab:2006}-\ref{tab:1919} in this Appendix, together with the composite direct requirements or coefficient matrix, $\pmb{N}^\texttt{t} = A$, for a comparison. These numerical results are also graphically presented in Fig.~\ref{fig:A_dist}-\ref{fig:t_dist}.

The sixth rows in the tables represent the coefficients for the service sector, $n_{6_k}^\texttt{*}$, $k=1,\ldots,7$. They are highlighted in the tables together with the corresponding rows in the figures for a convenience in comparing the results.

\begin{table}[h]
\centering
\caption{The US Requirements Tables for 2006}
\label{tab:2006}
\resizebox{.8\columnwidth}{!}{%
\begin{tabular}{lccccccc}
\toprule
%\multicolumn{2}{c}{Item} \\
%\cmidrule(r){1-2}
\multicolumn{8}{l}{Technical Coefficients Matrix ($A$)} \\
\midrule
 & 1 & 2 & 3 & 4 & 5 & 6 & 7 \\
 \midrule 
1 & 0.2403 & 0 & 0.0014 & 0.0345 & 0.0001 & 0.0018 & 0.0007 \\ 
2 & 0.0028 & 0.1307 & 0.0079 & 0.0756 & 0.031 & 0.0004 & 0.0066 \\ 
3 & 0.0035 & 0.0002 & 0.001 & 0.0019 & 0.0039 & 0.0072 & 0.0242 \\ 
4 & 0.1858 & 0.0959 & 0.2673 & 0.3311 & 0.0581 & 0.0558 & 0.1027 \\ 
5 & 0.0774 & 0.0379 & 0.1063 & 0.1003 & 0.0698 & 0.0329 & 0.0439 \\ 
\rowcolor{yellow}
6 & 0.0875 & 0.1298 & 0.1262 & 0.1239 & 0.1846 & 0.2889 & 0.2029 \\ 
7 & 0.0102 & 0.0096 & 0.0095 & 0.0233 & 0.0223 & 0.0192 & 0.0225 \\
\midrule
\multicolumn{8}{l}{Simple Direct Requirements Matrix ($N^\texttt{d}$)} \\
\midrule
 & 1 & 2 & 3 & 4 & 5 & 6 & 7 \\
\midrule
1 & 0.3212 &      0 & 0.0014 & 0.0551 & 0.0001 & 0.0026 & 0.0007 \\
2 & 0.0037 & 0.1531 & 0.0079 & 0.1207 & 0.0343 & 0.0006 & 0.0069 \\
3 & 0.0047 & 0.0002 & 0.0010 & 0.0030 & 0.0043 & 0.0106 & 0.0251 \\
4 & 0.2483 & 0.1124 & 0.2689 & 0.5288 & 0.0643 & 0.0818 & 0.1066 \\
5 & 0.1034 & 0.0444 & 0.1069 & 0.1602 & 0.0773 & 0.0482 & 0.0456 \\
\rowcolor{yellow}
6 & 0.1169 & 0.1521 & 0.1269 & 0.1979 & 0.2045 & 0.4235 & 0.2107 \\
7 & 0.0136 & 0.0112 & 0.0096 & 0.0372 & 0.0247 & 0.0281 & 0.0234 \\
\midrule
\multicolumn{8}{l}{Simple Indirect Requirements Matrix ($N^\texttt{i}$)} \\
\midrule
 & 1 & 2 & 3 & 4 & 5 & 6 & 7 \\
\midrule
1 & 0.0154 & 0.0101 & 0.0224 & 0.0184 & 0.0073 & 0.0074 & 0.0110 \\
2 & 0.0444 & 0.0185 & 0.0486 & 0.0262 & 0.0181 & 0.0155 & 0.0236 \\
3 & 0.0045 & 0.0033 & 0.0048 & 0.0051 & 0.0036 & 0.0014 & 0.0035 \\
4 & 0.1792 & 0.0940 & 0.1961 & 0.0684 & 0.0780 & 0.0619 & 0.1106 \\
5 & 0.0693 & 0.0379 & 0.0756 & 0.0411 & 0.0303 & 0.0236 & 0.0454 \\
\rowcolor{yellow}
6 & 0.1871 & 0.1278 & 0.2024 & 0.1849 & 0.1300 & 0.0425 & 0.1592 \\
7 & 0.0210 & 0.0127 & 0.0227 & 0.0153 & 0.0112 & 0.0061 & 0.0149 \\
\midrule
\multicolumn{8}{l}{Simple Transfer (Total) Requirements Matrix ($N^\texttt{t}$)} \\
\midrule
 & 1 & 2 & 3 & 4 & 5 & 6 & 7 \\
\midrule
1 & 0.3365 & 0.0101 & 0.0238 & 0.0735 & 0.0075 & 0.0101 & 0.0118 \\
2 & 0.0481 & 0.1716 & 0.0566 & 0.1470 & 0.0524 & 0.0161 & 0.0305 \\
3 & 0.0092 & 0.0036 & 0.0058 & 0.0081 & 0.0079 & 0.0120 & 0.0286 \\
4 & 0.4275 & 0.2064 & 0.4650 & 0.5972 & 0.1424 & 0.1437 & 0.2172 \\
5 & 0.1727 & 0.0823 & 0.1825 & 0.2013 & 0.1076 & 0.0718 & 0.0910 \\
\rowcolor{yellow}
6 & 0.3041 & 0.2799 & 0.3294 & 0.3828 & 0.3345 & 0.4660 & 0.3698 \\
7 & 0.0346 & 0.0239 & 0.0323 & 0.0525 & 0.0359 & 0.0342 & 0.0382 \\
\bottomrule
\end{tabular}
}
\end{table}
\begin{table}
\centering
\caption{The US Requirements Tables for 2002}
\label{tab:2002}
\resizebox{.8\columnwidth}{!}{%
\begin{tabular}{lccccccc}
\toprule
\multicolumn{8}{l}{Technical Coefficients Matrix ($A$)} \\
\midrule
 & 1 & 2 & 3 & 4 & 5 & 6 & 7 \\
 \midrule 
1 & 0.2638 & 0.002 & 0.0027 & 0.0379 & 0.0004 & 0.0008 & 0.0008\\ 
2 & 0.0032 & 0.0468 & 0.0099 & 0.0381 & 0.0236 & 0.0004 & 0.0042\\ 
3 & 0.0043 & 0.0359 & 0.0007 & 0.0032 & 0.0058 & 0.0081 & 0.0204\\ 
4 & 0.1491 & 0.0934 & 0.245 & 0.351 & 0.05 & 0.0472 & 0.0959\\ 
5 & 0.0852 & 0.064 & 0.0968 & 0.0913 & 0.0794 & 0.0254 & 0.0452\\ 
\rowcolor{yellow}
6 & 0.1333 & 0.2457 & 0.144 & 0.1386 & 0.1844 & 0.2682 & 0.2026\\ 
7 & 0.0087 & 0.0138 & 0.0073 & 0.015 & 0.0267 & 0.0162 & 0.0193 \\
\midrule
\multicolumn{8}{l}{Simple Direct Requirements Matrix ($N^\texttt{d}$)} \\
\midrule
 & 1 & 2 & 3 & 4 & 5 & 6 & 7 \\
\midrule
1 & 0.3635 & 0.0021 & 0.0027 & 0.0616 & 0.0004 & 0.0011 & 0.0008 \\
2 & 0.0044 & 0.0497 & 0.0100 & 0.0619 & 0.0263 & 0.0006 & 0.0043 \\
3 & 0.0059 & 0.0381 & 0.0007 & 0.0052 & 0.0065 & 0.0115 & 0.0210 \\
4 & 0.2055 & 0.0991 & 0.2469 & 0.5705 & 0.0557 & 0.0669 & 0.0989 \\
5 & 0.1174 & 0.0679 & 0.0975 & 0.1484 & 0.0885 & 0.0360 & 0.0466 \\
\rowcolor{yellow}
6 & 0.1837 & 0.2608 & 0.1451 & 0.2253 & 0.2054 & 0.3799 & 0.2090 \\
7 & 0.0120 & 0.0146 & 0.0074 & 0.0244 & 0.0297 & 0.0229 & 0.0199 \\
\midrule
\multicolumn{8}{l}{Simple Indirect Requirements Matrix ($N^\texttt{i}$)} \\
\midrule
 & 1 & 2 & 3 & 4 & 5 & 6 & 7 \\
\midrule
1 & 0.0146 & 0.0127 & 0.0238 & 0.0229 & 0.0072 & 0.0067 & 0.0111 \\
2 & 0.0199 & 0.0118 & 0.0222 & 0.0085 & 0.0068 & 0.0064 & 0.0108 \\
3 & 0.0068 & 0.0057 & 0.0069 & 0.0081 & 0.0051 & 0.0015 & 0.0046 \\
4 & 0.1657 & 0.1192 & 0.1856 & 0.0548 & 0.0707 & 0.0521 & 0.1006 \\
5 & 0.0621 & 0.0472 & 0.0667 & 0.0400 & 0.0255 & 0.0188 & 0.0385 \\
\rowcolor{yellow}
6 & 0.2017 & 0.1854 & 0.1990 & 0.1824 & 0.1246 & 0.0367 & 0.1481 \\
7 & 0.0176 & 0.0146 & 0.0176 & 0.0142 & 0.0086 & 0.0040 & 0.0116 \\
\midrule
\multicolumn{8}{l}{Simple Transfer (Total) Requirements Matrix ($N^\texttt{t}$)} \\
\midrule
 & 1 & 2 & 3 & 4 & 5 & 6 & 7 \\
\midrule
1 & 0.3781 & 0.0149 & 0.0265 & 0.0845 & 0.0076 & 0.0078 & 0.0120 \\
2 & 0.0243 & 0.0615 & 0.0322 & 0.0704 & 0.0331 & 0.0070 & 0.0151 \\
3 & 0.0128 & 0.0438 & 0.0076 & 0.0133 & 0.0115 & 0.0130 & 0.0257 \\
4 & 0.3711 & 0.2184 & 0.4325 & 0.6252 & 0.1264 & 0.1189 & 0.1996 \\
5 & 0.1795 & 0.1152 & 0.1642 & 0.1884 & 0.1140 & 0.0548 & 0.0851 \\
\rowcolor{yellow}
6 & 0.3854 & 0.4462 & 0.3441 & 0.4076 & 0.3300 & 0.4166 & 0.3571 \\
7 & 0.0296 & 0.0292 & 0.0250 & 0.0386 & 0.0383 & 0.0270 & 0.0315 \\
\bottomrule
\end{tabular}
}
\end{table}
\begin{table}
\centering
\caption{The US Requirements Tables for 1997}
\label{tab:1997}
\resizebox{.8\columnwidth}{!}{%
\begin{tabular}{lccccccc}
\toprule
\multicolumn{8}{l}{Technical Coefficients Matrix ($A$)} \\
\midrule
 & 1 & 2 & 3 & 4 & 5 & 6 & 7 \\
 \midrule 
1 & 0.2618 & 0.0001 & 0.0015 & 0.0401 & 0.0013 & 0.002 & 0.0008 \\ 
2 & 0.0017 & 0.115 & 0.0062 & 0.0306 & 0.0236 & 0.0003 & 0.0036 \\ 
3 & 0.0039 & 0.0002 & 0.0011 & 0.002 & 0.0052 & 0.006 & 0.0101 \\ 
4 & 0.174 & 0.1162 & 0.2372 & 0.3627 & 0.0758 & 0.0583 & 0.0424 \\ 
5 & 0.0731 & 0.0643 & 0.0975 & 0.098 & 0.0847 & 0.0288 & 0.0267 \\ 
\rowcolor{yellow}
6 & 0.111 & 0.257 & 0.1376 & 0.1232 & 0.2294 & 0.2146 & 0.0902 \\ 
7 & 0.0063 & 0.0181 & 0.0086 & 0.0177 & 0.0212 & 0.0169 & 0.0167 \\
\midrule
\multicolumn{8}{l}{Simple Direct Requirements Matrix ($N^\texttt{d}$)} \\
\midrule
 & 1 & 2 & 3 & 4 & 5 & 6 & 7 \\
\midrule
1 & 0.3612 & 0.0001 & 0.0015 & 0.0669 & 0.0015 & 0.0026 & 0.0008 \\
2 & 0.0023 & 0.1315 & 0.0062 & 0.0511 & 0.0267 & 0.0004 & 0.0037 \\
3 & 0.0054 & 0.0002 & 0.0011 & 0.0033 & 0.0059 & 0.0079 & 0.0103 \\
4 & 0.2400 & 0.1328 & 0.2384 & 0.6053 & 0.0856 & 0.0770 & 0.0434 \\
5 & 0.1008 & 0.0735 & 0.0980 & 0.1635 & 0.0957 & 0.0380 & 0.0273 \\
\rowcolor{yellow}
6 & 0.1531 & 0.2938 & 0.1383 & 0.2056 & 0.2592 & 0.2834 & 0.0922 \\
7 & 0.0087 & 0.0207 & 0.0086 & 0.0295 & 0.0240 & 0.0223 & 0.0171 \\
\midrule
\multicolumn{8}{l}{Simple Indirect Requirements Matrix ($N^\texttt{i}$)} \\
\midrule
 & 1 & 2 & 3 & 4 & 5 & 6 & 7 \\
\midrule
1 & 0.0184 & 0.0165 & 0.0254 & 0.0251 & 0.0117 & 0.0085 & 0.0060 \\
2 & 0.0202 & 0.0117 & 0.0207 & 0.0126 & 0.0102 & 0.0066 & 0.0052 \\
3 & 0.0040 & 0.0044 & 0.0040 & 0.0040 & 0.0031 & 0.0009 & 0.0014 \\
4 & 0.1960 & 0.1417 & 0.1980 & 0.0635 & 0.0993 & 0.0603 & 0.0521 \\
5 & 0.0696 & 0.0536 & 0.0703 & 0.0404 & 0.0342 & 0.0213 & 0.0199 \\
\rowcolor{yellow}
6 & 0.1721 & 0.1679 & 0.1710 & 0.1557 & 0.1194 & 0.0372 & 0.0599 \\
7 & 0.0178 & 0.0161 & 0.0176 & 0.0129 & 0.0111 & 0.0044 & 0.0056 \\
\midrule
\multicolumn{8}{l}{Simple Transfer (Total) Requirements Matrix ($N^\texttt{t}$)} \\
\midrule
 & 1 & 2 & 3 & 4 & 5 & 6 & 7 \\
\midrule
1 & 0.3796 & 0.0166 & 0.0269 & 0.0921 & 0.0131 & 0.0112 & 0.0068 \\
2 & 0.0226 & 0.1432 & 0.0269 & 0.0637 & 0.0369 & 0.0070 & 0.0089 \\
3 & 0.0094 & 0.0047 & 0.0051 & 0.0074 & 0.0089 & 0.0088 & 0.0117 \\
4 & 0.4360 & 0.2745 & 0.4364 & 0.6687 & 0.1850 & 0.1373 & 0.0954 \\
5 & 0.1705 & 0.1271 & 0.1683 & 0.2039 & 0.1299 & 0.0594 & 0.0472 \\
\rowcolor{yellow}
6 & 0.3252 & 0.4617 & 0.3093 & 0.3613 & 0.3785 & 0.3206 & 0.1521 \\
7 & 0.0264 & 0.0368 & 0.0263 & 0.0425 & 0.0350 & 0.0267 & 0.0226 \\
\bottomrule
\end{tabular}
}
\end{table}

\begin{table}
\centering
\caption{The US Requirements Tables for 1992}
\label{tab:1992}
\resizebox{.8\columnwidth}{!}{%
\begin{tabular}{lccccccc}
\toprule
\multicolumn{8}{l}{Technical Coefficients Matrix ($A$)} \\
\midrule
 & 1 & 2 & 3 & 4 & 5 & 6 & 7 \\
 \midrule 
1 & 0.2339 & 0.0003 & 0.0061 & 0.0419 & 0.0005 & 0.0036 & 0.0004\\ 
2 & 0.0018 & 0.1654 & 0.009 & 0.0329 & 0.0274 & 0.0002 & 0.003\\ 
3 & 0.0122 & 0.017 & 0.0009 & 0.0061 & 0.0208 & 0.0187 & 0.023\\ 
4 & 0.1667 & 0.0787 & 0.2992 & 0.3454 & 0.056 & 0.0673 & 0.0135\\ 
5 & 0.0914 & 0.081 & 0.1061 & 0.1057 & 0.1048 & 0.0427 & 0.016\\ 
\rowcolor{yellow}
6 & 0.09 & 0.1514 & 0.1139 & 0.0712 & 0.1555 & 0.2039 & 0.0134\\ 
7 & 0.0038 & 0.0105 & 0.0048 & 0.0119 & 0.0201 & 0.0112 & 0.0034 \\
\midrule
\multicolumn{8}{l}{Simple Direct Requirements Matrix ($N^\texttt{d}$)} \\
\midrule
 & 1 & 2 & 3 & 4 & 5 & 6 & 7 \\
\midrule
1 & 0.3107 & 0.0004 & 0.0062 & 0.0675 & 0.0006 & 0.0047 & 0.0004 \\
2 & 0.0024 & 0.2004 & 0.0091 & 0.0530 & 0.0317 & 0.0003 & 0.0030 \\
3 & 0.0162 & 0.0206 & 0.0009 & 0.0098 & 0.0240 & 0.0242 & 0.0231 \\
4 & 0.2214 & 0.0954 & 0.3035 & 0.5561 & 0.0647 & 0.0872 & 0.0136 \\
5 & 0.1214 & 0.0982 & 0.1076 & 0.1702 & 0.1211 & 0.0553 & 0.0161 \\
\rowcolor{yellow}
6 & 0.1196 & 0.1835 & 0.1155 & 0.1146 & 0.1797 & 0.2641 & 0.0135 \\
7 & 0.0050 & 0.0127 & 0.0049 & 0.0192 & 0.0232 & 0.0145 & 0.0034 \\
\midrule
\multicolumn{8}{l}{Simple Indirect Requirements Matrix ($N^\texttt{i}$)} \\
\midrule
 & 1 & 2 & 3 & 4 & 5 & 6 & 7 \\
\midrule
1 & 0.0177 & 0.0131 & 0.0316 & 0.0220 & 0.0097 & 0.0104 & 0.0026 \\
2 & 0.0230 & 0.0114 & 0.0291 & 0.0183 & 0.0126 & 0.0095 & 0.0033 \\
3 & 0.0118 & 0.0105 & 0.0134 & 0.0115 & 0.0072 & 0.0036 & 0.0016 \\
4 & 0.1741 & 0.1108 & 0.2169 & 0.0539 & 0.0830 & 0.0715 & 0.0255 \\
5 & 0.0780 & 0.0549 & 0.0933 & 0.0487 & 0.0347 & 0.0313 & 0.0118 \\
\rowcolor{yellow}
6 & 0.1141 & 0.1017 & 0.1272 & 0.0993 & 0.0738 & 0.0310 & 0.0175 \\
7 & 0.0118 & 0.0090 & 0.0136 & 0.0081 & 0.0053 & 0.0040 & 0.0016 \\
\midrule
\multicolumn{8}{l}{Simple Transfer (Total) Requirements Matrix ($N^\texttt{t}$)} \\
\midrule
 & 1 & 2 & 3 & 4 & 5 & 6 & 7 \\
\midrule
1 & 0.3284 & 0.0134 & 0.0378 & 0.0894 & 0.0103 & 0.0151 & 0.0030 \\
2 & 0.0254 & 0.2118 & 0.0383 & 0.0712 & 0.0443 & 0.0098 & 0.0063 \\
3 & 0.0280 & 0.0311 & 0.0143 & 0.0213 & 0.0312 & 0.0278 & 0.0247 \\
4 & 0.3956 & 0.2062 & 0.5204 & 0.6100 & 0.1478 & 0.1586 & 0.0391 \\
5 & 0.1994 & 0.1530 & 0.2009 & 0.2189 & 0.1559 & 0.0866 & 0.0279 \\
\rowcolor{yellow}
6 & 0.2336 & 0.2851 & 0.2428 & 0.2139 & 0.2535 & 0.2951 & 0.0309 \\
7 & 0.0168 & 0.0217 & 0.0184 & 0.0272 & 0.0286 & 0.0185 & 0.0050 \\
\bottomrule
\end{tabular}
}
\end{table}

\begin{table}
\centering
\caption{The US Requirements Tables for 1987}
\label{tab:1987}
\resizebox{.8\columnwidth}{!}{%
\begin{tabular}{lccccccc}
\toprule
\multicolumn{8}{l}{Technical Coefficients Matrix ($A$)} \\
\midrule
 & 1 & 2 & 3 & 4 & 5 & 6 & 7 \\
 \midrule 
1 & 0.3016 & 0.0002 & 0.0062 & 0.04 & 0.0003 & 0.004 & 0.0003\\ 
2 & 0.0023 & 0.0541 & 0.0093 & 0.0396 & 0.0204 & 0.0006 & 0.004\\ 
3 & 0.0076 & 0.0173 & 0.0006 & 0.0058 & 0.0206 & 0.0217 & 0.0292\\ 
4 & 0.1376 & 0.0715 & 0.2945 & 0.3419 & 0.0533 & 0.0836 & 0.0184\\ 
5 & 0.0834 & 0.0602 & 0.1029 & 0.095 & 0.1144 & 0.0461 & 0.0256\\ 
\rowcolor{yellow}
6 & 0.0933 & 0.1486 & 0.1118 & 0.0558 & 0.1446 & 0.2158 & 0.0123\\ 
7 & 0.0042 & 0.0095 & 0.0045 & 0.0143 & 0.0165 & 0.0124 & 0.0036\\
\midrule
\multicolumn{8}{l}{Simple Direct Requirements Matrix ($N^\texttt{d}$)} \\
\midrule
 & 1 & 2 & 3 & 4 & 5 & 6 & 7 \\
\midrule
1 & 0.4386 & 0.0002 & 0.0063 & 0.0638 & 0.0003 & 0.0053 & 0.0003 \\
2 & 0.0033 & 0.0577 & 0.0094 & 0.0632 & 0.0238 & 0.0008 & 0.0040 \\
3 & 0.0111 & 0.0185 & 0.0006 & 0.0093 & 0.0240 & 0.0286 & 0.0294 \\
4 & 0.2001 & 0.0763 & 0.2986 & 0.5453 & 0.0621 & 0.1100 & 0.0185 \\
5 & 0.1213 & 0.0642 & 0.1043 & 0.1515 & 0.1332 & 0.0607 & 0.0257 \\
\rowcolor{yellow}
6 & 0.1357 & 0.1586 & 0.1134 & 0.0890 & 0.1684 & 0.2841 & 0.0124 \\
7 & 0.0061 & 0.0101 & 0.0046 & 0.0228 & 0.0192 & 0.0163 & 0.0036 \\
\midrule
\multicolumn{8}{l}{Simple Indirect Requirements Matrix ($N^\texttt{i}$)} \\
\midrule
 & 1 & 2 & 3 & 4 & 5 & 6 & 7 \\
\midrule
1 & 0.0157 & 0.0115 & 0.0334 & 0.0289 & 0.0101 & 0.0138 & 0.0036 \\
2 & 0.0201 & 0.0095 & 0.0264 & 0.0085 & 0.0080 & 0.0108 & 0.0036 \\
3 & 0.0123 & 0.0090 & 0.0133 & 0.0107 & 0.0074 & 0.0041 & 0.0020 \\
4 & 0.1640 & 0.0941 & 0.2123 & 0.0497 & 0.0827 & 0.0854 & 0.0328 \\
5 & 0.0722 & 0.0439 & 0.0868 & 0.0456 & 0.0313 & 0.0358 & 0.0151 \\
\rowcolor{yellow}
6 & 0.1070 & 0.0813 & 0.1146 & 0.0888 & 0.0688 & 0.0323 & 0.0210 \\
7 & 0.0118 & 0.0074 & 0.0139 & 0.0067 & 0.0056 & 0.0048 & 0.0021 \\
\midrule
\multicolumn{8}{l}{Simple Transfer (Total) Requirements Matrix ($N^\texttt{t}$)} \\
\midrule
 & 1 & 2 & 3 & 4 & 5 & 6 & 7 \\
\midrule
1 & 0.4544 & 0.0117 & 0.0397 & 0.0927 & 0.0105 & 0.0191 & 0.0039 \\
2 & 0.0234 & 0.0672 & 0.0358 & 0.0717 & 0.0318 & 0.0116 & 0.0076 \\
3 & 0.0234 & 0.0275 & 0.0139 & 0.0200 & 0.0313 & 0.0327 & 0.0314 \\
4 & 0.3641 & 0.1704 & 0.5109 & 0.5950 & 0.1448 & 0.1955 & 0.0514 \\
5 & 0.1935 & 0.1081 & 0.1912 & 0.1971 & 0.1646 & 0.0965 & 0.0408 \\
\rowcolor{yellow}
6 & 0.2427 & 0.2399 & 0.2279 & 0.1778 & 0.2372 & 0.3163 & 0.0333 \\
7 & 0.0179 & 0.0176 & 0.0184 & 0.0295 & 0.0248 & 0.0211 & 0.0057 \\
\bottomrule
\end{tabular}
}
\end{table}

\begin{table}
\centering
\caption{The US Requirements Tables for 1982}
\label{tab:1982}
\resizebox{.8\columnwidth}{!}{%
\begin{tabular}{lccccccc}
\toprule
\multicolumn{8}{l}{Technical Coefficients Matrix ($A$)} \\
\midrule
 & 1 & 2 & 3 & 4 & 5 & 6 & 7 \\
 \midrule 
1 & 0.2853 & 0.0002 & 0.0019 & 0.0455 & 0.0005 & 0.0045 & 0.0047 \\ 
2 & 0.0025 & 0.0467 & 0.0078 & 0.078 & 0.0486 & 0.0006 & 0.0041 \\ 
3 & 0.0093 & 0.0247 & 0.001 & 0.005 & 0.0194 & 0.0238 & 0.0221 \\ 
4 & 0.1806 & 0.0637 & 0.3201 & 0.3505 & 0.0764 & 0.0877 & 0.0194 \\ 
5 & 0.0683 & 0.0414 & 0.0973 & 0.1042 & 0.1255 & 0.0454 & 0.0351 \\ 
\rowcolor{yellow}
6 & 0.0816 & 0.1561 & 0.098 & 0.0542 & 0.121 & 0.17 & 0.0071 \\ 
7 & 0.0035 & 0.0046 & 0.0043 & 0.015 & 0.0164 & 0.011 & 0.0043 \\ 
\midrule
\multicolumn{8}{l}{Simple Direct Requirements Matrix ($N^\texttt{d}$)} \\
\midrule
 & 1 & 2 & 3 & 4 & 5 & 6 & 7 \\
\midrule
1 & 0.4081 & 0.0002 & 0.0019 & 0.0749 & 0.0006 & 0.0056 & 0.0047 \\
2 & 0.0036 & 0.0499 & 0.0079 & 0.1284 & 0.0577 & 0.0007 & 0.0041 \\
3 & 0.0133 & 0.0264 & 0.0010 & 0.0082 & 0.0230 & 0.0296 & 0.0222 \\
4 & 0.2584 & 0.0680 & 0.3249 & 0.5771 & 0.0907 & 0.1090 & 0.0195 \\
5 & 0.0977 & 0.0442 & 0.0988 & 0.1716 & 0.1489 & 0.0564 & 0.0353 \\
\rowcolor{yellow}
6 & 0.1167 & 0.1667 & 0.0995 & 0.0892 & 0.1436 & 0.2113 & 0.0071 \\
7 & 0.0050 & 0.0049 & 0.0044 & 0.0247 & 0.0195 & 0.0137 & 0.0043 \\
\midrule
\multicolumn{8}{l}{Simple Indirect Requirements Matrix ($N^\texttt{i}$)} \\
\midrule
 & 1 & 2 & 3 & 4 & 5 & 6 & 7 \\
\midrule
1 & 0.0224 & 0.0124 & 0.0386 & 0.0315 & 0.0142 & 0.0155 & 0.0057 \\
2 & 0.0481 & 0.0178 & 0.0574 & 0.0185 & 0.0193 & 0.0221 & 0.0077 \\
3 & 0.0126 & 0.0084 & 0.0139 & 0.0140 & 0.0085 & 0.0041 & 0.0023 \\
4 & 0.2080 & 0.0991 & 0.2442 & 0.0694 & 0.1047 & 0.0954 & 0.0364 \\
5 & 0.0866 & 0.0435 & 0.0995 & 0.0530 & 0.0378 & 0.0397 & 0.0172 \\
\rowcolor{yellow}
6 & 0.0942 & 0.0633 & 0.1029 & 0.0920 & 0.0621 & 0.0314 & 0.0189 \\
7 & 0.0128 & 0.0067 & 0.0145 & 0.0070 & 0.0058 & 0.0050 & 0.0022 \\
\midrule
\multicolumn{8}{l}{Simple Transfer (Total) Requirements Matrix ($N^\texttt{t}$)} \\
\midrule
 & 1 & 2 & 3 & 4 & 5 & 6 & 7 \\
\midrule
1 & 0.4305 & 0.0126 & 0.0405 & 0.1064 & 0.0148 & 0.0211 & 0.0105 \\
2 & 0.0517 & 0.0676 & 0.0653 & 0.1469 & 0.0770 & 0.0228 & 0.0118 \\
3 & 0.0259 & 0.0348 & 0.0149 & 0.0222 & 0.0315 & 0.0337 & 0.0246 \\
4 & 0.4663 & 0.1671 & 0.5691 & 0.6464 & 0.1953 & 0.2044 & 0.0559 \\
5 & 0.1843 & 0.0877 & 0.1982 & 0.2246 & 0.1868 & 0.0961 & 0.0525 \\
\rowcolor{yellow}
6 & 0.2109 & 0.2299 & 0.2023 & 0.1812 & 0.2057 & 0.2427 & 0.0261 \\
7 & 0.0178 & 0.0116 & 0.0189 & 0.0317 & 0.0253 & 0.0187 & 0.0065 \\
\bottomrule
\end{tabular}
}
\end{table}

\begin{table}
\centering
\caption{The US Requirements Tables for 1977}
\label{tab:1977}
\resizebox{.8\columnwidth}{!}{%
\begin{tabular}{lccccccc}
\toprule
\multicolumn{8}{l}{Technical Coefficients Matrix ($A$)} \\
\midrule
 & 1 & 2 & 3 & 4 & 5 & 6 & 7 \\
 \midrule 
1 & 0.2463 & 0.0004 & 0.0035 & 0.047 & 0.0011 & 0.0052 & 0.0007\\ 
2 & 0.0022 & 0.0712 & 0.0093 & 0.0575 & 0.0288 & 0.0005 & 0.0048\\ 
3 & 0.0107 & 0.0375 & 0.0011 & 0.0064 & 0.019 & 0.0293 & 0.0193\\ 
4 & 0.2021 & 0.095 & 0.3722 & 0.3816 & 0.0645 & 0.0885 & 0.0126\\ 
5 & 0.0716 & 0.0478 & 0.1148 & 0.0855 & 0.1058 & 0.0498 & 0.0228\\ 
\rowcolor{yellow}
6 & 0.0864 & 0.0999 & 0.0724 & 0.0482 & 0.12 & 0.151 & 0.0083\\ 
7 & 0.0029 & 0.0049 & 0.0043 & 0.0141 & 0.0137 & 0.009 & 0.004\\
\midrule
\multicolumn{8}{l}{Simple Direct Requirements Matrix ($N^\texttt{d}$)} \\
\midrule
 & 1 & 2 & 3 & 4 & 5 & 6 & 7 \\
\midrule
1 & 0.3352 & 0.0004 & 0.0036 & 0.0812 & 0.0013 & 0.0063 & 0.0007 \\
2 & 0.0030 & 0.0780 & 0.0095 & 0.0994 & 0.0332 & 0.0006 & 0.0048 \\
3 & 0.0146 & 0.0411 & 0.0011 & 0.0111 & 0.0219 & 0.0355 & 0.0194 \\
4 & 0.2750 & 0.1040 & 0.3787 & 0.6594 & 0.0744 & 0.1072 & 0.0127 \\
5 & 0.0974 & 0.0523 & 0.1168 & 0.1477 & 0.1221 & 0.0603 & 0.0229 \\
\rowcolor{yellow}
6 & 0.1176 & 0.1094 & 0.0737 & 0.0833 & 0.1384 & 0.1830 & 0.0083 \\
7 & 0.0039 & 0.0054 & 0.0044 & 0.0244 & 0.0158 & 0.0109 & 0.0040 \\
\midrule
\multicolumn{8}{l}{Simple Indirect Requirements Matrix ($N^\texttt{i}$)} \\
\midrule
 & 1 & 2 & 3 & 4 & 5 & 6 & 7 \\
\midrule
1 & 0.0256 & 0.0164 & 0.0453 & 0.0281 & 0.0129 & 0.0160 & 0.0032 \\
2 & 0.0381 & 0.0171 & 0.0500 & 0.0143 & 0.0141 & 0.0171 & 0.0043 \\
3 & 0.0144 & 0.0086 & 0.0165 & 0.0141 & 0.0088 & 0.0045 & 0.0019 \\
4 & 0.2407 & 0.1333 & 0.3055 & 0.0685 & 0.1029 & 0.1106 & 0.0298 \\
5 & 0.0783 & 0.0460 & 0.0962 & 0.0448 & 0.0315 & 0.0362 & 0.0114 \\
\rowcolor{yellow}
6 & 0.0825 & 0.0529 & 0.0942 & 0.0690 & 0.0445 & 0.0288 & 0.0120 \\
7 & 0.0119 & 0.0065 & 0.0146 & 0.0051 & 0.0046 & 0.0048 & 0.0014 \\
\midrule
\multicolumn{8}{l}{Simple Transfer (Total) Requirements Matrix ($N^\texttt{t}$)} \\
\midrule
 & 1 & 2 & 3 & 4 & 5 & 6 & 7 \\
\midrule
1 & 0.3608 & 0.0169 & 0.0489 & 0.1093 & 0.0142 & 0.0223 & 0.0039 \\
2 & 0.0411 & 0.0951 & 0.0595 & 0.1137 & 0.0473 & 0.0177 & 0.0091 \\
3 & 0.0289 & 0.0497 & 0.0176 & 0.0252 & 0.0308 & 0.0400 & 0.0213 \\
4 & 0.5157 & 0.2374 & 0.6842 & 0.7279 & 0.1773 & 0.2179 & 0.0425 \\
5 & 0.1757 & 0.0983 & 0.2130 & 0.1925 & 0.1536 & 0.0966 & 0.0344 \\
\rowcolor{yellow}
6 & 0.2001 & 0.1623 & 0.1679 & 0.1522 & 0.1830 & 0.2118 & 0.0204 \\
7 & 0.0158 & 0.0118 & 0.0190 & 0.0295 & 0.0204 & 0.0157 & 0.0054 \\
\bottomrule
\end{tabular}
}
\end{table}
\begin{table}
\centering
\caption{The US Requirements Tables for 1972}
\label{tab:1972}
\resizebox{.8\columnwidth}{!}{%
\begin{tabular}{lccccccc}
\toprule
\multicolumn{8}{l}{Technical Coefficients Matrix ($A$)} \\
\midrule
 & 1 & 2 & 3 & 4 & 5 & 6 & 7 \\
 \midrule 
1 & 0.3141 & 0.0003 & 0.0028 & 0.0542 & 0.001 & 0.0053 & 0.0012 \\ 
2 & 0.0019 & 0.0542 & 0.0091 & 0.0296 & 0.016 & 0.0002 & 0.002 \\ 
3 & 0.0069 & 0.0282 & 0.0003 & 0.0043 & 0.0156 & 0.0263 & 0.0166 \\ 
4 & 0.1436 & 0.0943 & 0.3522 & 0.3771 & 0.0407 & 0.0892 & 0.0078 \\ 
5 & 0.0616 & 0.0481 & 0.1043 & 0.0786 & 0.098 & 0.0442 & 0.0202 \\ 
\rowcolor{yellow}
6 & 0.0865 & 0.1471 & 0.0686 & 0.0591 & 0.1157 & 0.1621 & 0.0105 \\ 
7 & 0.0023 & 0.0063 & 0.0042 & 0.0117 & 0.0118 & 0.0096 & 0.0033 \\
\midrule
\multicolumn{8}{l}{Simple Direct Requirements Matrix ($N^\texttt{d}$)} \\
\midrule
 & 1 & 2 & 3 & 4 & 5 & 6 & 7 \\
\midrule
1 & 0.4684 & 0.0003 & 0.0028 & 0.0916 & 0.0011 & 0.0065 & 0.0012 \\
2 & 0.0028 & 0.0578 & 0.0092 & 0.0500 & 0.0181 & 0.0002 & 0.0020 \\
3 & 0.0103 & 0.0301 & 0.0003 & 0.0073 & 0.0177 & 0.0323 & 0.0167 \\
4 & 0.2142 & 0.1006 & 0.3563 & 0.6375 & 0.0461 & 0.1095 & 0.0078 \\
5 & 0.0919 & 0.0513 & 0.1055 & 0.1329 & 0.1110 & 0.0542 & 0.0203 \\
\rowcolor{yellow}
6 & 0.1290 & 0.1569 & 0.0694 & 0.0999 & 0.1310 & 0.1989 & 0.0105 \\
7 & 0.0034 & 0.0067 & 0.0042 & 0.0198 & 0.0134 & 0.0118 & 0.0033 \\
\midrule
\multicolumn{8}{l}{Simple Indirect Requirements Matrix ($N^\texttt{i}$)} \\
\midrule
 & 1 & 2 & 3 & 4 & 5 & 6 & 7 \\
\midrule
1 & 0.0229 & 0.0200 & 0.0523 & 0.0436 & 0.0114 & 0.0198 & 0.0031 \\
2 & 0.0154 & 0.0086 & 0.0234 & 0.0063 & 0.0051 & 0.0084 & 0.0017 \\
3 & 0.0102 & 0.0086 & 0.0115 & 0.0100 & 0.0061 & 0.0029 & 0.0013 \\
4 & 0.1838 & 0.1255 & 0.2692 & 0.0530 & 0.0726 & 0.0993 & 0.0214 \\
5 & 0.0585 & 0.0423 & 0.0799 & 0.0375 & 0.0216 & 0.0308 & 0.0084 \\
\rowcolor{yellow}
6 & 0.0788 & 0.0646 & 0.0948 & 0.0684 & 0.0412 & 0.0283 & 0.0106 \\
7 & 0.0087 & 0.0061 & 0.0115 & 0.0044 & 0.0034 & 0.0038 & 0.0010 \\
\midrule
\multicolumn{8}{l}{Simple Transfer (Total) Requirements Matrix ($N^\texttt{t}$)} \\
\midrule
 & 1 & 2 & 3 & 4 & 5 & 6 & 7 \\
\midrule 
1 & 0.4913 & 0.0204 & 0.0551 & 0.1353 & 0.0125 & 0.0263 & 0.0043 \\
2 & 0.0183 & 0.0664 & 0.0326 & 0.0563 & 0.0232 & 0.0087 & 0.0037 \\
3 & 0.0205 & 0.0387 & 0.0118 & 0.0173 & 0.0237 & 0.0352 & 0.0179 \\
4 & 0.3979 & 0.2260 & 0.6255 & 0.6904 & 0.1187 & 0.2088 & 0.0292 \\
5 & 0.1503 & 0.0936 & 0.1854 & 0.1703 & 0.1326 & 0.0850 & 0.0286 \\
\rowcolor{yellow}
6 & 0.2078 & 0.2215 & 0.1642 & 0.1683 & 0.1723 & 0.2272 & 0.0212 \\
7 & 0.0121 & 0.0128 & 0.0157 & 0.0242 & 0.0167 & 0.0155 & 0.0043 \\
\bottomrule
\end{tabular}
}
\end{table}
\begin{table}
\centering
\caption{The US Requirements Tables for 1967}
\label{tab:1967}
\resizebox{.8\columnwidth}{!}{%
\begin{tabular}{lccccccc}
\toprule
\multicolumn{8}{l}{Technical Coefficients Matrix ($A$)} \\
\midrule
 & 1 & 2 & 3 & 4 & 5 & 6 & 7 \\
 \midrule 
1 & 0.3016 & 0 & 0.0025 & 0.0508 & 0.0061 & 0.0022 & 0.0177 \\ 
2 & 0.0022 & 0.0515 & 0.009 & 0.028 & 0.0085 & 0.0001 & 0.0044 \\ 
3 & 0.0095 & 0.0229 & 0.0003 & 0.0041 & 0.0248 & 0.0088 & 0.0534 \\ 
4 & 0.136 & 0.0935 & 0.3634 & 0.3894 & 0.0418 & 0.1577 & 0.2452 \\ 
5 & 0.1225 & 0.1726 & 0.1221 & 0.0834 & 0.1432 & 0.1438 & 0.2661 \\ 
\rowcolor{yellow}
6 & 0.0278 & 0.0228 & 0.0526 & 0.0325 & 0.0548 & 0.0694 & 0.0703 \\ 
7 & 0.0183 & 0.0962 & 0.0088 & 0.0408 & 0.0444 & 0.0286 & 0.0455 \\
\midrule
\multicolumn{8}{l}{Simple Direct Requirements Matrix ($N^\texttt{d}$)} \\
\midrule
 & 1 & 2 & 3 & 4 & 5 & 6 & 7 \\
\midrule
1 & 0.4423 & 0 & 0.0025 & 0.0899 & 0.0075 & 0.0024 & 0.0195 \\
2 & 0.0032 & 0.0549 & 0.0091 & 0.0495 & 0.0105 & 0.0001 & 0.0048 \\
3 & 0.0139 & 0.0244 & 0.0003 & 0.0073 & 0.0305 & 0.0098 & 0.0588 \\
4 & 0.1994 & 0.0997 & 0.3688 & 0.6890 & 0.0515 & 0.1748 & 0.2702 \\
5 & 0.1796 & 0.1841 & 0.1239 & 0.1476 & 0.1763 & 0.1594 & 0.2932 \\
\rowcolor{yellow}
6 & 0.0408 & 0.0243 & 0.0534 & 0.0575 & 0.0675 & 0.0769 & 0.0775 \\
7 & 0.0268 & 0.1026 & 0.0089 & 0.0722 & 0.0547 & 0.0317 & 0.0501 \\
\midrule
\multicolumn{8}{l}{Simple Indirect Requirements Matrix ($N^\texttt{i}$)} \\
\midrule
 & 1 & 2 & 3 & 4 & 5 & 6 & 7 \\
\midrule
1 & 0.0241 & 0.0274 & 0.0554 & 0.0439 & 0.0174 & 0.0304 & 0.0548 \\
2 & 0.0155 & 0.0117 & 0.0236 & 0.0059 & 0.0061 & 0.0132 & 0.0221 \\
3 & 0.0135 & 0.0170 & 0.0147 & 0.0148 & 0.0058 & 0.0116 & 0.0160 \\
4 & 0.2132 & 0.1878 & 0.3215 & 0.0803 & 0.1128 & 0.1755 & 0.3035 \\
5 & 0.1115 & 0.1225 & 0.1384 & 0.1039 & 0.0550 & 0.0904 & 0.1545 \\
\rowcolor{yellow}
6 & 0.0414 & 0.0434 & 0.0505 & 0.0331 & 0.0192 & 0.0318 & 0.0593 \\
7 & 0.0371 & 0.0344 & 0.0497 & 0.0262 & 0.0147 & 0.0303 & 0.0518 \\
\midrule
\multicolumn{8}{l}{Simple Transfer (Total) Requirements Matrix ($N^\texttt{t}$)} \\
\midrule
 & 1 & 2 & 3 & 4 & 5 & 6 & 7 \\
\midrule
1 & 0.4664 & 0.0274 & 0.0579 & 0.1337 & 0.0249 & 0.0328 & 0.0743 \\
2 & 0.0188 & 0.0666 & 0.0328 & 0.0555 & 0.0166 & 0.0133 & 0.0270 \\
3 & 0.0274 & 0.0414 & 0.0150 & 0.0221 & 0.0363 & 0.0213 & 0.0748 \\
4 & 0.4126 & 0.2876 & 0.6903 & 0.7693 & 0.1642 & 0.3504 & 0.5737 \\
5 & 0.2911 & 0.3066 & 0.2624 & 0.2514 & 0.2313 & 0.2498 & 0.4477 \\
\rowcolor{yellow}
6 & 0.0822 & 0.0677 & 0.1039 & 0.0906 & 0.0867 & 0.1087 & 0.1368 \\
7 & 0.0639 & 0.1370 & 0.0586 & 0.0984 & 0.0694 & 0.0620 & 0.1019 \\
\bottomrule
\end{tabular}
}
\end{table}
\begin{table}
\centering
\caption{The US Requirements Tables for 1963}
\label{tab:1963}
\resizebox{.8\columnwidth}{!}{%
\begin{tabular}{lccccccc}
\toprule
\multicolumn{8}{l}{Technical Coefficients Matrix ($A$)} \\
\midrule
 & 1 & 2 & 3 & 4 & 5 & 6 & 7 \\
 \midrule 
1 & 0.31 & 0 & 0.0038 & 0.0574 & 0.0087 & 0.0006 & 0.0322 \\ 
2 & 0.0022 & 0.0553 & 0.0086 & 0.0314 & 0.0085 & 0.0002 & 0.0077 \\ 
3 & 0.0099 & 0.0202 & 0.0003 & 0.003 & 0.0315 & 0.0093 & 0.055 \\ 
4 & 0.133 & 0.0812 & 0.37 & 0.3983 & 0.0401 & 0.1496 & 0.2574 \\ 
5 & 0.1054 & 0.1935 & 0.133 & 0.0807 & 0.1415 & 0.1544 & 0.257 \\ 
\rowcolor{yellow}
6 & 0.0246 & 0.0143 & 0.0429 & 0.0267 & 0.0473 & 0.0604 & 0.0662 \\ 
7 & 0.0198 & 0.0982 & 0.0075 & 0.036 & 0.0439 & 0.034 & 0.038 \\ 
\midrule
\multicolumn{8}{l}{Simple Direct Requirements Matrix ($N^\texttt{d}$)} \\
\midrule
 & 1 & 2 & 3 & 4 & 5 & 6 & 7 \\
\midrule
1 & 0.4619 & 0 & 0.0039 & 0.1029 & 0.0107 & 0.0007 & 0.0352 \\
2 & 0.0033 & 0.0593 & 0.0087 & 0.0563 & 0.0104 & 0.0002 & 0.0084 \\
3 & 0.0148 & 0.0217 & 0.0003 & 0.0054 & 0.0387 & 0.0102 & 0.0601 \\
4 & 0.1982 & 0.0871 & 0.3760 & 0.7140 & 0.0493 & 0.1637 & 0.2813 \\
5 & 0.1570 & 0.2074 & 0.1352 & 0.1447 & 0.1739 & 0.1689 & 0.2808 \\
\rowcolor{yellow}
6 & 0.0367 & 0.0153 & 0.0436 & 0.0479 & 0.0581 & 0.0661 & 0.0723 \\
7 & 0.0295 & 0.1053 & 0.0076 & 0.0645 & 0.0540 & 0.0372 & 0.0415 \\
\midrule
\multicolumn{8}{l}{Simple Indirect Requirements Matrix ($N^\texttt{i}$)} \\
\midrule
 & 1 & 2 & 3 & 4 & 5 & 6 & 7 \\
\midrule
1 & 0.0281 & 0.0334 & 0.0668 & 0.0538 & 0.0221 & 0.0352 & 0.0721 \\
2 & 0.0170 & 0.0128 & 0.0272 & 0.0068 & 0.0072 & 0.0146 & 0.0255 \\
3 & 0.0140 & 0.0196 & 0.0160 & 0.0163 & 0.0056 & 0.0135 & 0.0186 \\
4 & 0.2133 & 0.1842 & 0.3326 & 0.0786 & 0.1170 & 0.1783 & 0.3228 \\
5 & 0.1050 & 0.1209 & 0.1377 & 0.1015 & 0.0551 & 0.0911 & 0.1582 \\
\rowcolor{yellow}
6 & 0.0333 & 0.0377 & 0.0430 & 0.0279 & 0.0164 & 0.0279 & 0.0508 \\
7 & 0.0333 & 0.0322 & 0.0474 & 0.0263 & 0.0138 & 0.0286 & 0.0513 \\
\midrule
\multicolumn{8}{l}{Simple Transfer (Total) Requirements Matrix ($N^\texttt{t}$)} \\
\midrule
 & 1 & 2 & 3 & 4 & 5 & 6 & 7 \\
\midrule
1 & 0.4900 & 0.0334 & 0.0706 & 0.1566 & 0.0328 & 0.0359 & 0.1073 \\
2 & 0.0203 & 0.0721 & 0.0359 & 0.0631 & 0.0176 & 0.0148 & 0.0339 \\
3 & 0.0288 & 0.0412 & 0.0163 & 0.0217 & 0.0443 & 0.0237 & 0.0787 \\
4 & 0.4115 & 0.2713 & 0.7087 & 0.7925 & 0.1663 & 0.3419 & 0.6040 \\
5 & 0.2620 & 0.3283 & 0.2729 & 0.2461 & 0.2290 & 0.2600 & 0.4391 \\
\rowcolor{yellow}
6 & 0.0699 & 0.0530 & 0.0866 & 0.0758 & 0.0745 & 0.0940 & 0.1232 \\
7 & 0.0628 & 0.1375 & 0.0551 & 0.0908 & 0.0678 & 0.0658 & 0.0928 \\
\bottomrule
\end{tabular}
}
\end{table}
\begin{table}
\centering
\caption{The US Requirements Tables for 1958}
\label{tab:1958}
\resizebox{.8\columnwidth}{!}{%
\begin{tabular}{lccccccc}
\toprule
\multicolumn{8}{l}{Technical Coefficients Matrix ($A$)} \\
\midrule
 & 1 & 2 & 3 & 4 & 5 & 6 & 7 \\
 \midrule 
1 & 0.2954 & 0 & 0.0034 & 0.0703 & 0.0095 & 0.0003 & 0.0414\\ 
2 & 0.0019 & 0.0616 & 0.0109 & 0.0374 & 0.0077 & 0.0005 & 0.0084\\ 
3 & 0.0116 & 0.0006 & 0.0001 & 0.0021 & 0.0357 & 0.0124 & 0.068\\ 
4 & 0.1158 & 0.0794 & 0.3828 & 0.3802 & 0.0422 & 0.2247 & 0.2935\\ 
5 & 0.1122 & 0.1611 & 0.1368 & 0.0877 & 0.142 & 0.1387 & 0.2581\\ 
\rowcolor{yellow}
6 & 0.023 & 0.0232 & 0.0428 & 0.0245 & 0.0451 & 0.0662 & 0.0714\\ 
7 & 0.0207 & 0.1067 & 0.0055 & 0.0392 & 0.043 & 0.0292 & 0.0414\\
\midrule
\multicolumn{8}{l}{Simple Direct Requirements Matrix ($N^\texttt{d}$)} \\
\midrule
 & 1 & 2 & 3 & 4 & 5 & 6 & 7 \\
\midrule
1 & 0.4323 & 0 & 0.0035 & 0.1239 & 0.0117 & 0.0003 & 0.0457 \\
2 & 0.0028 & 0.0666 & 0.0111 & 0.0659 & 0.0095 & 0.0006 & 0.0093 \\
3 & 0.0170 & 0.0006 & 0.0001 & 0.0037 & 0.0440 & 0.0137 & 0.0751 \\
4 & 0.1695 & 0.0858 & 0.3897 & 0.6703 & 0.0520 & 0.2483 & 0.3243 \\
5 & 0.1642 & 0.1741 & 0.1393 & 0.1546 & 0.1751 & 0.1533 & 0.2852 \\
\rowcolor{yellow}
6 & 0.0337 & 0.0251 & 0.0436 & 0.0432 & 0.0556 & 0.0732 & 0.0789 \\
7 & 0.0303 & 0.1153 & 0.0056 & 0.0691 & 0.0530 & 0.0323 & 0.0457 \\
\midrule
\multicolumn{8}{l}{Simple Indirect Requirements Matrix ($N^\texttt{i}$)} \\
\midrule
 & 1 & 2 & 3 & 4 & 5 & 6 & 7 \\
\midrule
1 & 0.0312 & 0.0399 & 0.0815 & 0.0615 & 0.0275 & 0.0561 & 0.0954 \\
2 & 0.0180 & 0.0140 & 0.0329 & 0.0081 & 0.0093 & 0.0226 & 0.0336 \\
3 & 0.0154 & 0.0223 & 0.0178 & 0.0192 & 0.0064 & 0.0159 & 0.0215 \\
4 & 0.1985 & 0.1864 & 0.3392 & 0.0928 & 0.1318 & 0.2350 & 0.3710 \\
5 & 0.1034 & 0.1198 & 0.1485 & 0.1090 & 0.0583 & 0.1120 & 0.1819 \\
\rowcolor{yellow}
6 & 0.0318 & 0.0364 & 0.0437 & 0.0308 & 0.0177 & 0.0319 & 0.0554 \\
7 & 0.0329 & 0.0322 & 0.0523 & 0.0294 & 0.0153 & 0.0370 & 0.0593 \\
\midrule
\multicolumn{8}{l}{Total Requirements ($N^\texttt{t}$)} \\
\midrule
 & 1 & 2 & 3 & 4 & 5 & 6 & 7 \\
\midrule
1 & 0.4635 & 0.0399 & 0.0850 & 0.1854 & 0.0393 & 0.0565 & 0.1411 \\
2 & 0.0208 & 0.0806 & 0.0440 & 0.0740 & 0.0188 & 0.0231 & 0.0429 \\
3 & 0.0324 & 0.0230 & 0.0179 & 0.0229 & 0.0504 & 0.0296 & 0.0966 \\
4 & 0.3680 & 0.2722 & 0.7289 & 0.7631 & 0.1838 & 0.4833 & 0.6953 \\
5 & 0.2676 & 0.2939 & 0.2877 & 0.2636 & 0.2334 & 0.2653 & 0.4671 \\
\rowcolor{yellow}
6 & 0.0655 & 0.0615 & 0.0873 & 0.0740 & 0.0734 & 0.1050 & 0.1343 \\
7 & 0.0632 & 0.1475 & 0.0579 & 0.0986 & 0.0683 & 0.0693 & 0.1050 \\
\bottomrule
\end{tabular}
}
\end{table}
\begin{table}
\centering
\caption{The US Requirements Tables for 1947}
\label{tab:1947}
\resizebox{.8\columnwidth}{!}{%
\begin{tabular}{lccccccc}
\toprule
\multicolumn{8}{l}{Technical Coefficients Matrix ($A$)} \\
\midrule
 & 1 & 2 & 3 & 4 & 5 & 6 & 7 \\
\midrule 
1 & 0.3272 & 0 & 0.0031 & 0.1212 & 0.0146 & 0.0053 & 0.0141 \\ 
2 & 0.001 & 0.0835 & 0.0094 & 0.0334 & 0.0098 & 0.0013 & 0.0032 \\ 
3 & 0.0122 & 0.0015 & 0.0002 & 0.0026 & 0.044 & 0.0081 & 0.0639 \\ 
4 & 0.0949 & 0.098 & 0.3795 & 0.3733 & 0.0522 & 0.1709 & 0.2733 \\ 
5 & 0.108 & 0.1091 & 0.1462 & 0.0735 & 0.1212 & 0.1254 & 0.3043 \\ 
\rowcolor{yellow}
6 & 0.0081 & 0.0088 & 0.0436 & 0.0161 & 0.0387 & 0.066 & 0.0433 \\ 
7 & 0.0011 & 0.0046 & 0.007 & 0.0232 & 0.0389 & 0.0338 & 0.0145 \\ 
\midrule
\multicolumn{8}{l}{Simple Direct Requirements Matrix ($N^\texttt{d}$)} \\
\midrule
 & 1 & 2 & 3 & 4 & 5 & 6 & 7 \\
\midrule
1 & 0.5054 & 0 & 0.0032 & 0.2085 & 0.0175 & 0.0058 & 0.0148 \\
2 & 0.0015 & 0.0919 & 0.0096 & 0.0575 & 0.0118 & 0.0014 & 0.0034 \\
3 & 0.0188 & 0.0017 & 0.0002 & 0.0045 & 0.0528 & 0.0088 & 0.0672 \\
4 & 0.1466 & 0.1079 & 0.3866 & 0.6421 & 0.0627 & 0.1864 & 0.2872 \\
5 & 0.1668 & 0.1201 & 0.1490 & 0.1264 & 0.1455 & 0.1368 & 0.3198 \\
\rowcolor{yellow}
6 & 0.0125 & 0.0097 & 0.0444 & 0.0277 & 0.0465 & 0.0720 & 0.0455 \\
7 & 0.0017 & 0.0051 & 0.0071 & 0.0399 & 0.0467 & 0.0369 & 0.0152 \\
\midrule
\multicolumn{8}{l}{Simple Indirect Requirements Matrix ($N^\texttt{i}$)} \\
\midrule
 & 1 & 2 & 3 & 4 & 5 & 6 & 7 \\
\midrule
1 & 0.0392 & 0.0429 & 0.1356 & 0.1077 & 0.0442 & 0.0755 & 0.1267 \\
2 & 0.0137 & 0.0091 & 0.0298 & 0.0084 & 0.0088 & 0.0165 & 0.0284 \\
3 & 0.0125 & 0.0099 & 0.0186 & 0.0172 & 0.0052 & 0.0153 & 0.0244 \\
4 & 0.1447 & 0.1061 & 0.3169 & 0.0780 & 0.1249 & 0.1833 & 0.3144 \\
5 & 0.0654 & 0.0511 & 0.1249 & 0.0924 & 0.0551 & 0.0851 & 0.1430 \\
\rowcolor{yellow}
6 & 0.0180 & 0.0132 & 0.0299 & 0.0178 & 0.0125 & 0.0188 & 0.0386 \\
7 & 0.0174 & 0.0128 & 0.0304 & 0.0116 & 0.0077 & 0.0184 & 0.0357 \\
\midrule
\multicolumn{8}{l}{Simple Transfer (Total) Requirements Matrix ($N^\texttt{t}$)} \\
\midrule
 & 1 & 2 & 3 & 4 & 5 & 6 & 7 \\
\midrule
1 & 0.5446 & 0.0429 & 0.1388 & 0.3161 & 0.0617 & 0.0813 & 0.1415 \\
2 & 0.0152 & 0.1010 & 0.0394 & 0.0658 & 0.0206 & 0.0179 & 0.0318 \\
3 & 0.0313 & 0.0116 & 0.0188 & 0.0217 & 0.0581 & 0.0241 & 0.0916 \\
4 & 0.2912 & 0.2140 & 0.7036 & 0.7201 & 0.1875 & 0.3697 & 0.6016 \\
5 & 0.2322 & 0.1712 & 0.2739 & 0.2188 & 0.2007 & 0.2219 & 0.4628 \\
\rowcolor{yellow}
6 & 0.0305 & 0.0229 & 0.0743 & 0.0455 & 0.0589 & 0.0908 & 0.0841 \\
7 & 0.0191 & 0.0179 & 0.0375 & 0.0515 & 0.0544 & 0.0552 & 0.0510 \\
\bottomrule
\end{tabular}
}
\end{table}
\begin{table}
\centering
\caption{The US Requirements Tables for 1939}
\label{tab:1939}
\resizebox{.8\columnwidth}{!}{%
\begin{tabular}{lccccccc}
\toprule
\multicolumn{8}{l}{Technical Coefficients Matrix ($A$)} \\
\midrule
 & 1 & 2 & 3 & 4 & 5 & 6 & 7 \\
\midrule 
1 & 0.1074 & 0 & 0.0788 & 0.0802 & 0.0209 & 0.0146 & 0.0232 \\ 
2 & 0.0032 & 0.2228 & 0.1804 & 0.0294 & 0.0247 & 0.002 & 0.0203 \\ 
3 & 0.0214 & 0.0085 & 0 & 0.0115 & 0.0304 & 0.0636 & 0.1379 \\ 
4 & 0.1593 & 0.0561 & 0.2187 & 0.2319 & 0.1837 & 0.1724 & 0.3142 \\ 
5 & 0.2352 & 0.2575 & 0.0304 & 0.2653 & 0.1129 & 0.0017 & 0.046 \\ 
\rowcolor{yellow}
6 & 0.0386 & 0.0029 & 0.0003 & 0.0123 & 0.0217 & 0.0217 & 0.0419 \\ 
7 & 0.0594 & 0.1764 & 0.0083 & 0.1762 & 0.2443 & 0.1732 & 0.0426 \\
\midrule
\multicolumn{8}{l}{Simple Direct Requirements Matrix ($N^\texttt{d}$)} \\
\midrule
 & 1 & 2 & 3 & 4 & 5 & 6 & 7 \\
\midrule
1 & 0.1293 & 0 & 0.0836 & 0.1433 & 0.0295 & 0.0154 & 0.0307 \\
2 & 0.0039 & 0.3042 & 0.1915 & 0.0525 & 0.0349 & 0.0021 & 0.0268 \\
3 & 0.0258 & 0.0116 & 0 & 0.0205 & 0.0429 & 0.0669 & 0.1824 \\
4 & 0.1917 & 0.0766 & 0.2322 & 0.4144 & 0.2593 & 0.1814 & 0.4155 \\
5 & 0.2831 & 0.3516 & 0.0323 & 0.4740 & 0.1594 & 0.0018 & 0.0608 \\
\rowcolor{yellow}
6 & 0.0465 & 0.0040 & 0.0003 & 0.0220 & 0.0306 & 0.0228 & 0.0554 \\
7 & 0.0715 & 0.2408 & 0.0088 & 0.3148 & 0.3449 & 0.1822 & 0.0563 \\
\midrule
\multicolumn{8}{l}{Simple Indirect Requirements Matrix ($N^\texttt{i}$)} \\
\midrule
 & 1 & 2 & 3 & 4 & 5 & 6 & 7 \\
\midrule
1 & 0.0743 & 0.0845 & 0.0753 & 0.0592 & 0.0864 & 0.0720 & 0.0990 \\
2 & 0.0726 & 0.0611 & 0.0944 & 0.0806 & 0.0781 & 0.0660 & 0.0972 \\
3 & 0.0763 & 0.1003 & 0.0615 & 0.1054 & 0.0864 & 0.0636 & 0.0286 \\
4 & 0.3800 & 0.4388 & 0.3222 & 0.3724 & 0.3742 & 0.3118 & 0.3275 \\
5 & 0.2508 & 0.2520 & 0.3090 & 0.1855 & 0.2523 & 0.2135 & 0.3078 \\
\rowcolor{yellow}
6 & 0.0351 & 0.0454 & 0.0331 & 0.0470 & 0.0358 & 0.0292 & 0.0243 \\
7 & 0.2744 & 0.2747 & 0.2581 & 0.2330 & 0.1731 & 0.1729 & 0.2661 \\
\midrule
\multicolumn{8}{l}{Simple Transfer (Total) Requirements Matrix ($N^\texttt{t}$)} \\
\midrule
 & 1 & 2 & 3 & 4 & 5 & 6 & 7 \\
\midrule
1 & 0.2035 & 0.0845 & 0.1590 & 0.2025 & 0.1159 & 0.0873 & 0.1297 \\
2 & 0.0765 & 0.3653 & 0.2859 & 0.1331 & 0.1130 & 0.0681 & 0.1241 \\
3 & 0.1021 & 0.1119 & 0.0615 & 0.1260 & 0.1293 & 0.1305 & 0.2110 \\
4 & 0.5717 & 0.5154 & 0.5544 & 0.7868 & 0.6335 & 0.4932 & 0.7431 \\
5 & 0.5339 & 0.6035 & 0.3412 & 0.6595 & 0.4117 & 0.2153 & 0.3686 \\
\rowcolor{yellow}
6 & 0.0816 & 0.0494 & 0.0334 & 0.0690 & 0.0664 & 0.0521 & 0.0797 \\
7 & 0.3459 & 0.5156 & 0.2669 & 0.5478 & 0.5180 & 0.3551 & 0.3225 \\
\bottomrule
\end{tabular}
}
\end{table}
\begin{table}
\centering
\caption{The US Requirements Tables for 1929}
\label{tab:1929}
\resizebox{.8\columnwidth}{!}{%
\begin{tabular}{lccccccc}
\toprule
\multicolumn{8}{l}{Technical Coefficients Matrix ($A$)} \\
\midrule
 & 1 & 2 & 3 & 4 & 5 & 6 & 7 \\
\midrule 
1 & 0.344 & 0.0057 & 0.0439 & 0.0882 & 0.0168 & 0.0067 & 0.0178 \\ 
2 & 0.0009 & 0.0794 & 0.1693 & 0.0516 & 0.0514 & 0.0098 & 0.0169 \\ 
3 & 0.0006 & 0.0045 & 0 & 0.0077 & 0.025 & 0 & 0.0718 \\ 
4 & 0.0949 & 0.0755 & 0.2443 & 0.259 & 0.2188 & 0.063 & 0.2553 \\ 
5 & 0.06 & 0.1971 & 0 & 0.028 & 0.0194 & 0.0143 & 0.0679 \\ 
\rowcolor{yellow}
6 & 0.0022 & 0 & 0.0146 & 0.0007 & 0 & 0.0228 & 0.014 \\ 
7 & 0.0676 & 0.2903 & 0.1179 & 0.2708 & 0.1991 & 0.5214 & 0 \\
\midrule
\multicolumn{8}{l}{Simple Direct Requirements Matrix ($N^\texttt{d}$)} \\
\midrule
 & 1 & 2 & 3 & 4 & 5 & 6 & 7 \\
\midrule
1 & 0.5441 & 0.0065 & 0.0454 & 0.1437 & 0.0183 & 0.0069 & 0.0216 \\
2 & 0.0014 & 0.0908 & 0.1752 & 0.0840 & 0.0561 & 0.0101 & 0.0205 \\
3 & 0.0009 & 0.0051 & 0 & 0.0125 & 0.0273 & 0 & 0.0870 \\
4 & 0.1501 & 0.0864 & 0.2529 & 0.4219 & 0.2387 & 0.0652 & 0.3095 \\
5 & 0.0949 & 0.2255 & 0 & 0.0456 & 0.0212 & 0.0148 & 0.0823 \\
\rowcolor{yellow}
6 & 0.0035 & 0 & 0.0151 & 0.0011 & 0 & 0.0236 & 0.0170 \\
7 & 0.1069 & 0.3321 & 0.1220 & 0.4411 & 0.2172 & 0.5394 & 0 \\
\midrule
\multicolumn{8}{l}{Simple Indirect Requirements Matrix ($N^\texttt{i}$)} \\
\midrule
 & 1 & 2 & 3 & 4 & 5 & 6 & 7 \\
\midrule
1 & 0.0375 & 0.0830 & 0.1125 & 0.0975 & 0.0939 & 0.0815 & 0.0899 \\
2 & 0.0353 & 0.0533 & 0.0587 & 0.0340 & 0.0521 & 0.0509 & 0.0546 \\
3 & 0.0234 & 0.0473 & 0.0351 & 0.0416 & 0.0335 & 0.0554 & 0.0073 \\
4 & 0.1867 & 0.3208 & 0.2943 & 0.2070 & 0.2689 & 0.3247 & 0.1985 \\
5 & 0.0357 & 0.0574 & 0.0989 & 0.0761 & 0.0698 & 0.0771 & 0.0383 \\
\rowcolor{yellow}
6 & 0.0041 & 0.0087 & 0.0065 & 0.0089 & 0.0073 & 0.0110 & 0.0024 \\
7 & 0.1347 & 0.1833 & 0.2578 & 0.0864 & 0.1874 & 0.1541 & 0.2121 \\
\midrule
\multicolumn{8}{l}{Simple Transfer (Total) Requirements Matrix ($N^\texttt{t}$)} \\
\midrule
 & 1 & 2 & 3 & 4 & 5 & 6 & 7 \\
\midrule
1 & 0.5816 & 0.0895 & 0.1579 & 0.2412 & 0.1123 & 0.0884 & 0.1114 \\
2 & 0.0367 & 0.1442 & 0.2339 & 0.1181 & 0.1082 & 0.0610 & 0.0751 \\
3 & 0.0243 & 0.0524 & 0.0351 & 0.0541 & 0.0608 & 0.0554 & 0.0944 \\
4 & 0.3368 & 0.4072 & 0.5472 & 0.6288 & 0.5076 & 0.3898 & 0.5079 \\
5 & 0.1306 & 0.2829 & 0.0989 & 0.1217 & 0.0910 & 0.0919 & 0.1206 \\
\rowcolor{yellow}
6 & 0.0076 & 0.0087 & 0.0217 & 0.0101 & 0.0073 & 0.0346 & 0.0194 \\
7 & 0.2416 & 0.5155 & 0.3798 & 0.5275 & 0.4047 & 0.6935 & 0.2121 \\
\bottomrule
\end{tabular}
}
\end{table}
\begin{table}
\centering
\caption{The US Requirements Tables for 1919}
\label{tab:1919}
\resizebox{.8\columnwidth}{!}{%
\begin{tabular}{lccccccc}
\toprule
\multicolumn{8}{l}{Technical Coefficients Matrix ($A$)} \\
\midrule
 & 1 & 2 & 3 & 4 & 5 & 6 & 7 \\
\midrule 
1 & 0.4009 & 0.0091 & 0.0802 & 0.1469 & 0.0129 & 0.0079 & 0.0397 \\ 
2 & 0.0006 & 0.0716 & 0.198 & 0.038 & 0.0811 & 0.0124 & 0.017 \\ 
3 & 0 & 0 & 0 & 0 & 0 & 0 & 0.0581 \\ 
4 & 0.0746 & 0.0693 & 0.3189 & 0.2275 & 0.253 & 0.0034 & 0.3359 \\ 
5 & 0.0441 & 0.1969 & 0 & 0.0158 & 0.0158 & 0.009 & 0.0622 \\ 
\rowcolor{yellow}
6 & 0.0009 & 0 & 0.0274 & 0.0008 & 0 & 0 & 0.0108 \\ 
7 & 0.035 & 0.401 & 0.105 & 0.2928 & 0.207 & 0.513 & 0 \\
\midrule
\multicolumn{8}{l}{Simple Direct Requirements Matrix ($N^\texttt{d}$)} \\
\midrule
 & 1 & 2 & 3 & 4 & 5 & 6 & 7 \\
\midrule
1 & 0.7039 & 0.0104 & 0.0824 & 0.2404 & 0.0141 & 0.0080 & 0.0505 \\
2 & 0.0011 & 0.0821 & 0.2035 & 0.0622 & 0.0886 & 0.0125 & 0.0216 \\
3 &  0 &  0 &  0 &  0 &  0 &  0 & 0.0739 \\
4 & 0.1310 & 0.0794 & 0.3278 & 0.3722 & 0.2763 & 0.0034 & 0.4272 \\
5 & 0.0774 & 0.2256 &  0 & 0.0259 & 0.0173 & 0.0091 & 0.0791 \\
\rowcolor{yellow}
6 & 0.0016 &  0 & 0.0282 & 0.0013 & 0 & 0 & 0.0137 \\
7 & 0.0614 & 0.4596 & 0.1079 & 0.4791 & 0.2260 & 0.5175 & 0 \\
\midrule
\multicolumn{8}{l}{Simple Indirect Requirements Matrix ($N^\texttt{i}$)} \\
\midrule
 & 1 & 2 & 3 & 4 & 5 & 6 & 7 \\
\midrule
1 & 0.0518 & 0.1961 & 0.2729 & 0.2061 & 0.1951 & 0.1445 & 0.2080 \\
2 & 0.0262 & 0.0640 & 0.0649 & 0.0312 & 0.0463 & 0.0427 & 0.0549 \\
3 & 0.0105 & 0.0403 & 0.0280 & 0.0326 & 0.0274 & 0.0388 & 0 \\
4 & 0.1567 & 0.4566 & 0.4016 & 0.2640 & 0.3298 & 0.3520 & 0.2268 \\
5 & 0.0228 & 0.0654 & 0.1121 & 0.0746 & 0.0747 & 0.0660 & 0.0388 \\
\rowcolor{yellow}
6 & 0.0025 & 0.0092 & 0.0061 & 0.0074 & 0.0065 & 0.0087 & 0.0028 \\
7 & 0.1191 & 0.2334 & 0.3744 & 0.0817 & 0.2451 & 0.1512 & 0.2719 \\
\midrule
\multicolumn{8}{l}{Simple Transfer (Total) Requirements Matrix ($N^\texttt{t}$)} \\
\midrule
 & 1 & 2 & 3 & 4 & 5 & 6 & 7 \\
\midrule
1 & 0.7557 & 0.2065 & 0.3554 & 0.4464 & 0.2092 & 0.1524 & 0.2585 \\
2 & 0.0273 & 0.1460 & 0.2684 & 0.0934 & 0.1349 & 0.0552 & 0.0765 \\
3 & 0.0105 & 0.0403 & 0.0280 & 0.0326 & 0.0274 & 0.0388 & 0.0739 \\
4 & 0.2876 & 0.5360 & 0.7294 & 0.6362 & 0.6061 & 0.3555 & 0.6541 \\
5 & 0.1002 & 0.2910 & 0.1121 & 0.1005 & 0.0920 & 0.0751 & 0.1179 \\
\rowcolor{yellow}
6 & 0.0040 & 0.0092 & 0.0343 & 0.0087 & 0.0065 & 0.0087 & 0.0165 \\
7 & 0.1805 & 0.6929 & 0.4824 & 0.5608 & 0.4711 & 0.6686 & 0.2719 \\
\bottomrule
\end{tabular}
}
\end{table}

\newpage

\end{document}